\ifdefined\pdfoutput\pdfoutput=1\fi
\documentclass[prd, twocolumn, superscriptaddress, tightenlines, longbibliography, nofootinbib, eqsecnum, amsfonts, amsmath, floatfix, notitlepage]{revtex4-2}

\usepackage{amsmath,amssymb,mathtools}
\usepackage{graphicx}
\usepackage{bm}
\usepackage{booktabs}
\usepackage{array}
\usepackage{tabularx}
\usepackage[usenames,dvipsnames,svgnames,table]{xcolor}
\definecolor{linkcolor}{rgb}{0.6,0,0}
\definecolor{citecolor}{rgb}{0,0,0.75}
\definecolor{urlcolor}{rgb}{0.12,0.46,0.7}
\usepackage{hyperref}

\usepackage[capitalize]{cleveref}
\Crefname{equation}{Equation}{Equations}
\crefname{equation}{Eq.}{Eqs.}
\Crefname{figure}{Figure}{Figures}
\crefname{figure}{Fig.}{Figs.}
\Crefname{table}{Table}{Tables}
\crefname{table}{Table}{Tables}
\Crefname{section}{Section}{Sections}
\crefname{section}{Sec.}{Secs.}
\Crefname{appendix}{Appendix}{Appendices}

\newcolumntype{Z}[1]{>{\hsize=#1\hsize\raggedright\arraybackslash}X}
\newcommand{\peakscale}{s_{\rm peak}}

\newcommand{\Sussex}{Department of Physics \& Astronomy, University of Sussex, Brighton BN1 9QH, UK}

\newcommand{\code}[1]{\texttt{\detokenize{#1}}}
\newcommand{\dd}{\mathrm{d}}
\newcommand{\SK}{S_K}
\newcommand{\CK}{C_K}
\newcommand{\cotK}{\cot{}_K}
\newcommand{\jl}{j_l}
\newcommand{\abs}[1]{\left|#1\right|}
\newcommand{\order}{\mathcal{O}}

\newcommand{\sgn}{\operatorname{sgn}}
\newcommand{\atanTwo}{\operatorname{atan2}}
\newcommand{\sech}{\operatorname{sech}}

\providecommand{\apj}{Astrophys. J.}
\providecommand{\apjs}{Astrophys. J. Suppl.}

\providecommand{\mnras}{Mon. Not. Roy. Astron. Soc.}
\providecommand{\prd}{Phys. Rev. D}

\providecommand{\jcap}{JCAP}
\providecommand{\aap}{Astron. Astrophys.}

\begin{document}

\title{CAMB~2: cosmological power spectra for high-precision surveys}

\author{Antony Lewis}
\affiliation{\Sussex}

\begin{abstract}
Upcoming cosmic microwave background (CMB) and large-scale-structure surveys
require theoretical power spectra with numerical errors well below their
observational uncertainties over the scales that carry most of the constraining
power. We describe a substantial update to \textsc{CAMB} designed to provide
fast, high-precision predictions for two key outputs:
the lensed CMB and matter power spectra. The central development is a new
treatment of the hyperspherical Bessel functions used for line-of-sight
integration in non-flat cosmologies. A leading-order Olver construction maps
the curved radial equation onto the flat spherical Bessel equation by matching
their actions through the turning point. The resulting approximation is exact
in the flat limit, remains smooth through the turning point, and reduces near
flatness to a simple rescaling of the flat Bessel argument and amplitude.
Occasional evaluations of this map also provide accurate starting values for
efficient Numerov propagation between neighbouring source points. We
additionally describe updated integrators, a recalibrated fast recombination model,
updated approximations for massive neutrinos, stabilized parameterized post-Friedmann
dark-energy evolution, and improvements to CMB lensing accuracy.
Numerical convergence is assessed by comparing unboosted default results
against more converged calculations across a grid of more than one hundred
cosmological configurations. The defaults meet conservative $10^{-3}$
pointwise convergence targets over the main lensed-CMB and quasi-linear
matter-power ranges relevant for future surveys.
Errors measured in typical runs are substantially smaller than this.
We also describe the convention dependence of the nominally linear matter power spectrum
when a homogeneous calculation attempts to represent the effects of reionization heating.
Essentially all of the new algorithms and code were developed with LLMs or AI agents
under human supervision. In particular, we introduce \emph{custom-precision} numerical routines: AI-optimized
implementations calibrated to the accuracy required by the physical problem,
rather than to the fixed precision of general-purpose libraries.
\end{abstract}

\maketitle

\section{Introduction}
\label{sec:intro}

The cosmic microwave background (CMB) and the large-scale matter distribution are
the two pillars of precision cosmology. For a given model, Einstein--Boltzmann
codes such as \textsc{CAMB}~\cite{Lewis:1999bs}\footnote{The
\textsc{CAMB} v2.x source code is available at \url{https://github.com/cmbant/CAMB};
the Python package can be installed from PyPI with
\href{https://pypi.org/project/camb/}{\texttt{pip install camb}}; this paper is based on v2.0.0.} and
\textsc{CLASS}~\cite{Blas:2011rf} predict both: they solve the linearized
perturbation equations and project them onto observable angular power spectra by
line-of-sight integration~\cite{Seljak:1996is}. \textsc{CAMB} descends directly
from \textsc{CMBFAST}~\cite{Seljak:1996is}, which introduced the line-of-sight
method as a fast replacement for the explicit integration of the full Boltzmann
hierarchy used in earlier codes such as
\textsc{COSMICS}~\cite{Ma:1995ey,Bertschinger:1995er} and the closely related
all-sky FRW treatments of the 1990s~\cite{Hu:1998mn}.
Ref.~\cite{Zaldarriaga:1997va} later generalized \textsc{CMBFAST} to non-flat
geometries using hyperspherical Bessel functions, and \textsc{CAMB} itself began
as a rewrite supporting closed as well as open and flat FRW
models~\cite{Lewis:1999bs}, built on the covariant CMB formalism of
Refs.~\cite{Hu:1998mn,Challinor:2000as}.

The accuracy these codes must reach is set by the data. \textit{Planck} analyses
were well served by default settings tuned to roughly the $0.1\%$ level at
$\ell\lesssim 2000$~\cite{Howlett:2012mh}. Higher-resolution ground-based CMB
experiments such as the Simons Observatory and its planned Advanced SO
extension~\cite{SO:2018sim}, together with forthcoming higher-precision galaxy
lensing and large-scale-structure density probes, push to both higher precision
and a wider range of scales. Over this whole range the theoretical spectra must
stay accurate well below the measurement errors, and meeting that demand cheaply
is the goal of the present work.

Ref.~\cite{Howlett:2012mh} gave the last systematic description of
\textsc{CAMB}'s accuracy and sampling machinery, in the context of
\textit{Planck}. Much has changed since, and the most substantial changes have no
published description; this paper fills that gap. It is not a review of the
underlying physics, which is standard and documented
elsewhere~\cite{Challinor:2000as,Lewis:2006fu,Lewis:1999bs}: we reproduce
established results only as far as needed to state a numerical problem precisely,
and otherwise concentrate on what is new.

We focus on the two outputs that dominate practical analyses, the lensed
CMB spectra and the matter power spectrum. \textsc{CAMB} of course also computes
cross-correlations, number-count and lensing observables, and a range of
extended models; these are not the focus here, although the line-of-sight and
accuracy improvements we describe benefit them as well.

The main innovation is a new treatment of the hyperspherical Bessel
functions used in curved-space line-of-sight integration
(\cref{sec:bessel}), which is the technical heart of the paper. Instead of
mapping the curved radial equation onto an Airy function near its turning point,
as in the uniform WKB treatments of
Ref.~\cite{Kosowsky:1998nc}, we map it onto the
\emph{flat} spherical Bessel equation. The comparison solution is then the
ordinary flat spherical Bessel function, evaluated by a fast dedicated
routine; the curved problem then reduces to a coordinate
map and an amplitude. The flat-Bessel mapping is very similar in spirit to the
mapping approximation used by Refs.~\cite{tram2013,lesgourgues-tram2013}, but more general and implemented differently.
We also describe a new fast fit to recombination models, direct fits for the massive-neutrino background and
momentum sampling (\cref{sec:recomb,sec:neutrinos}), a stabilized parameterized
post-Friedmann (PPF) dark-energy evolution (\cref{sec:de}), and the lensing
accuracy work (\cref{sec:lensing}).

The second contribution is methodological, and has two related parts. The first
is a design philosophy: a move away from general-purpose special-function
libraries with their fixed accuracy/speed trade-off, towards
\emph{custom-precision} numerical methods, fitted and explicitly calibrated
to exactly the accuracy a given problem needs on the grid it actually samples. The
hyperspherical Bessel routine is the clearest example, but the same principle --
spend accuracy where it matters, and demonstrate it with a reproducible test -- runs
through all of the changes. The second part is the process: we developed almost
all of this work with AI tools. For the more analytic problems we used LLMs to
suggest and critique candidate algorithms and approximations, with iterative
testing and cross-LLM and human validation. For accuracy-stability failures and
bugs, agents with a human in the loop drove a deterministic validation harness to
isolate and fix each issue (\cref{sec:validation,sec:agentic}). The agents made it
practical to fit and systematically validate so many small numerical changes
and alternative methods, a labour that would otherwise be prohibitive.

We begin with conventions and code notation (\cref{sec:conventions}) and an
overview of how the perturbation equations are evolved (\cref{sec:evolution}),
then describe the line-of-sight and hyperspherical-Bessel calculation
(\cref{sec:bessel}), recombination, reionization and background evolution
(\cref{sec:recomb}), massive-neutrino background and momentum sampling
(\cref{sec:neutrinos}), stabilized dark-energy evolution (\cref{sec:de}), 
lensed-CMB and matter-power accuracy (\cref{sec:lensing}). We then describe the validation methodology
(\cref{sec:validation}), the agentic development approach (\cref{sec:agentic}),
performance (\cref{sec:performance}), and conclusions.
\Cref{app:bessel} collects the detailed Bessel mathematics
and \cref{app:accuracy} the accuracy-parameter reference.

\section{Conventions and the code}
\label{sec:conventions}

\textsc{CAMB} works in conformal time and megaparsec units, propagating the
covariant perturbation equations in the zero-acceleration (CDM) frame, which is
equivalent to the synchronous gauge. The variables, the multipole hierarchy, the
scalar/vector/tensor harmonic expansion and the $C_\ell$ normalization follow
Refs.~\cite{Challinor:2000as,Lewis:2004ef} and are documented in the long-standing
\textsc{CAMB} working notes that accompany the code~\cite{camb_notes}; we do not
repeat them. For multipoles we write $\ell$ for the CMB power spectra, $L$ for
the lensing-potential spectrum, and $l$ for the mathematical multipole order of
the radial functions and Boltzmann hierarchies. The curvature conventions used
by the Bessel routines are given at the start of \cref{sec:bessel}.

A reader who needs to relate \textsc{CAMB}'s internal variables to equations
written in another gauge can use the \code{camb.symbolic} module, which expresses
the linearized scalar perturbation equations symbolically in the covariant
notation, projects them into the Newtonian or synchronous gauge, constructs
gauge-invariant combinations, and records how each symbolic quantity maps onto the
corresponding code variable. The same module can convert a symbolic source
expression into compiled code for custom line-of-sight sources, which is the
mechanism behind \textsc{CAMB}'s custom-source and time-evolution interfaces.
This makes the source and gauge conventions of the code programmatically explicit,
which is useful both for adapting external results and for checking the
implementation.

\section{Evolving the perturbation equations}
\label{sec:evolution}

\textsc{CAMB} computes its observables from the linearized
Einstein--Boltzmann system, evolved in conformal time independently for each
wavenumber. The photon and neutrino phase-space distributions are expanded in
angular multipoles, giving for each species a coupled hierarchy for the density,
velocity, anisotropic stress and higher
moments~\cite{Ma:1995ey,Hu:1998mn,Challinor:2000as}; massive neutrinos carry a
separate hierarchy for each sampled momentum (\cref{sec:neutrinos}). The
hierarchies are truncated at a finite multipole with a free-streaming closure
that approximates the coupling to the first unevolved moment, so that the cut
does not reflect spurious power back into the retained
multipoles~\cite{Ma:1995ey}. Initial conditions are set deep in radiation
domination from the regular series solution of the
equations~\cite{Bucher:1999re}, adiabatic by default. The evolved variables are
combined into the line-of-sight
sources~\cite{Seljak:1996is,Zaldarriaga:1997va,Challinor:2000as}, stored on a
sampled time grid, and projected against the radial functions of
\cref{sec:bessel} to give the transfer functions and hence the $C_\ell$.

Before recombination, rapid Thomson scattering couples the photons and baryons
so tightly that the photon hierarchy is stiff: the scattering rate is far larger
than the expansion and acoustic-oscillation rates. Rather than integrate this
stiff system directly, \textsc{CAMB} evolves a reduced tight-coupling
approximation, in which the photon--baryon velocity slip and the leading photon
anisotropy are expanded to second order in the scattering
time~\cite{CyrRacine:2010bk,Blas:2011rf}. The approximation is switched off mode
by mode -- and the full hierarchy switched on -- once the scattering is no longer
fast compared with the mode frequency and the expansion rate, after which the
system is non-stiff for the rest of its evolution.

This non-stiff system is advanced with an adaptive embedded Runge--Kutta
integrator that sets its own step size from a built-in error estimate. The one
substantive change here is the integrator itself: the long-standing
\textsc{dverk} Runge--Kutta--Verner routine has been replaced by a
Dormand--Prince 5(4) pair~\cite{DormandPrince:1980}, keeping the same
step-control and interrupt interface. The Dormand--Prince pair reaches the same tolerances with fewer
right-hand-side evaluations per step, and reuses the final stage of an accepted
step as the first stage of the next; each right-hand-side evaluation dominates the cost, since
each one sweeps the full multipole hierarchy. The integration tolerance scales
with \code{IntTolBoost} (\cref{app:accuracy}).

By construction the explicit integrator is only ever applied where the
equations are non-stiff. The early stiff radiation era is handled by the
tight-coupling approximation above; the separately stiff early phase of the
recombination network is integrated with an implicit Rosenbrock method before
handing back to the same Dormand--Prince solver (\cref{sec:recomb-model}); and
the stiff high-$k/\mathcal H$ limit of the PPF dark-energy equation is
regularized analytically rather than integrated (\cref{sec:de}). Stiffness is
thus removed by a method matched to each regime, leaving the explicit adaptive
integrator a well-conditioned problem wherever it runs.

\section{Line-of-sight integration and hyperspherical Bessel functions}
\label{sec:bessel}

The line-of-sight solution writes each transfer function as an integral of a
source against a radial Bessel function~\cite{Seljak:1996is,Zaldarriaga:1997va,Challinor:2000as}. In
a flat universe the radial function is the ordinary spherical Bessel function
$\jl(k\chi)$; in a curved Friedmann--Robertson--Walker (FRW) universe it is the
hyperspherical (ultraspherical) Bessel function $\phi_l^\nu(K,\chi)$, the
non-flat generalization of $\jl$ \cite{abbott-schaefer,Zaldarriaga:1997va,Kosowsky:1998nc,tram2013}.
\textsc{CAMB} evaluates it for many source times, wavenumbers and multipoles, so a
direct special-function call at every source point would be far too slow. This
section describes how we avoid that. The detailed formulae are collected in
\cref{app:bessel}; here we give the construction and the results.

\Cref{sec:bessel-eq} sets up the radial equation and its turning
point. \Cref{sec:bessel-olver,sec:bessel-amp} then build the main
approximation: an Olver action map that relates $\phi_l^\nu$ to the ordinary
flat-space Bessel function $\jl$ and does most of the work at general $l$.
\Cref{sec:bessel-shifted} gives a cheaper near-flat limit of this map, which
covers entire source integration ranges from shifted flat-Bessel lookups rather
than a fresh evaluation at each point. \Cref{sec:bessel-gates} sets out accuracy
gates that choose the appropriate approximation.
When the near-flat approximations cannot be used, evaluating
the map at every source point would be slow, so \cref{sec:bessel-numerov}
shows how two evaluations can be used to seed an integration that
propagates the solution to neighbouring points. \Cref{sec:bessel-recur}
handles fallback seed evaluation with an exact
recurrence for low $l$ and a few invalid corners. We close with optimized multipole sampling and the residual curvature
numerical overhead (\cref{sec:bessel-lsampling,sec:bessel-curvature}).

\subsection{The radial function and its turning point}
\label{sec:bessel-eq}

We denote the dimensional spatial curvature by $\mathcal K$. Distances are
rescaled by $R_K=\abs{\mathcal K}^{-1/2}$, so that $\chi=r/R_K$ and only the sign
$K=\sgn(\mathcal K)=+1,0,-1$ (closed, flat, open) remains. The radial metric
function and its derivative are
\begin{equation}
  \begin{aligned}
  \SK(\chi)&=\{\sin\chi,\;\chi,\;\sinh\chi\},\\
  \CK(\chi)&=\{\cos\chi,\;1,\;\cosh\chi\},
  \end{aligned}
\end{equation}
for $K=\{+1,0,-1\}$, and $\cotK\chi\equiv\CK/\SK$. For $K=0$ there is no
curvature scale and $R_K$ can be taken to be any convenient reference length:
only the combination $\nu\chi=qr$ then enters, so setting $K=0$ at finite $\nu$
and $\chi$ is equivalent to the physical flat limit of a curved model,
$R_K\to\infty$ at fixed $qr$ (i.e.\ $\nu\to\infty$ and $\chi\to0$ with
$\nu\chi$ fixed). The dimensionless radial
eigenvalue is $\nu=qR_K$, where $q$ is the transfer-grid wavenumber; the
primordial power is evaluated at the physical wavenumber $k$ with
$(kR_K)^2=\nu^2-K$ for scalars and $\nu^2-3K$ for tensors. The hyperspherical
construction below depends only on $\nu$ and $l$, so it is common to all mode types;
only this eigenvalue relation and the spin-dependent source projection differ. In the
flat limit $q=k=\nu/R_K$.

It is convenient to work with the reduced function
$u_l^\nu=\SK\,\phi_l^\nu$, which obeys a Schr\"odinger-form equation with no
first-derivative term,
\begin{equation}
  u''=\left[\frac{l(l+1)}{\SK^2(\chi)}-\nu^2\right]u ,
  \label{eq:reduced}
\end{equation}
with primes denoting $\dd/\dd\chi$. The regular solution is fixed by
\begin{equation}
\phi_l^\nu\sim
\big[\textstyle\prod_{j=1}^l b_j\big]\chi^l/(2l+1)!!
\end{equation}
 as $\chi\to0$, with
$b_j(K,\nu)=\sqrt{\nu^2-Kj^2}$; for $K=0$ this is the usual
$\jl(\nu\chi)\sim(\nu\chi)^l/(2l+1)!!$. For $K=+1$, $\nu$ is a positive integer
and the regular spectrum requires $\nu>l$.

Writing $\hat\ell=\sqrt{l(l+1)}$ for the exact centrifugal multipole and
$\alpha=\nu/\hat\ell$, \cref{eq:reduced} becomes
$u''+\hat\ell^2[\alpha^2-\SK^{-2}]u=0$, with a turning point where the bracket
vanishes,
\begin{equation}
  \SK(\chi_t)=\frac1\alpha=\frac{\hat\ell}{\nu}.
  \label{eq:turning}
\end{equation}
For $\chi<\chi_t$ the solution is on the evanescent, non-oscillatory branch; for
$\chi>\chi_t$ it oscillates. The turning point is what makes accurate evaluation
delicate: a naive WKB amplitude diverges there even though the true function
stays finite, so any construction has to remain uniform through the transition.
The uniform construction we develop below (\cref{sec:bessel-olver,sec:bessel-amp})
does exactly this and is the method the code uses for general $l$; only for
$l\le2$ and a few exceptional corners does it fall back on the exact recurrence of
\cref{sec:bessel-recur}.

For most cosmologies of interest the curvature radius is far larger than the
horizon, so every source point has $\chi\lesssim\chi_0=\eta_0\sqrt{\abs{\mathcal
K}}\ll1$, where $\eta_0$ is the conformal time today. The Bessel factor for multipole $l$ peaks near $\SK(\chi)\simeq\hat\ell/\nu$,
so the contributing modes have $\alpha\simeq1/\SK(\chi)\gtrsim1/\chi_0\gg1$,
i.e.\ $\nu\gg l$ except in low-$l$, endpoint, or deep-tail regions. The near-flat
shortcuts below live in exactly this comfortable regime.

\subsection{A flat-Bessel comparison: the Olver action map}
\label{sec:bessel-olver}

Olver (uniform Liouville--Green) asymptotics map a second-order equation with a
large parameter onto a comparison equation by matching the action integral from
the turning point and including the corresponding amplitude
factor~\cite{olver1958uniform,olver}. The textbook choice is the Airy equation, giving the uniform
WKB/Langer treatment used for hyperspherical Bessel functions by
Ref.~\cite{Kosowsky:1998nc}. Our choice is different
and better adapted to the problem: we use the \emph{flat} spherical Bessel
equation as the comparison equation. Define the regular flat comparison
solution
\begin{equation}
  v_0(z)\equiv z\jl(\nu z),
\end{equation}
which satisfies
\begin{equation}
  \frac{\dd^2v_0}{\dd z^2}=\left[\frac{l(l+1)}{z^2}-\nu^2\right]v_0,
  \label{eq:flat-comparison}
\end{equation}
which differs from \cref{eq:reduced} only by replacing $\SK(\chi)$ with $z$. We seek a
change of coordinate $z=z(\chi)$ and amplitude $A(\chi)$ such that the exact
curved reduced radial function is
\begin{equation}
  u_l^\nu(\chi)=A(\chi)\,V\!\big(z(\chi)\big),
  \label{eq:olver-ansatz}
\end{equation}
where at leading order the transformed solution is approximated by the flat
comparison solution, $V\simeq v_0$. The resulting approximation is
\begin{equation}
  \begin{aligned}
  u_l^\nu(K,\chi)&\simeq A(\chi)\,z(\chi)\,\jl\!\big(\nu z(\chi)\big),\\
  \phi_l^\nu&\simeq \frac{A(\chi)z(\chi)}{\SK(\chi)}\jl\!\big(\nu z(\chi)\big).
  \end{aligned}
  \label{eq:basic-olver}
\end{equation}
For closed models ($K=+1$) the argument is first folded into
$0\le\chi\le\pi/2$ using the parity relations \eqref{eq:closed-parity} of
\cref{app:bessel-recur}, whose accumulated parity sign is restored after
evaluation; on the folded interval $\SK$ is monotonic and $\CK\ge0$. The
approximation computes $z$ and $A$, calls the flat
Bessel routine for $\jl(\nu z)$, and then multiplies by $Az$ (and divides by $\SK$ for the unreduced
value if needed).

The coordinate $z(\chi)$ is fixed by equality of the action measured from the
turning point. With $f_K=\alpha^2-\SK^{-2}$ and $f_0=\alpha^2-z^{-2}$, and the
dimensionless flat action $I_0$ defined in \cref{eq:flat-action}, the map is
\begin{equation}
  Q_K(\chi)=I_0(\alpha z),\qquad z_t=\frac1\alpha=\frac{\hat\ell}{\nu},
  \label{eq:action-map}
\end{equation}
where $Q_K$ is the curved action integral \cref{eq:curved-action}. Both actions
are positive and measured away from their respective turning points. On corresponding branches of the two problems,
\begin{equation}
  \frac{\dd Q_K}{\dd\chi}=s\sqrt{\abs{f_K(\chi)}},\qquad
  \frac{\dd I_0(\alpha z)}{\dd z}=s\sqrt{\abs{f_0(z)}},
\end{equation}
where $s=-1$ on the evanescent side and $s=+1$ on the oscillatory side.
Differentiating the action map \cref{eq:action-map} therefore gives
\begin{equation}
  z'\equiv\frac{\dd z}{\dd\chi}=\sqrt{\frac{\abs{f_K}}{\abs{f_0}}}
  =\sqrt{\frac{f_K}{f_0}},
  \label{eq:zprime}
\end{equation}
because $f_K$ and $f_0$ have the same sign on corresponding branches.
Equivalently, $f_K=(z')^2f_0$; this relation extends continuously through the
turning point. The map is
inverted on the branch corresponding to the original point, $\alpha z<1$ for
$\chi<\chi_t$ and $\alpha z>1$ for $\chi>\chi_t$, which maps the evanescent and
oscillatory regions onto their flat-space counterparts and makes $z(\chi)$
unique. The intuition
is the usual WKB one: away from the turning point the regular solution behaves as
$|f|^{-1/4}\exp(\pm\hat\ell Q)$ on the evanescent side and
$|f|^{-1/4}\cos(\hat\ell Q-\pi/4)$ on the oscillatory side, with $\hat\ell Q$ the action.
Two equations with the same action measured from their turning points have
solutions that stay in step everywhere: they have the same leading phase, node
structure, and evanescent suppression, so $\jl(\nu z(\chi))$ already carries the curved phase and
tunnelling. Because the turning point maps to the turning point ($Q_K=0\to
I_0=0$), the construction is automatically smooth there, unlike a bare WKB
amplitude. The curved action $Q_K$ has closed analytic forms
(\cref{app:bessel-action}), so no quadrature is needed in the hot loop; inverting
the flat action for $z$ uses stable branch parameterizations, also given in the
appendix.

We evaluate the resulting ordinary spherical Bessel function $\jl(\nu z)$ with a fast custom-precision routine (better than $10^{-5}$ peak accuracy at high $l$; for details of the LLM-optimized approximations and fits see~\cite{cambflatbessel}).
For flat and near-flat cases, the function is splined from precomputed tables.

\subsection{Amplitude, the flat limit, and the turning point}
\label{sec:bessel-amp}

Matching the action fixes the coordinate map, while the amplitude is determined
by requiring the transformed equation to retain Schr\"odinger form. Substituting
$u_l^\nu=A(\chi)V(z(\chi))$ gives
\begin{equation}
  u''=A(z')^2V_{zz}+\left(2A'z'+Az''\right)V_z+A''V.
\end{equation}
The flat comparison equation contains no first derivative, so we use the
standard Liouville transformation and set
\begin{equation}
  2A'z'+Az''=0.
\end{equation}
This gives
\begin{equation}
  A(\chi)=C\,[z'(\chi)]^{-1/2}
  =C\left|\frac{f_0}{f_K}\right|^{1/4},
  \label{eq:amplitude-general}
\end{equation}
where $C$ is independent of $\chi$. Equivalently, this choice makes the
Wronskians satisfy
\begin{equation}
  W_\chi[u_1,u_2]=A^2z'\,W_z[V_1,V_2]
  =C^2W_z[V_1,V_2].
\end{equation}
The transformed equation is independent of $C$, which only determines how the
overall normalization is divided between $A$ and $V$. Once we approximate the
transformed regular solution by the conventionally normalized flat solution
$v_0$, however, this constant fixes the normalization of the resulting
approximation.

The canonical Wronskian-preserving convention sets C=1, for which the coordinate transformation
 reduces exactly to $A=1$ in flat space.
 As shown in \cref{app:bessel-smallchi}, this convention reproduces the exact origin normalization through $\order(h)$ and its leading large-$l$ contribution at $\order(h^2)$; the remaining finite-$l$ mismatch is included in the calibrated approximation error.
  The transformed equation is then
\begin{equation}
  V_{zz}+\left[\hat\ell^2 f_0(z)+\Psi(z)\right]V=0,
\end{equation}
where $\Psi$ is the Schwarzian residual given explicitly in
\cref{app:bessel-action}. The leading-order Olver construction neglects $\Psi$
and approximates the regular transformed solution by
$V(z)\simeq v_0(z)=z\jl(\nu z)$, recovering
\cref{eq:basic-olver}. The construction is therefore explicit: action matching fixes $z$,
cancellation of $V_z$ fixes the $\chi$ dependence of $A$, the canonical
Wronskian-preserving convention sets $C=1$, and neglecting $\Psi$ gives
the leading-order approximation.

At the turning point both $f_0$ and $f_K$ vanish, but the limit is finite: with
$C_t=\CK(\chi_t)$,
\begin{equation}
  z-z_t\simeq C_t^{1/3}(\chi-\chi_t),\qquad A(\chi_t)=C_t^{-1/6}.
\end{equation}
Thus the apparent $0/0$ in $f_0/f_K$ is removable: $z'(\chi_t)=C_t^{1/3}$ is
finite, and so is the amplitude. For $K=0$ we have $\SK=\chi$, $z=\chi$, $A=1$,
and $\Psi=0$, so the construction reduces \emph{exactly} to $\jl(\nu\chi)$.

\subsection{The near-flat small-$\chi$ approximations}
\label{sec:bessel-shifted}
Very near the origin, inversion of the full action map subtracts nearly equal
quantities, so we instead use the local expansion of the differentiated map
given in \cref{app:bessel-smallchi}. Formally this is a joint expansion in
$a=K\chi^2$ and $h=K/\alpha^2$. Its leading, $\chi$-independent part gives the
cheapest useful approximation. We denote the cubic truncation used by the
implementation by
\begin{equation}
  F_0=
  1-\frac{h}{6}-\frac{13h^2}{360}
  -\frac{737h^3}{45360},
  \qquad
  h=K\frac{l(l+1)}{\nu^2}.
  \label{eq:F0}
\end{equation}
At the origin the coordinate map and its derivative have the same expansion,
so $z\simeq F_0\chi$ and $z'\simeq F_0$. The Liouville amplitude
\cref{eq:amplitude-general}, with $C=1$, is therefore
$A\simeq F_0^{-1/2}$. Consequently the prefactor $Az$ in
\cref{eq:basic-olver} becomes $\sqrt{F_0}\chi$, while
$\nu z\simeq\nu F_0\chi\equiv\nu_{\rm eff}\chi$, giving
\begin{equation}
  \phi_l^\nu(K,\chi)\simeq
  \sqrt{F_0}\,\frac{\chi}{\SK(\chi)}
  \jl(\nu_{\rm eff}\chi),
  \qquad
  \nu_{\rm eff}=\nu F_0 .
  \label{eq:shifted-nu}
\end{equation}

This is as cheap as a flat-table lookup: a flat $\jl$ evaluated at a shifted
argument $\nu_{\rm eff}\chi$, multiplied by $\sqrt{F_0}$ and the geometrical factor
$\chi/\SK$. Expanding, $\nu_{\rm eff}^2=\nu^2[1-h/3-2h^2/45-\dots]$, whose leading
correction $-\nu^2 h/3=-Kl(l+1)/3$ is exactly the leading constant curvature shift
obtained by expanding $1/\SK^2$ directly; the $F_0$ form retains the next two
constant Olver terms at (almost) no extra cost. This near-flat behaviour is what makes the
flat comparison so natural: in the near-flat regime the Olver coordinate map is
almost a constant rescaling of the flat argument.

\subsection{Accuracy gates and the approximation hierarchy}
\label{sec:bessel-gates}

The approximations above are used in a controlled hierarchy rather than as
unconditional replacements for the hyperspherical Bessel function. Although the
full action map is exact for $K=0$, for $K=\pm1$ it is a leading-order comparison
approximation obtained by neglecting the Schwarzian residual $\Psi$ in
\cref{eq:olver-residual}. Its use must therefore be restricted to regions where
the resulting error is sufficiently small.

The gates are calibrated using peak-normalized Bessel errors, since a pointwise
relative error would be meaningless near Bessel zeros. They keep the raw-Olver
error envelope near $10^{-4}$ over most of the open and closed validation grid,
with a small number of accepted cases reaching $2\times10^{-4}$. Here ``raw
Olver'' means \cref{eq:basic-olver} evaluated with the full action map and
without any fallback.

The broad structure of the gates is motivated by the asymptotics. In the
near-flat oscillatory overlap region, $1/\alpha\ll\chi\ll1$, the Schwarzian
residual is suppressed relative to the leading transformed coefficient as
\begin{equation}
  \left|
  \frac{\Psi}{\hat\ell^2 f_0}
  \right|
  \simeq
  \frac{1}{30\hat\ell^2\alpha^4},
  \qquad
  \alpha=\frac{\nu}{\hat\ell},
  \label{eq:olver-relative-residual}
\end{equation}
so the full action map becomes more accurate as either $\alpha$ or $l$
increases. This scaling explains the principal dependence of the gates but does
not determine their numerical boundaries. The constants in the implemented
inequalities are empirical calibration parameters, obtained by comparison with
the recurrence-based reference evaluation over a wide grid in
$(K,l,\nu,\chi)$.

The pointwise dispatcher \code{phi_olver} first tests the local small-$\chi$
map and then, where appropriate, the raw-Olver value. If neither is sufficiently
accurate, it classifies the remaining cases using endpoint and spectral metrics,
tries validated Airy and open small-$\nu$ fallbacks~\cite{cambhypersphericalbessel}, and finally uses the stable
recurrence evaluation. We refer to this complete operation as a \emph{pointwise evaluation}. The precise decision order, numerical thresholds,
and source-integration gates are given in \cref{app:bessel-gates}.

\subsection{Source integration: Numerov stepping and re-seeding}
\label{sec:bessel-numerov}

Inside the line-of-sight integral the Bessel function is needed at many
neighbouring times. Rather than perform an independent pointwise evaluation at
each one, we use \code{phi_olver} only to supply starting values and then
advance the reduced equation by Numerov integration~\cite{Numerov:1924,Numerov:1927}. The reduced equation has the special form $y''=Q(\chi)y$ with
$Q=l(l+1)/\SK^2-\nu^2$ and no first-derivative term, which is exactly the case for
which Numerov's method attains an $\order(\delta\chi^6)$ local finite-difference residual and fourth-order accumulated solution error:
\begin{multline}
  y_{n+1}=
  \Big[2\big(1+\tfrac{5(\delta\chi)^2Q_n}{12}\big)y_n
  -\big(1-\tfrac{(\delta\chi)^2Q_{n-1}}{12}\big)y_{n-1}\Big] \\
  {}\times\left(1-\tfrac{(\delta\chi)^2Q_{n+1}}{12}\right)^{-1},
  \label{eq:numerov}
\end{multline}
where $\delta\chi$ is the (signed) substep. Being a two-step recurrence it needs two
consecutive accurate values to start; these come from two pointwise evaluations near
the turning point, after which \cref{eq:numerov} propagates the solution across the
range, and the stored value at each source grid point is $\phi_l^\nu=y_n/\SK$.

Two details keep the loop cheap and stable. The metric factors $\SK,\CK$ are not
recomputed with fresh transcendental calls at each substep but stepped forward by
the fixed substep using angle-addition (a two-term linear update with precomputed
coefficients), and after a fixed re-bootstrap interval -- of order a hundred
steps, reduced at higher accuracy -- the code re-seeds by re-evaluating the
two Numerov starting values from fresh pointwise evaluations.
This bounds the accumulated drift of both recursions without paying for a
pointwise evaluation at every step. The substep is a fraction of $1/\nu$ so the oscillation is well
sampled, with accuracy-boost factors scaling the step and the re-bootstrap
interval, and the integration stops early in the exponentially small tails.

The Numerov substep is capped independently of the source time-grid spacing, so
a single interval between source points can require many substeps in significantly
non-flat models. One pointwise evaluation costs approximately as much as 13 Numerov
substeps. Accounting for the additional stepping and re-seeding overhead, benchmarks
favour direct evaluation when the required number of substeps per source interval,
$n_{\rm sub}$, exceeds 10. In this regime the code evaluates $\phi_l^\nu$ directly
by pointwise evaluation at each source-grid point, avoiding both the
intermediate Numerov grid and the associated accumulation of propagation error.
This gate reduces the cost of non-flat source integration by roughly one third for
closed models. It is essentially inactive for open models and for tensor sources,
whose denser time grids and lower maximum wavenumbers keep $n_{\rm sub}$ well
below 10.

Before stepping the ODE, the code first tries
the near-flat constant-map (small-$\chi$) approximation above and the fuller local small-$\chi$ map of \cref{app:bessel-smallchi},
which fill an entire range directly from the (shifted) flat Bessel
table when their accuracy gates pass. These avoid stepping the curved ODE at all and
vary continuously into the exact flat result as $\mathcal K\to0$,
avoiding a discontinuous switch between flat and non-flat treatments -- the same
continuity goal as the \textsc{CLASS} flat-rescaling
approximation~\cite{lesgourgues-tram2013}, here achieved with the local Olver
coefficients rather than an empirical amplitude.

\subsection{Stable recurrence fallback}
\label{sec:bessel-recur}

For low $l$, for $\alpha$ below the leading-Olver regime, and in a few invalid
corners, \textsc{CAMB} uses a fallback sequence. The first fallback is a
second-order one-point Airy approximation in validated high-$l$ regions; open
models also have a narrow small-$\nu$ approximation for the remaining very low
$\nu/l$ tail. If neither applies, the code evaluates $\phi_l^\nu$ by direct
recurrence. This recurrence machinery is essentially
standard: exact $l=0,1$ seeds, upward recurrence in calibrated stable regions, and
Miller downward recurrence elsewhere started either from a continued-fraction
logarithmic derivative (flat and open) or from the finite closed-spectrum
endpoint, with a closed-space Gegenbauer identity supplying the start near that
endpoint. It follows the hyperspherical-Bessel relations of Abbott \&
Schaefer~\cite{abbott-schaefer} and the stable-recursion improvements of
Refs.~\cite{tram2013,lesgourgues-tram2013}, and a direct comparison with
\textsc{CLASS} v3.3.4 shows the same seeds, recurrences and closed-space start.
Because the recurrence is standard, we relegate the formulae to \cref{app:bessel-recur}.

\subsection{Multipole sampling and template interpolation}
\label{sec:bessel-lsampling}

The transfer functions, and hence the $C_\ell$, are computed at a sparse set of
multipoles and interpolated to every $\ell$, as in
Ref.~\cite{Howlett:2012mh}. The sampling is dense at low $\ell$ and coarsens to a
step of about $50$ in $\ell$ at high $\ell$, up to a transition multipole (default
$5000$) above which the unlensed spectra are smoothly damping and a widening
logarithmic spacing is used; computing every $\ell$ is available as a reference
limit. Interpolating the $C_\ell$ directly at a $\Delta\ell\sim50$ spacing would
not give quite enough accuracy at all the interpolated points, so for $TT$, $EE$ and $TE$ we instead
spline-interpolate the \emph{difference} between the computed $C_\ell$ and a smooth
fiducial template, adding the accurately interpolated template back afterwards;
the residual is small and featureless, so the coarse sampling suffices.

The ingredient we add here is to shift the template to the model's peak scale
before taking the difference. The code uses the sampling proxy
\begin{equation}
  \vartheta_{\rm samp}=\frac{r_s(z_{\rm drag})}{D_M(\eta_0)},\qquad
  \peakscale=\frac{\vartheta_{\rm samp}^{\rm P18}}{\vartheta_{\rm samp}},
\end{equation}
where \(D_M(\eta_0)\) is the comoving angular-diameter distance corresponding to
the full present conformal time. This is not the usual acoustic scale
\(\theta_*\): it is only a cheap proxy for how the CMB peaks move relative to the
Planck-2018 template. When the proxy differs from unity by more than one per
cent, the fiducial template is evaluated at \(\ell/\peakscale\) so that its
acoustic peaks line up with the model's. The \(\ell\)-sampling gate is one-sided.
If \(\peakscale\ge1-0.03\), the fiducial grid is kept;
wider-spaced peaks are then simply oversampled. If \(\peakscale<1-0.03\), the
peaks are compressed enough that the sampled \(\ell\)-grid is made denser, and the
near-flat shortcuts that reuse the flat Bessel tables are not assumed. In
ordinary CMB parameter fits \(\peakscale\) stays close to unity because the
acoustic scale is tightly measured.

\subsection{Curvature overhead}
\label{sec:bessel-curvature}

The non-flat integration is now only moderately more expensive than flat, rather
than a separate slow path. On the validation curvature sweep
$\Omega_K=\{\pm0.03,\pm0.02,\pm0.01,\pm10^{-3},\pm10^{-5}\}$ the overhead is
modest -- the near-flat $|\Omega_K|\le10^{-3}$ models cost little more than the
exactly flat case -- and the source-integration and lensing work parallelizes
well over the 4--8 threads typical of interactive analysis runs. Current CMB
constraints concentrate close to flatness, where the overhead is smallest.
Detailed timings, including the large speed-up over the previous \textsc{CAMB}
version at matched lensing accuracy and the \textsc{CLASS} comparison, are
collected in \cref{sec:performance}.

\section{Recombination and the background}
\label{sec:recomb}

\subsection{BBN and helium abundance}
\label{sec:bbn}

The primordial helium abundance is not an independent late-time parameter in the
minimal model. Big Bang nucleosynthesis predicts it from the baryon density, the
early expansion rate (usually parameterized by \(N_{\rm eff}\)), the neutron
lifetime and the nuclear reaction network~\cite{Hamann:2007sb,Pisanti:2007hk,
Consiglio:2017pot,Pitrou:2018cgg}. This matters for the CMB because helium
changes the number of free electrons per baryon before and during recombination,
and hence the visibility function and diffusion damping scale; the CMB can also
measure a free helium abundance directly, but much less accurately than BBN
predicts it in the standard model~\cite{Trotta:2003xg,PCP2018}. The other
light-element abundances are not direct inputs to the standard CMB recombination
calculation: deuterium, \(^{3}{\rm He}\) and lithium have number fractions many
orders of magnitude below hydrogen and helium, so they make a negligible
contribution to the electron density, baryon mass density and Thomson opacity.
They are important external BBN observables, especially deuterium, but not CMB
recombination inputs.

In its BBN-consistency mode, \textsc{CAMB} interpolates a two-dimensional
prediction in \(\Omega_{\rm b}h^2\) and \(\Delta N_{\rm eff}\), reads the BBN
helium nucleon fraction \(Y_p^{\rm BBN}\), and converts it to the mass fraction
\(Y_{\rm He}\) used by the recombination and reionization modules.\footnote{In
the Python interface this is selected by leaving \code{YHe} unset.} Older
\textsc{CAMB} releases used PArthENoPE-based calculations~\cite{Pisanti:2007hk,
Consiglio:2017pot}; the current default CAMB interpolation table is generated with \textsc{PRIMAT}, whose
weak-rate treatment changed the helium prediction; later updates mainly affected
deuterium through improved nuclear rates, with smaller helium shifts from the
updated neutron lifetime and convention choices~\cite{Pitrou:2018cgg,
Pitrou:2020etk}. For the release described here we use the September 2024
\textsc{PRIMAT} calculation as the default BBN prediction.\footnote{The
corresponding \textsc{CAMB} interpolation file is
\code{PRIMAT_Yp_DH_ErrorMC_2024.dat}.}

The differences between the supported BBN predictions are small for the CMB. For a
Planck-like model with \(\Omega_{\rm b}h^2=0.02237\) and
\(\Delta N_{\rm eff}=0\), the 2024 \textsc{PRIMAT} calculation gives
\(Y_{\rm He}=0.245656\); the older PArthENoPE calculation with
\(\tau_n=880.2\,{\rm s}\) is lower by \(2.6\times10^{-4}\), and the 2021
\textsc{PRIMAT} calculation is higher by \(2.0\times10^{-4}\).
The corresponding deuterium changes
can be larger -- for example the old standard PArthENoPE prediction is about \(5.7\%\)
high in D/H at the same point -- but this only matters when using an external BBN
abundance likelihood.

\subsection{Recombination model and calibration}
\label{sec:recomb-model}

The CMB anisotropies are sensitive to the free-electron fraction
$x_e(z)\equiv n_e/n_{\rm H}$, the number of free electrons per hydrogen
nucleus, and to the baryon temperature through the visibility function, the
baryon sound speed and the photon diffusion scale.
$f_{\rm He}\equiv n_{\rm He}/n_{\rm H}=Y_{\rm He}/[4(1-Y_{\rm He})]$ is the
helium-to-hydrogen number ratio, so fully ionized hydrogen plus singly ionized
helium corresponds to $x_e=1+f_{\rm He}$ and fully ionized helium to
$x_e=1+2f_{\rm He}$. Future CMB data require at least per-mille-level predictions
for the relevant spectra, so small recombination-history errors in the damping
tail can no longer be treated as harmless implementation details. The statistical
weight of many high-$\ell$ modes means that coherent $10^{-3}$-level residuals
can give non-negligible shifts in spectra and likelihoods. This subsection
describes physical calibration of the effective model to more complete analyses and
some implementation details.

The basic physical bottleneck has been understood since the classic effective
atom treatments of Peebles and of Zel'dovich, Kurt and
Sunyaev~\cite{Peebles:1968ja,Zeldovich:1968}: direct recombination to the ground
state mostly re-ionizes another atom, so the net rate is controlled by escape
from Ly$\alpha$ due to redshifting and by the two-photon $2s\to1s$ channel.
Modern CMB calculations
need this picture refined to the percent-to-subpercent level in $x_e$, including
excited states, feedback between lines, two-photon and Raman processes, and
radiative-transfer effects.

The most complete public recombination calculations used for this purpose are
\textsc{CosmoRec}, \textsc{HyRec} and
\textsc{HyRec-2}~\cite{Chluba:2010ca,Ali-Haimoud:2010tlj,AliHaimoud:2010dx,
Lee:2020obi}. \textsc{CosmoRec} and full \textsc{HyRec} replace the simplest
few-level picture by effective multilevel-atom calculations of the highly
excited states, using precomputed effective rates to a small set of interface
states rather than solving a brute-force many-level atom at runtime. Their main
difference is how this backbone is coupled to radiative transfer. \textsc{HyRec}
uses a compact helium module, treating He\,{\sc iii} recombination in Saha
equilibrium and He\,{\sc ii}\(\to\)He\,{\sc i} with a Peebles-style ODE including
the \(2^1S\) two-photon channel, escape in the 584\,\AA\ singlet line,
H\,{\sc i} continuum opacity, the 591\,\AA\ intercombination line, and feedback
between the main helium lines in an on-the-spot approximation. \textsc{CosmoRec}
can instead solve the radiative-transfer problem for helium photons explicitly,
including overlapping singlet, triplet and quadrupole series, line scattering,
electron scattering, H\,{\sc i} absorption, and feedback between helium
lines~\cite{Chluba:2011hw}. \textsc{HyRec-2} further accelerates the hydrogen
calculation by tabulating a cosmology-dependent correction to the Ly\(\alpha\)
net decay rate calibrated on full \textsc{HyRec}~\cite{Lee:2020obi}, but does not
introduce a similar helium correction. For the helium radiative-transfer
modelling tested here, this makes \textsc{CosmoRec} the more complete baseline;
\textsc{HyRec-2} remains a valuable independent high-accuracy comparison during the
most important hydrogen ionization phase.

\begin{figure*}[t!]
  \centering
  \includegraphics[width=0.82\textwidth]{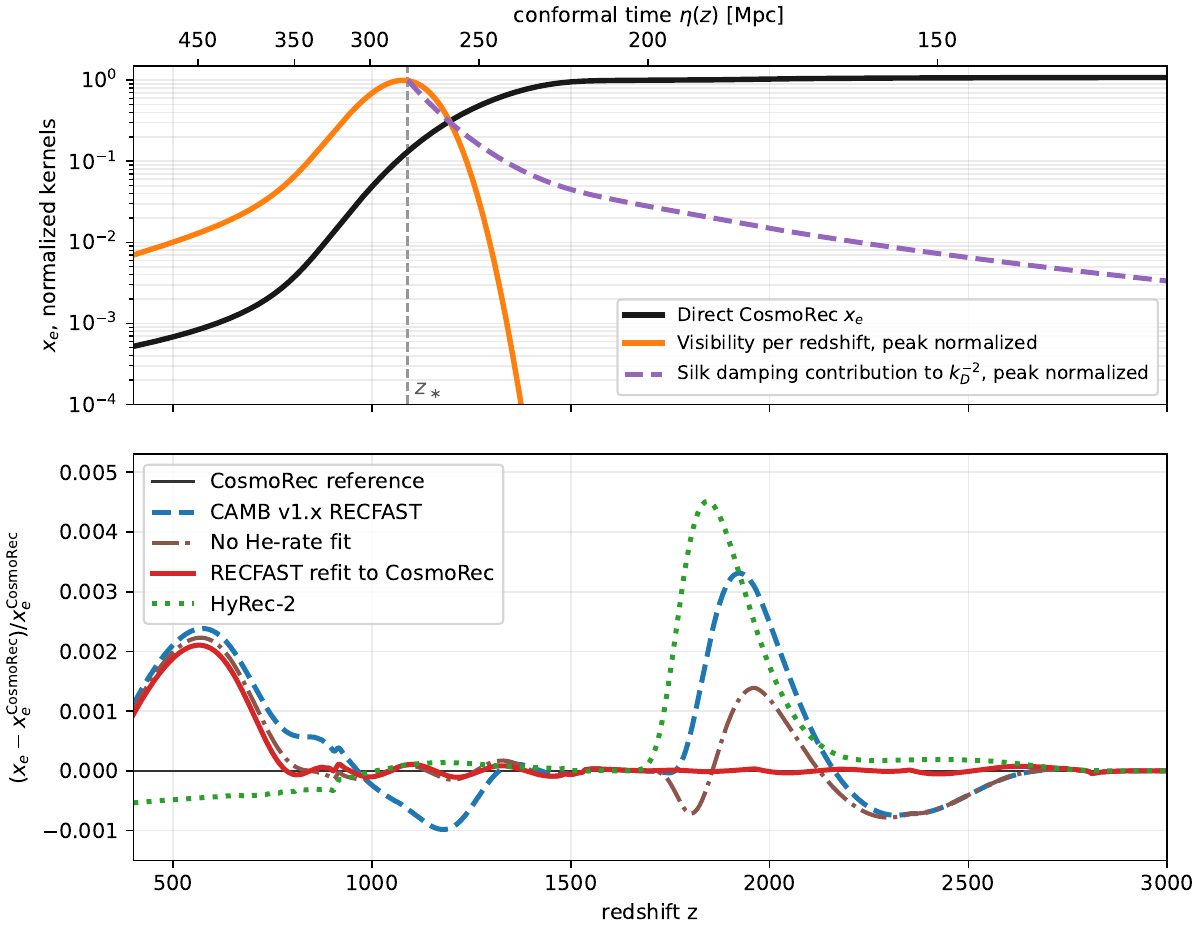}
  \caption{Upper panel: the direct \textsc{CosmoRec} reference $x_e(z)$ for a $\Lambda$CDM model, and the
  peak-normalized visibility density per unit redshift and Silk damping
  integrand. The visibility curve is
  obtained from the \textsc{CAMB} conformal-time visibility by multiplying by
  $|d\eta/dz|=H(z)^{-1}$ before normalization. The damping curve is the positive
  redshift integrand whose integral gives the inverse squared photon diffusion
  scale $k_D^{-2}(z_\ast)$, plotted for $z\ge z_\ast$ and normalized to its
  maximum. The grey dashed vertical line marks $z_\ast$. The upper horizontal
  axis gives the conformal time $\eta(z)$ in Mpc. Lower panel: fractional
  ionization-history residuals relative to the direct \textsc{CosmoRec}
  reference used for the updated
  \textsc{RECFAST} approximation. The reference is
  \textsc{CosmoRec} with \code{accuracy=6}, ten hydrogen shells, \code{n_s_2gamma=8},
  \code{n_s_raman=7} and two diffusion iterations. The ``No He-rate fit'' curve
  uses the same \textsc{RECFAST} refit but without the new multiplicative
  He\,{\sc i} rate correction.}
  \label{fig:recfast-cosmorec-fit}
\end{figure*}

\textsc{CAMB}'s default model remains based on the \textsc{RECFAST} effective
few-level approximation~\cite{Seager:1999bc,Seager:1999km}. \textsc{RECFAST}
evolves separate hydrogen and helium ionization fractions plus the matter
temperature, using Saha solutions while they are accurate and then switching to
effective rate equations. Its helium treatment uses effective rates and
phenomenological correction factors to approximate the main continuum-opacity and
triplet-channel effects, and its hydrogen history is corrected by calibrated
Gaussian terms~\cite{Wong:2007ym}. This model remains useful because it is fast,
robust, and easy to vary inside larger parameter scans. The price is that the
effective parameters must be calibrated to a more complete
calculation~\cite{Shaw:2011wq}. We keep the usual \textsc{RECFAST} structure,
but update the calibration against a high-accuracy \textsc{CosmoRec} reference,
using the same default \textsc{CAMB} time sampling as in ordinary spectrum
calculations. Refitting only the existing hydrogen Gaussians and helium
normalization already gives very small CMB-spectrum residuals, but leaves a
coherent feature through He\,{\sc i} recombination, where \textsc{CosmoRec}'s
explicit helium radiative transfer is doing more than can be represented by a
single overall helium normalization.

We therefore add a small phenomenological correction to the \textsc{RECFAST}
He\,{\sc i} rate equation. After summing the usual singlet and triplet
contributions to \(dx_{\rm He}/dz\), the updated fit applies
\begin{equation}
  \begin{aligned}
    \left.\frac{dx_{\rm He}}{dz}\right|_{\rm fit}
    &= F_{\rm He}\left.\frac{dx_{\rm He}}{dz}\right|_{\rm RECFAST},\\
    F_{\rm He}
    &= 1+\frac{a_0+a_1u+a_2u^2}{1+d_2u^2+d_4u^4},\\
    u&=\frac{z_{\rm scale}-z_0}{w}.
  \end{aligned}
\end{equation}
Here \(z_{\rm scale}\equiv (T_{\rm CMB}/T_{\rm COBE})(1+z)-1\) with
\(T_{\rm COBE}=2.7255\,{\rm K}\), so the correction
tracks photon temperature rather than raw redshift, and the factor is used only
over the He\,{\sc i} recombination range \(1500\lesssim z_{\rm scale}\lesssim
3000\). The constants in \(F_{\rm He}\) were fitted to the total
\(x_e(z)\) residual relative to \textsc{CosmoRec}, not to a pointwise inferred
helium derivative, and then the existing hydrogen Gaussians and overall helium
normalization \code{RECFAST_fudge_He} were jointly refitted at fixed
\code{RECFAST_fudge}. Each stage is a bounded nonlinear least-squares fit for $x_e$
over a redshift grid
spanning the recombination window (\(800\lesssim z\lesssim3000\)) against the
fixed high-accuracy \textsc{CosmoRec} reference (the reference settings are those
given in the caption of \cref{fig:recfast-cosmorec-fit}); the helium-rate stage
additionally weights the residual towards the redshifts where the helium
recombination rate is largest, anchoring the correction where the helium
radiative-transfer difference is physically significant.\footnote{The rounded helium-rate constants in
\(F_{\rm He}\) are \((a_0,a_1,a_2,d_2,d_4)=(0.0781,\,0.899,\,-3.06,\,3.75,\,18.1)\)
with \((z_0,w)=(1968,\,774)\). The associated fitted \textsc{CAMB} \code{Recfast}
parameters are \code{RECFAST_fudge}=1.125, \code{RECFAST_fudge_He}=0.84724,
\code{RECFAST_He_rate_correction}=T, and
\((\code{AGauss1},\code{AGauss2},\code{zGauss1},\code{zGauss2},\code{wGauss1},\code{wGauss2})
=(-0.1395,\,0.07299,\,7.281,\,6.767,\,0.1639,\,0.2786)\), with the default
\textsc{RECFAST} switches.}
This extra factor is not needed to meet the default \(C_\ell\) accuracy target:
the no-He-rate fit is already well within that tolerance. It is nevertheless
cheap to evaluate, removes the visible helium-history residual, and keeps the
fast \textsc{RECFAST} approximation consistent with the chosen \textsc{CosmoRec}
reference to a level well below the remaining physical and numerical
uncertainties in the detailed recombination calculation\footnote{We use this version of CosmoRec \url{https://github.com/cmbant/CosmoRec/tree/camb}.}.

The \textsc{HyRec-2} strategy of tabulating a cosmology-dependent effective
correction to the hydrogen recombination rate is arguably a more physical fast
approximation, and is faster than direct effective multilevel-atom evolution. We
do not adopt that form for the default \textsc{CAMB} update because a
\textsc{RECFAST}-based fit preserves easy reproducibility of older calculations,
including Planck-era analyses, while reducing the dominant residuals with only a
minimal extension of the existing effective model.

The fitted corrections are cosmology-independent, so they do not capture the
small cosmology dependence of the higher-order recombination corrections
themselves. We checked that the same \textsc{RECFAST} fit remains consistent
with direct \textsc{CosmoRec} in early-dark-energy stress tests, and the
expected impact of the neglected cosmology dependence is small; this is consistent
with the \textsc{HyRec-2} result that its tabulated hydrogen radiative-transfer
correction gives no noticeable parameter bias even for an ideal
cosmic-variance-limited CMB experiment to $\ell=5000$~\cite{Lee:2020obi}.
\textsc{CosmoRec} and \textsc{HyRec-2} can also be linked to \textsc{CAMB} recombination
classes for full recombination evaluation in other extended models if required.

\begin{figure}[t!]
  \centering
  \includegraphics[width=\columnwidth]{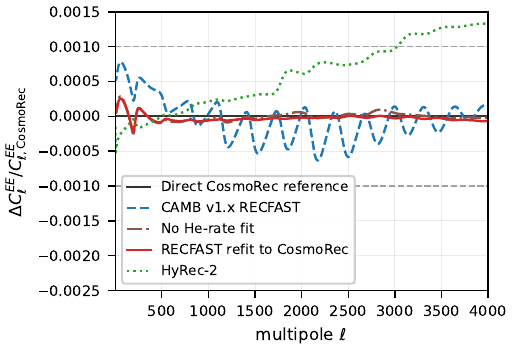}
  \caption{Fractional differences in the lensed $EE$ spectrum relative to the
  direct \textsc{CosmoRec} reference used for the updated \textsc{RECFAST}
  calibration, computed to $\ell_{\rm max}=4000$. The dashed horizontal lines show
  $\pm10^{-3}$, the default interior tolerance scale used for lensed $TT$ and
  $EE$ spectra in the \textsc{CAMB} accuracy checks.}
  \label{fig:recfast-ee-fit}
\end{figure}

\Cref{fig:recfast-cosmorec-fit} shows the resulting ionization-history
residuals. The upper panel shows the reference ionization history, the
visibility density per unit redshift, and the positive contribution per unit
redshift to the photon diffusion damping scale (Silk damping)~\cite{Silk:1968}. The
visibility illustrates that $800<z<1600$ contains the main last-scattering event,
while the damping kernel shows why earlier helium recombination still matters
for small-scale spectra: the damping scale accumulates the Thomson mean free
path before decoupling, and therefore retains a tail through the helium
recombination epoch even when the visibility is negligible. In the hydrogen
recombination window the maximum fractional residual relative to direct
\textsc{CosmoRec} falls from $9.82\times10^{-4}$ for the Planck-era \textsc{RECFAST}
parameters to $1.14\times10^{-4}$ for the updated fit.
In the helium window, $1600<z<3000$, where the helium
radiative-transfer correction matters most, the maximum residual falls from
$3.31\times10^{-3}$ for the Planck-era parameters, and $1.39\times10^{-3}$ for
that no-He-rate fit, to $7.75\times10^{-5}$. The figure also
shows direct \textsc{HyRec-2}. Its maximum difference from the \textsc{CosmoRec}
reference is $3.46\times10^{-4}$ in the hydrogen window and $4.51\times10^{-3}$
in the helium window, consistent with the two codes agreeing well on the main
hydrogen recombination physics while differing mainly in their helium
radiative-transfer approximations.

\Cref{fig:recfast-ee-fit} shows the corresponding effect on the lensed
$EE$ damping tail. The Planck-era \textsc{RECFAST} parameters remain within the
default $10^{-3}$ interior tolerance used in the \textsc{CAMB} accuracy checks,
but produce a coherent high-$\ell$ residual. The no-He-rate \textsc{CosmoRec}
fit and the updated helium-rate fit are both comfortably below this tolerance;
the latter mainly improves the physical consistency of the helium part of
\(x_e(z)\), while leaving the spectrum-level residual negligible at the target
accuracy. \textsc{HyRec-2} is marginally outside the $10^{-3}$ band at
$\ell\gtrsim3000$, again reflecting the residual modelling difference relative
to the more detailed \textsc{CosmoRec} helium radiative-transfer calculation
used as the reference. If we take \textsc{CosmoRec}'s helium as the more complete model,
it is straightforward to recalibrate \textsc{HyRec}'s helium equation to match it closely
across the full redshift range, although the \textsc{HyRec-2} results shown here use the code as distributed.

As a quantitative CMB-impact diagnostic, we use the SO-like Gaussian mean
likelihood of the accuracy-checking code (defined in \cref{app:accuracy}),
applied to the lensed spectra over $2\le \ell\le4000$.
Against the direct \textsc{CosmoRec} reference the resulting $\Delta\chi^2$
values are 0.56 for the Planck-era \textsc{RECFAST} parameters, 0.02 for the
seven-parameter \textsc{CosmoRec} fit without the He-rate correction, 0.03 for
the updated helium-rate fit, and 3.8 for direct \textsc{HyRec-2}. The updated
helium-rate correction was fitted to improve the physical helium part of
\(x_e(z)\), not to minimize this spectrum-level distance, and the small reversal
between the two fitted \textsc{RECFAST} variants has no practical
\(C_\ell\)-accuracy significance. The \textsc{HyRec-2} number is its mismatch to
the chosen \textsc{CosmoRec} reference, dominated by the less detailed helium
radiative-transfer modelling. For the matter power spectrum all models agree at the
$< 2\times 10^{-5}$ level.
The residual physical uncertainty is therefore not set by the fast
\textsc{RECFAST} fit, but by effects that are approximated or neglected,
together with remaining uncertainties in atomic rates, helium radiative transfer,
and any energy injection, varying constants, or other non-standard
physics not included in the baseline model.

For the \textsc{RECFAST} default model, the coupled hydrogen, helium and matter-temperature
system is mildly stiff in its early phase, where a purely explicit integration
wastes work. An alternative is the post-Saha quasi-equilibrium expansion used in
\textsc{HyRec}, which solves algebraically for the small departure from Saha
equilibrium before starting the effective ODEs~\cite{AliHaimoud:2010dx}. Here
\textsc{CAMB} instead keeps the original \textsc{RECFAST} equations and
integrates the early phase with the two-stage, second-order, L-stable Rosenbrock
method ROS2~\cite{Rosenbrock:1963,Verwer:1999}, a linearly implicit scheme with
stabilizing parameter $\gamma = 1 + 1/\sqrt{2}$ that uses the analytic Jacobian
of the right-hand side together with its explicit time-derivative term, and an
embedded lower-order solution for adaptive step control. Once the ionization
fraction has fallen enough that the system is safely
non-stiff (by default at $x_{\rm H}\simeq 0.976$), it hands off to the explicit
Dormand--Prince integrator used for the perturbations (\cref{sec:evolution}).
The Rosenbrock and Dormand--Prince tolerances and the
recombination sampling scale with the global accuracy settings, so the
recombination history converges smoothly as the overall calculation is tightened.
The default overall recombination calculation is under a millisecond, very similar
to \textsc{HyRec-2}, and highly subdominant to the perturbation calculation time
except when only the background is being calculated.

\subsection{Reionization}
\label{sec:reion}

After recombination the first luminous sources reionize the intergalactic medium,
and the free electrons rescatter a fraction of the CMB photons. The main
consequences are a near-uniform suppression of the primary anisotropies by
$e^{-2\tau}$ on the small scales that dominate the high-$\ell$ spectra, and a
large-scale reionization bump in the $E$-mode polarization. The suppression
depends only on the total Thomson optical depth $\tau$; the large-scale $E$-mode
bump carries some additional information about the shape of $x_e(z)$, but for the
smooth low-redshift histories of interest here it too is controlled mainly by
$\tau$~\cite{Mortonson:2007hq}. \textsc{CAMB} therefore models reionization with a
smooth,
low-dimensional ionization fraction that joins onto the residual recombination
history of \cref{sec:recomb} at high redshift and is integrated on the same time
grid.

The default model is the phenomenological $\tanh$ parameterization of
Ref.~\cite{Lewis:2008wr}, which places a smooth step in $x_e$,
\begin{equation}
  \label{eq:reion-tanh}
  \begin{aligned}
  x_e(z)&=x_e^{\rm rec} + \frac{f - x_e^{\rm rec}}{2}
  \left[1 + \tanh\frac{y_{\rm re}-y}{\Delta y}\right],\\
  y&\equiv (1+z)^{3/2},
  \end{aligned}
\end{equation}
centred on $y_{\rm re}=(1+z_{\rm re})^{3/2}$ with width
$\Delta y = \tfrac{3}{2}(1+z_{\rm re})^{1/2}\,\Delta z$, where
$x_e\equiv n_e/n_{\rm H}$, $x_e^{\rm rec}$ is the residual electron fraction
left by recombination, and
$f_{\rm He}\equiv n_{\rm He}/n_{\rm H}=Y_{\rm He}/[4(1-Y_{\rm He})]$. The
default main reionization step assumes that hydrogen and the first ionization of
helium occur together, so its upper plateau is $f=1+f_{\rm He}$. Writing the
step in the variable $(1+z)^{3/2}$ is convenient because, for a fixed midpoint
$z_{\rm re}$, it gives the same $\tau$ as an infinitely sharp transition during
matter domination, so $\tau$ and $z_{\rm re}$ map onto each other almost
analytically. \textsc{CAMB} nonetheless obtains $z_{\rm re}(\tau)$ by a monotonic
root find, so any other monotonic shape can be substituted without changing the
interface. The second, full ionization of helium near $z\simeq3.5$ is added as a
separate narrow $\tanh$ of fixed width; its precise placement has a negligible
effect on the spectra.

The symmetric $\tanh$ shape is a poor description of the actual history. Non-CMB
probes show that hydrogen reionization ends late and inhomogeneously: large-scale
scatter in the Ly$\alpha$ forest opacity persists to $z\simeq5.3$, pointing to a
late and patchy end to reionization~\cite{Bosman:2021obf,Fan:2006dp}, while the
intergalactic medium is still substantially neutral at $z\gtrsim7$, from quasar
damping wings~\cite{Davies:2018yfp} and the declining visibility of Ly$\alpha$
emitters~\cite{Mason:2017eqr}. The inferred history is therefore markedly
asymmetric in redshift, rising gradually at early times and completing rapidly
near $z\simeq6$. Because the CMB constrains mainly the integral $\tau$ rather than
the shape, this mismatch has little effect on parameter estimation at current
sensitivity; but it biases reionization histories inferred under the $\tanh$
assumption~\cite{Qin:2020xrg,Planck:2016mks}, and the residual shape information in
the large-scale polarization is relevant for future tests of the history.

\textsc{CAMB} therefore also provides an exponential model
(\code{ExpReionization}) closer to the data-motivated asymmetric form. Here the
ionization fraction saturates to full ionization below a fixed completion
redshift $z_{\rm end}$ and decays above it,
\begin{equation}
  \label{eq:reion-exp}
  \begin{aligned}
  x_e(z)={}&x_e^{\rm rec} + (f - x_e^{\rm rec})\\
  &{}\times\exp\!\left[-\lambda\,(z-z_{\rm end})^{p}\,s(z)\right],
  \qquad z > z_{\rm end},
  \end{aligned}
\end{equation}
with $x_e=f$ for $z\le z_{\rm end}$. The decay rate
$\lambda=\ln 2/(z_{\rm re}-z_{\rm end})^{p}$ sets the midpoint, so that $x_e$ falls
to half its ionized value at $z_{\rm re}$ up to the small effect of the endpoint
smoothing (equivalently the model is set through $\tau$); the power $p$ controls
the steepness (default $p=1$); and the numerical smoothing factor
$s(z)=[1+\epsilon_{\rm end}/(z-z_{\rm end})^2]^{-1}$, with
$\epsilon_{\rm end}$ set by \code{reion_exp_smooth_width}, makes the derivative
continuous at $z_{\rm end}$. Pinning the completion redshift near
$z_{\rm end}\simeq6$, consistent with the forest data, gives the model a non-zero
floor on the optical depth ($\tau\gtrsim0.04$ for $z_{\rm end}=6.1$), unlike the
$\tanh$ model which extends smoothly to $\tau\to0$. This form is similar in spirit
to the asymmetric parameterizations of Refs.~\cite{Douspis:2015nca,Qin:2020xrg},
but it is meant only to capture the gross asymmetry of the history as seen by the
CMB, and is not intended as a precision fit to non-CMB data. In particular it does
not model the detailed residual ionization structure at $z<z_{\rm end}$ --- the
slow final approach of $x_e$ to unity probed by the Ly$\alpha$ forest --- which
has a negligible effect on the CMB. Both models use the same $\tau$ mapping and
time-sampling machinery, so either can be driven by an input optical depth or by a
midpoint redshift, and further reionization parameterizations are
straightforward to add as new model classes sharing the same interface.

For a linear code, what exactly is meant by the effect of reionization on the gas temperature
and hence Jeans scale after reionization is somewhat ambiguous. This is discussed further in \cref{app:reion-heating},
where we also describe a crude (disabled-by-default) diagnostic for the convention dependence of the low-redshift linear matter power under homogeneous reionization heating.

\section{Massive neutrinos}
\label{sec:neutrinos}

By default \textsc{CAMB} approximates the neutrino masses by a single massive
eigenstate carrying the full $\sum m_\nu = 0.06\,{\rm eV}$, leaving the remaining
relativistic degrees of freedom massless. This minimal-mass single-eigenstate model
is the standard default: for a fixed small total mass the way it is divided between
eigenstates has a very small effect on the CMB, so resolving the individual masses
is unnecessary. For work that does need the physical mass splittings, \textsc{CAMB}
also provides explicit \code{normal} and \code{inverted} hierarchy settings. In the
physically allowed regime these approximate the three masses by two eigenstates---a
near-degenerate pair and a single state---fixing the masses from the total
$\sum m_\nu$ and the measured squared-mass splittings $\Delta m^2_{21}$ and
$\Delta m^2_{31}$, and reverting to a single eigenstate below the minimum total mass
allowed by oscillation data. We use the default single-eigenstate model throughout
this paper.

The thermal massive-neutrino background density and pressure are needed throughout
the evolution and were previously interpolated from a precomputed table. We now use
direct low-order fits in the intermediate regime, with small- and large-mass
series limits outside it. Here \(\tilde m_\nu\equiv m_\nu c^2/(k_{\rm B}T_{\nu0})\) is
the dimensionless mass used by the code, so the expansion variable is
\(a\tilde m_\nu\). The fitted variables are not the raw physical density and pressure:
\textsc{CAMB} uses the massless-neutrino-normalized, scale-factor-weighted
quantities
\begin{equation}
  \bar\rho_\nu=\frac{a^4\rho_\nu^{\rm phys}}{\rho_{\nu0}^{(0)}},\qquad
  \bar P_\nu=\frac{a^4P_\nu^{\rm phys}}{\rho_{\nu0}^{(0)}},
\end{equation}
where \(\rho_{\nu0}^{(0)}\) is the present density of one massless eigenstate. A
massless eigenstate therefore has \(\bar\rho_\nu=1\) and
\(\bar P_\nu=1/3\). For these barred variables the raw physical continuity
equation implies
\begin{equation}
  \dot{\bar\rho}_\nu=\mathcal H D_\nu,\qquad
  D_\nu\equiv\bar\rho_\nu-3\bar P_\nu .
\end{equation}
The density and pressure use a smooth fit for \(0.42<a\tilde m_\nu<70\), while
\(D_\nu\), which controls the derivative used by the integrator, uses a separate
fit over \(0.3<a\tilde m_\nu<70\). Against direct Fermi--Dirac quadrature these fits
reach a maximum relative error below \(10^{-4}\) in \(\bar\rho_\nu\),
\(\bar P_\nu\) and \(D_\nu\), which is amply sufficient.

The perturbations are the more delicate part. Massive neutrinos free-stream at a
momentum-dependent velocity, so their phase-space distribution is evolved as a
separate Boltzmann hierarchy for each sampled comoving momentum $q$, with the
density, heat flux and anisotropic stress that source the metric obtained as
Fermi--Dirac-weighted integrals over $q$~\cite{Ma:1995ey,Lewis:2002nc}. Sampling
$q$ finely would be costly, so \textsc{CAMB} uses a sparse quadrature: a handful of
momenta and weights chosen to integrate the Fermi--Dirac moment kernels accurately,
as detailed in the appendix of Ref.~\cite{Howlett:2012mh}. The default three-point
rule is adequate for the minimal-mass neutrino case. With the targeted-accuracy
settings used here, \textsc{CAMB} automatically raises this to four points
(per non-degenerate eigenstate) once the
largest eigenstate has dimensionless mass
$\tilde m_\nu\gtrsim600$ (roughly $0.1\,{\rm eV}$), and to five
points above $\tilde m_\nu\gtrsim4300$ for the heaviest transfer
cases.
Gauss--Laguerre abscissae are a useful guide at high
$q$~\cite{Lesgourgues:2011rh}, but the four- and five-point rules used here are
free-node fits for the \textsc{CAMB} kernels: the four-point rule is a
least-squares refit, and the five-point rule keeps the relativistic moment
constraints while using the remaining freedom to improve velocity-dependent
integrals as the neutrinos slow down.

There are three distinct approximation switches. The first is the perturbatively relativistic \textsc{CAMB} expansion documented
by Howlett et al.~\cite{Howlett:2012mh}. While a massive eigenstate remains
highly relativistic, its hierarchy is independent of momentum at leading order
and coincides with the massless hierarchy, with the leading mass dependence
captured by a perturbative correction of order
$(a\tilde m_\nu/q)^2$. \textsc{CAMB} therefore evolves a common relativistic
hierarchy and its first mass correction before initializing the full set of
momentum-dependent hierarchies once the species becomes sufficiently
non-relativistic. This delays the expensive momentum-resolved evolution and
allows it to start with a smaller multipole hierarchy depth.
The second switch is the late-time non-relativistic
velocity-integrated hierarchy of Ref.~\cite{Lewis:2002nc}, evolving four
perturbation variables associated with the density,
pressure, heat flux and anisotropic stress. Contributions from higher velocity
moments are included through their non-relativistic redshifting expansion and
the corresponding hierarchy closure.
The third switch applies to the massless species: the high-$k\eta$ subhorizon
closure described for \textsc{CLASS}~\cite{Blas:2011rf}, in which the
ultra-relativistic fluid approximation replaces the full radiation hierarchy
once $k\eta$ is large, and the late radiation hierarchy is then switched off
once the multipoles are dynamically negligible. These schemes are
standard; we summarize them only to record the momentum sampling and switching
used in the runs validated below.

\section{Dark energy}
\label{sec:de}

\textsc{CAMB}'s default dark energy is a fluid with equation of state $w(a)$ and a
rest-frame sound speed $c_s$. This fluid description is singular where $w$ crosses
$-1$, so for general $w(a)$ we use the parameterized post-Friedmann (PPF)
prescription of Hu and of Fang, Hu \& Lewis~\cite{Hu:2008zd,Fang:2008sn}, which
stays regular through the phantom divide. PPF does not model a physical field: it
parameterizes the dark-energy momentum density in terms of the other species,
with a single transition scale $c_\Gamma k/\mathcal H$ separating the
super-horizon regime, where dark energy can support perturbations, from small
scales, where it cannot. We summarize the equations because the small-scale
behaviour is where the numerical problem and our fix lie; the full derivation and
the meaning of the variables are in Refs.~\cite{Hu:2008zd,Fang:2008sn}.

Write $\mathcal H$ for the conformal Hubble rate and, following the code, absorb a
factor $8\pi G a^2$ into density and pressure variables (so $\bar\rho\equiv8\pi G
a^2\rho$ and $\bar h\equiv\bar\rho+\bar p$ is an enthalpy density). The PPF
dynamics are carried by a single variable $\Gamma$ with source
\begin{equation}
  S_\Gamma=\frac{\bar h_{de}\,(V_T+\sigma)}{2 k\,\mathcal H},
  \label{eq:ppf-source}
\end{equation}
where $V_T$ and $\sigma$ are the velocity and shear of the total non-dark-energy
fluid. With $c_\Gamma=0.4\,c_s$ and the transition variable
$\beta\equiv(c_\Gamma k/\mathcal H)^2$, the PPF evolution equation is
\begin{equation}
  \dot\Gamma=\mathcal H\left[\frac{S_\Gamma}{1+\beta}-(1+\beta)\,\Gamma\right].
  \label{eq:ppf-gamma}
\end{equation}
For a fixed source this is a relaxation equation: $\Gamma$ is driven at the rate
$(1+\beta)\mathcal H$ towards the algebraic quasi-static value obtained by
setting $\dot\Gamma=0$,
\begin{equation}
  \Gamma_{\rm qs}=\frac{S_\Gamma}{(1+\beta)^2}.
  \label{eq:ppf-gamma-qs}
\end{equation}
The dark-energy momentum and density perturbations are reconstructed algebraically
from $\Gamma$,
\begin{align}
  \bar h_{de}V_{de}
  &=-\frac{2k\mathcal H}{F}\Big(S_\Gamma-\frac{\dot\Gamma}{\mathcal H}-\Gamma\Big)
  \nonumber\\
  &\quad{}+\bar h_{de}V_T,
  \nonumber\\
  \bar{\delta\rho}_{de}&=-2k^2 k_f\,\Gamma-\frac{3\mathcal H}{k}\,\bar h_{de}V_{de},
  \label{eq:ppf-close}
\end{align}
with $F=1+3\,\bar h_{\rm noDE}/(2k^2 k_f)$ and $k_f$ the curvature factor
($k_f=1$ for a flat model), closing the system~\cite{Hu:2008zd}.

The difficulty is that the relaxation rate $(1+\beta)\mathcal H$ grows as
$k^2/\mathcal H$. For high-$k/\mathcal H$ modes  \cref{eq:ppf-gamma} therefore becomes
numerically stiff. At high redshift the dark-energy enthalpy and its observable
gravitational effect are generally tiny, but reconstructing perturbations
normalized by this small background can produce very large formal
dark-energy variables and strong sensitivity to the integration tolerance.
Historically this was avoided by setting $\Gamma=0$ above a
high threshold, but a hard cutoff is discontinuous and, for some extended models,
leaves the dark energy far from a sensible fluid value.

We instead keep the same quasi-static fixed point but cap the relaxation rate. For
$\beta>\beta_\star$ with $\beta_\star=30$ we replace \cref{eq:ppf-gamma} by
\begin{equation}
  \dot\Gamma=(1+\beta_\star)\,\mathcal H\,(\Gamma_{\rm qs}-\Gamma),
  \label{eq:ppf-fix}
\end{equation}
which is \cref{eq:ppf-gamma} with the relaxation rate frozen at its threshold
value, so it is continuous at $\beta=\beta_\star$ and leaves the equation
unchanged below it. This removes the stiffness while keeping $\Gamma$ close to
$\Gamma_{\rm qs}$, and avoids the matter-power and $C_\ell$ artefacts of a hard
cutoff.

We have checked that the cap is harmless at the observable level and that
$\beta_\star=30$ is not finely tuned. For dark-energy models ranging from
near-$\Lambda$CDM to phantom-crossing and non-flat crossing cases, the default
capped code reproduces the unregularized \cref{eq:ppf-gamma}, integrated at a
much tighter tolerance, to $\lesssim10^{-5}$ in the lensed $C_\ell$, matter
power and lensing potential---well below the target accuracy. Values of
$\beta_\star$ between $10$ and $300$ give similarly converged results.
\footnote{Using the PPF fluid with $c_s^2=1$: $w_0=-0.98$;
$(w_0,w_a)=(-0.9,\,0.3)$ (non-crossing);
$(-0.9,\,-0.5)$ (crossing); $(-1.15,\,0.4)$ (strongly phantom); and
$\Omega_K=-0.03$ with $(-0.9,\,-0.5)$. Against the unregularized reference
(\cref{eq:ppf-gamma} with the perturbation tolerance tightened by
$e^5\approx150$), the default $\beta_\star=30$ code agrees to
$<6\times10^{-6}$ in $TT$, $<2\times10^{-6}$ in $EE$, $<10^{-6}$ in the
linear matter power for $k\le2\,h\,\mathrm{Mpc}^{-1}$, and $<10^{-6}$ in
$C_\ell^{\phi\phi}$, at the run-to-run numerical floor. Varying
$\beta_\star\in\{10,30,100,300\}$ changes these observables by
$\lesssim4\times10^{-5}$.}

At the default tolerance, direct integration of the unregularized equation
instead changes the lensed $TT$ spectrum by $\sim2\times10^{-3}$ relative to
the tightly integrated result and is strongly tolerance-dependent. The cap
removes this sensitivity by more than two orders of magnitude, as well as
avoiding the very small integration steps that the stiff equation would
otherwise require. This cap is the new default PPF behaviour.

We stress that PPF is a phenomenological parameterization, not a physical model of
dark-energy perturbations. The variable $\Gamma$, and especially its small-scale
behaviour, is a device for giving the correct large-scale momentum and a smooth
$w=-1$ crossing; the stabilized high-$k/\mathcal H$ evolution
(\cref{eq:ppf-fix}) is a regularization in a regime where dark-energy
perturbations have no observable effect, and the dark-energy ``perturbations''
there should not be over-interpreted as physical.

Beyond the fluid and PPF parameterizations, the dark-energy sector is modular:
each model is a user-facing Python class mirrored by a Fortran implementation
that carries the numerics, so alternative models can be added by deriving a new
class that supplies $w(a)$, the background density and the perturbation evolution
(and, for field models, the potential). \textsc{CAMB} ships a few such models in
addition to the default fluid. The \texttt{AxionEffectiveFluid} model implements
an axion-like (ultralight-field) early-dark-energy fluid~\cite{Poulin:2018dzj},
and \texttt{EarlyQuintessence} integrates a single scalar field in an axion-like
potential $V(\phi)=m^2f^2\,[1-\cos(\phi/f)]^n$~\cite{Smith:2019ihp}, both
motivated by attempts to ease the Hubble tension; a general \texttt{Quintessence}
base class provides the field sampling and spline machinery for other
single-field models. These dark-energy implementations can serve as worked
examples of how to add a custom dark-energy class.

\section{Lensed CMB and matter power}
\label{sec:lensing}

\textsc{CAMB} lenses the CMB with the curved-sky correlation-function
method~\cite{Challinor:2005jy,Lewis:2006fu}: it forms the lensed correlation
functions from the unlensed spectra and the lensing-potential spectrum, using the
non-perturbative isotropic factor $\exp[-\ell(\ell+1)\sigma^2/4]$ together with a
second-order expansion in the deflection correlation $C_{\mathrm{gl},2}$, and
transforms back to the lensed spectra. Two implementations are available, chosen
automatically according to whether accurate $B$-mode polarization is requested.
When accurate $BB$ is not required the angular transform is truncated to small
separations, where the lensing correction is concentrated, which is much faster;
the truncation is apodized with a new $C^2$-continuous (quintic) taper that brings
the correlation function smoothly to zero over a narrow window, suppressing the
ringing that a sharp cut would alias into the spectra (the taper replaces a cruder
earlier treatment).\footnote{At low multipoles the reionization bump in $EE$ can
cause instabilities in the apodized truncated correlation-function calculation. We
therefore taper the low-$\ell$ ($2\le\ell\lesssim20$) $EE$ that enters the short-range
lensing-correction integrand; the full unlensed spectrum, bump included, is still added
back to form the lensed output, so only the negligible lensing \emph{correction} to the
bump is dropped, not the bump itself. Consistently, we do not attempt to model the tiny
linear signal of lensed reionization rescattering, which the full angular-range
calculation slightly overestimates because the rescattered signal originates at lower
redshift than the primary CMB.} When accurate $BB$ is required -- for tensor $B$-mode analysis for
example -- the correlation integrals are instead evaluated over the full angular
range by Gauss--Legendre quadrature, accumulating the curved-sky Wigner-$d$
combinations directly.

Lensing moves power between multipoles, so the lensed spectra near the requested
maximum multipole depend on unlensed power somewhat beyond it. To keep the output
accurate up to the requested $\ell_{\max}$, the code computes the unlensed spectra
to $\ell_{\max}$ plus a margin (by default $200$), and extends the convolution
integrands to higher multipoles still using a rescaled fiducial high-$\ell$
template, over a range that grows automatically with $\ell_{\max}$ and the lensing
accuracy boost. The lensed output therefore reaches the requested $\ell_{\max}$
without the user having to compute to a higher multipole and truncate to avoid
convolution edge effects (cf.\ the \textsc{CLASS} comparison in
\cref{sec:performance}).

High-accuracy lensing requires accurate unlensed CMB spectra and lensing-potential
power over the multipole ranges that contribute to the lensing convolution. This
in turn requires the perturbation sources to be evolved to a sufficiently large
maximum wavenumber $k_{\max}$. \textsc{CAMB} now chooses $k_{\max}$
automatically from the requested $\ell_{\max}$, so that both the lensed CMB
spectra and $C_L^{\phi\phi}$ reach the target accuracy without the user having to
set it by hand. In practice, $C_L^{\phi\phi}$ converges more slowly with
increasing $k_{\max}$ than the lensed CMB spectra. The lensing-potential
requirement therefore determines the automatic cutoff and imposes a minimum
$k_{\max}$ even when the requested $\ell_{\max}$ is relatively low.\footnote{The
corresponding input is \code{lens_potential_accuracy}. If left unset, it follows
the automatic prescription of \cref{app:accuracy}; the user can instead specify
it explicitly, with $0$ recovering the earlier Planck-era low-$k_{\max}$
default.}

The lensing deflection field is computed in the standard single
(background-orthogonal) source-plane approximation: photons are lensed along
the unperturbed line of sight to an effective last-scattering source, rather
than integrating over the finite width of the recombination era or including
corrections due to the perturbed photon path. Ref.~\cite{Hadzhiyska:2017nqe}
shows that relaxing the single-emission-time assumption shifts the lensed
spectra by well under $0.1\%$ on relevant scales. Ref.~\cite{Lewis:2017ans} computes
emission-angle and time-delay effects, and Ref.~\cite{Pratten:2016dsm} calculates
post-Born corrections to the deflection field (including rotation, which induces an additional $B$-mode signal);
these effects are also small, but are available separately via \textsc{CAMB}'s \code{emission_angle} and \code{postborn} modules.
\textsc{CAMB} can also calculate the lensed gradient spectra required for lensing reconstruction responses~\cite{Lewis:2011fk} (in the
flat-sky approximation, which is sufficient for typical lensing reconstruction accuracy).

For the matter power spectrum the accuracy is set by the transfer-function
wavenumber sampling and, for nonlinear spectra, by the nonlinear model
(e.g. \textsc{halofit}~\cite{Takahashi:2012em},
\textsc{hmcode}~\cite{Mead:2020vgs}, or another \code{NonLinear} class implementation).
The selected nonlinear prescription also supplies the nonlinear correction to the
Weyl-potential power entering $C_L^{\phi\phi}$, which is related to the nonlinear
matter power by the standard GR Poisson relation; it therefore modifies the lensed
CMB spectra, especially at high multipoles.
The validation work of
\cref{sec:validation} drove a number of small, physically scoped fixes to the
sampling and hierarchy settings -- for example the massive-neutrino multipole
hierarchy depth at high mass -- which we do not enumerate individually; they are
recorded in the code history and, more usefully, are captured by the validation
suite, which now passes at the target tolerances across the model grid. The
linear matter power in particular is numerically very stable at sampled points, comfortably inside
tolerance up to interpolation errors at sparse sampling. The practical accuracy floor is set by nonlinear modelling uncertainty rather than by the linear numerics. Baryonic feedback on the
small-scale matter power can now also be included with the \textsc{SP}(k)
model~\cite{Salcido:2023etq}, contributed by its authors.

\section{Validating accuracy}
\label{sec:validation}

The CMB-spectrum and matter-power accuracy claims in this paper rest on a deterministic validation harness,
\code{camb.check_accuracy}, which compares a default run to a reference run that
doubles the main accuracy boosts (described in \cref{app:accuracy}), and reports
the fractional differences per output, multipole or wavenumber range, avoiding
fractional comparisons for spectra that cross zero; in particular the
$TE$ cross-spectrum, which changes sign, is measured as
$|\Delta C_\ell^{TE}|/\sqrt{C_\ell^{TT}C_\ell^{EE}}$ (the normalized $TE$ error
used below) rather than as a relative difference. The default,
unboosted settings are designed so that the interior lensed CMB spectra and the
quasi-linear matter power agree with this reference at the $10^{-3}$ level --
the accuracy relevant for Advanced SO and the matter-power surveys -- with
looser tolerances at the spectrum ends, in the high-$\ell$ tail, and in the
nonlinear regime; \cref{tab:tol} in \cref{app:accuracy} lists the full set of
per-output acceptance tolerances.

\begin{figure*}[t!]
  \begin{center}
  \includegraphics[width=0.88\textwidth]{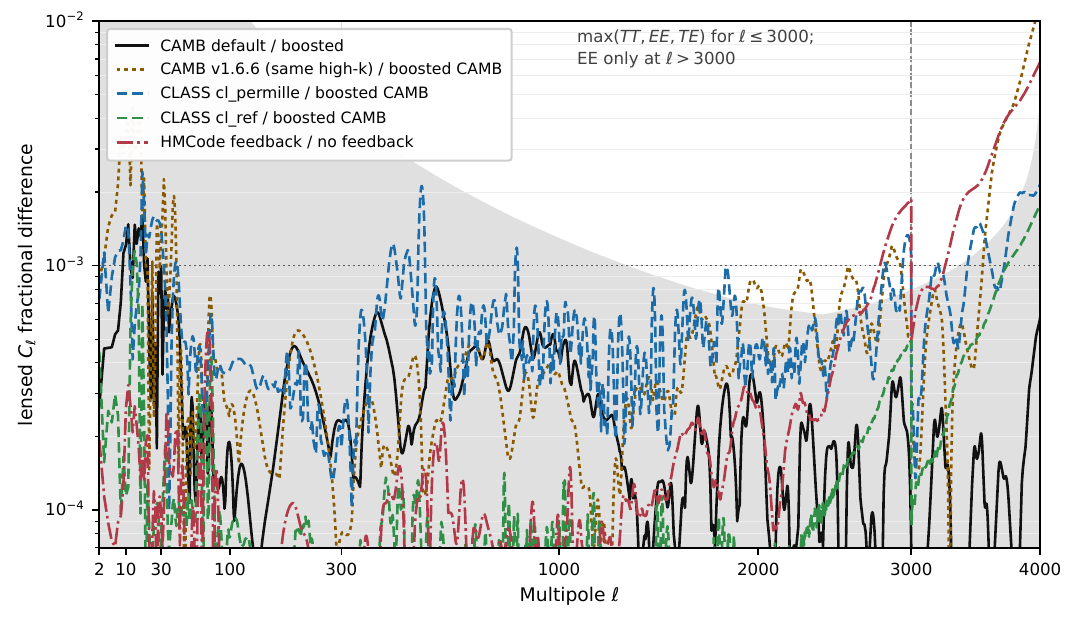}
  \caption{Lensed CMB numerical validation for a flat \(\Lambda\)CDM model with
  \(\ell_{\max}=4000\), using a strict-reference boosted \textsc{CAMB} run as the
  reference (the stricter \code{check_accuracy} preset, with the main accuracy boosts
  raised to three rather than the default two, and denser high-\(\ell\) sampling). Up to \(\ell=3000\) the plotted quantity is the envelope
  \(\max(|\Delta TT|/TT,|\Delta EE|/EE,|\Delta TE|/\sqrt{TT\,EE})\); above
  \(\ell=3000\), where foregrounds dominate temperature, the plot shows \(|\Delta EE|/EE\). The
  \textsc{CLASS} comparisons use
  version 3.3.4 with \code{cl_permille} and \code{cl_ref} precision settings; to avoid lensing-convolution edge effects both \textsc{CLASS} curves are computed to
  \(\ell=8000\) and then truncated to the plotted range. Both
  codes use Planck-era RECFAST with fixed \(Y_{\rm He}=0.245\), one massive
  neutrino eigenstate with \(\sum m_\nu=0.06\,{\rm eV}\), standard HMCode
  2020 with no feedback, and matched physical CMB lensing source cutoff, so the comparison as far as possible tests
  numerical like-for-like accuracy rather than physical accuracy. The grey band
  is the corresponding idealized \(1\sigma\)
  uncertainty on an overall spectral-amplitude parameter for an SO-like
  measurement in a bin of width \(\Delta \ell=\ell/2\) centred on each \(\ell\).
  The HMCode baryonic-feedback curve uses
  the \code{mead2020_feedback} model with
  \(\log_{10}(T_{\rm AGN}/{\rm K})=7.8\), relative to the no-feedback HMCode
  result, to indicate the size of possible physical modelling uncertainties.
  The dotted curve shows the last previous-generation release,
  \textsc{CAMB} v1.6.6, run with the same physical model and the same high-\(k\)
  lensing support as the current default (\code{lens_potential_accuracy}=5),
  relative to the same boosted reference; at its own historical default
  lensing cutoff the old code is substantially less accurate still (see text).}
  \label{fig:lensed-cl-validation}
  \end{center}
\end{figure*}

Running this harness over a grid of more than a hundred parameter files --
spanning curvature, massive neutrinos and their mass hierarchies, additional
dark radiation ($N_{\rm eff}$), dynamical dark energy, tensors,
recombination and reionization variants, nonlinear and transfer-only
configurations, and a range of $\ell_{\max}$ -- the current default settings pass
at the target tolerances\footnote{In unusual, source-count or $21\,$cm configurations not included in the test grid, there can be minor failures, for example at the very edge of a requested $\ell_{\max}$.}. The grid itself is generated programmatically
by a script distributed with the code (in \code{fortran/tests}) and is easily extended if needed.

The SO-like likelihood proxy described in \cref{app:accuracy} translates this
grid-level agreement into its impact on parameter inference. For each grid model
containing CMB power spectra, we evaluate the Gaussian $\Delta\chi^2$ between the
default run and the boosted reference over $2\le\ell\le\min(4000,\,\ell_{\max})$.
Number-counts, tensor-only and pure matter-transfer files are left
out of this proxy. Over the resulting set of roughly one hundred
CMB models the grid has median
$\Delta\chi^2\simeq0.4$, consistent with the single-model value quoted in
\cref{app:accuracy}, and the maximum over the grid is $\Delta\chi^2\simeq1.0$
(the $\ell_{\max}=4000$ dark-energy $\theta_\star$/PPF and heavy
three-degenerate-neutrino models). Since coherent errors at the tolerance
boundaries of \cref{tab:tol} could give $\Delta\chi^2\simeq34$
(\cref{app:accuracy}), the measured grid values confirm that in practice the
default-accuracy differences sit more than an order of magnitude below the
acceptance envelope on the likelihood scale.

This convergence check is complemented by a fixed-reference regression test. The
standard \textsc{CAMB} test suite generates a broad grid of scalar, tensor, lensed,
transfer and matter-power outputs and compares them against precomputed saved
reference results generated at boosted numerical accuracy. The two checks are
complementary: \code{check_accuracy} bounds the distance to a more converged
calculation, while the regression suite guards against unintended changes and
reproduces the released outputs across the model grid. The released defaults
pass both.

The new non-flat approximations are built so that the curved calculation varies
continuously into the flat one as $\Omega_K\to0$, with no step at the switch
between the flat and non-flat treatments. The validation grid checks this
directly: it includes a sequence of near-flat models straddling zero curvature
(down to $|\Omega_K|\sim10^{-6}$), and sweeping $\Omega_K$ through zero at fixed
angular acoustic scale the lensed $C_\ell$ and the other spectra vary smoothly
and approach the flat result to well within tolerance.

Comparison against the \textsc{CLASS} code provides a
complementary external check. Such cross-code tests have a long history: with
matched physical assumptions and strongly boosted settings, \textsc{CAMB} and
\textsc{CLASS} agree at the $\sim10^{-4}$ level for flat
$\Lambda$CDM~\cite{Lesgourgues:2011rg}, and later work established the boosted
settings both codes need for Stage-IV lensed-CMB and galaxy-survey
applications~\cite{McCarthy:2021lfp,Euclid:2023pxu,Bolliet:2023sst}; the
defaults described here are designed to reach the per-mille accuracy those
applications require without such strong boosting.
\Cref{fig:lensed-cl-validation} shows a representative high-resolution lensed
CMB test at \(\ell_{\max}=4000\). To isolate numerical accuracy from changes in
physical modelling, both \textsc{CAMB} and \textsc{CLASS} are run with the same
late-time nonlinear model (standard HMCode 2020 with no feedback), the same
physical CMB lensing source cutoff, fixed \(Y_{\rm He}=0.245\), and the
older Planck-era RECFAST recombination settings rather than the updated
recombination fit introduced in this paper. The comparison is therefore a
like-for-like numerical test, not a statement that the shared physical model is
the most accurate description of the Universe. The same figure also overlays a
plausible baryonic-feedback change in
\textsc{CAMB}, showing the scale of physical modelling uncertainty in the
nonlinear lensing contribution, which dominates at very high $\ell$.

The figure also quantifies the accuracy gain over the previous generation of
\textsc{CAMB}: the dotted curve is the last v1.x release (v1.6.6) run with the
same physical model and the same high-$k$ lensing support as the current
default (\code{lens_potential_accuracy}=5, matching the timing comparison of
\cref{tab:performance-timings}). With matched support the old code stays below
$10^{-3}$ up to $\ell\sim1900$, but then exceeds the SO-like sensitivity band
over much of $\ell\sim1900$--$3000$ and degrades rapidly beyond, reaching
$2.6\%$ in $TT$ and $1.1\%$ in $EE$ at $\ell=4000$ ($14\%$ in the lensed
$BB$). On the SO-like likelihood scale of \cref{app:accuracy} (evaluated over
$2\le\ell\le4000$ against the boosted reference) this corresponds to
$\Delta\chi^2\simeq10^2$, compared to $\Delta\chi^2\simeq0.4$ for the current
defaults on the same model. With default lensing settings (not shown), which
use a low lensing source cutoff $\max(k\eta_0)=10375$ tuned for
Planck-era requirements, v1.6.6 is far less accurate: it leaves the
sensitivity band above $\ell\sim1500$ and reaches $1.5\%$ error in $TT$ by
$\ell=3000$, with $\Delta\chi^2\simeq6\times10^3$.

\section{Agentic testing and tuning}
\label{sec:agentic}

The custom-precision approach of this paper is laborious to build and, more so, to
validate, and almost all of that CAMB-specific work was carried out by AI agents with a human in
the loop, against the deterministic \code{check_accuracy} harness of
\cref{sec:validation} as the validation target for final \textsc{CAMB} outputs.

For developing new low-level numerical approximations (e.g. for Bessel functions) the AI agents worked
directly: suggesting and deriving the required approximations, generating diagnostic scripts that compare options against a high-accuracy reference, scanning the regions of approximation validity with timings, and refining final code with gated fallbacks to more accurate or alternative approximations where needed. Most of the hyperspherical Bessel calibration was done this way -- for example tuning the continuous \(\nu/l\) cutoff for the raw-Olver fast path to the
point where a handful of accepted errors between \(10^{-4}\) and \(2\times10^{-4}\)
buy a large fraction of fast-path evaluations, a measured tradeoff chosen for the task in hand. Initial passes were later optimized by the agents for code efficiency (e.g. using cheap Horner-form polynomial fits where possible).

For accuracy target violations, a skill/subagent workflow separated the roles so
that code changes stay targeted and physically reasonable, and to guard against
fixes that satisfy the diagnostic but not the underlying cause. A subagent reproduces one failure in a disposable copy of the code and drills from the broad
accuracy control down to the responsible code lines. An orchestrator then develops a patch to fix one distinct failure at a time. An independent read-only subagent audits the result for scope, gating and physical motivation. The
governing rule is locality -- a fix must match the measured extent of the problem
and be gated in its natural coordinate (\(k\), \(\ell\), redshift, or \(\nu/l\)),
neither broader than the runtime cost justifies nor narrower than correctness
allows. Where higher accuracy is numerically cheap, the tolerance is exceeded by a safe margin.

The agents were consistently good at isolating a minimal change that restores
a test harness pass, but markedly less reliable at finding the \emph{right} one: a patch could
meet the target tolerances while resting on a numerical coincidence rather than a
physical mechanism. Deciding whether the mechanism was correct required physical
judgement, which current models are not yet reliably good at, so the work was iterative --
multiple rounds of search, with the human rejecting passing-but-unphysical
candidates and redirecting the next pass.

The markdown files defining the \code{isolate-accuracy-fix} skill and subagents,
along with a \code{classy-comparisons} skill guiding \textsc{CLASS} installation and like-with-like runs, are provided with the source code.
The source also includes a \code{devcontainer} environment configuration and related scripts, and \code{AGENTS.md} files for easy future development.

\section{Performance}
\label{sec:performance}

The revised defaults are intended to give next-generation CMB accuracy without
making ordinary likelihood calculations substantially more expensive.  They
automatically adjust the lensing convolution margin and the lensing source wavenumber
range, as described in \cref{sec:lensing}.  \Cref{tab:performance-timings}
summarizes representative code timing, showing how the new version obtains more accurate results at only slightly higher cost.
The flat runs use a Planck-2018 \(\Lambda\)CDM model with \(m_\nu=0.06\,{\rm eV}\) and
lensed spectra requested to \(\ell_{\max}=4000\).  The current default uses
\(\max(k\eta_0)=90000\), or \(k_{\max}=6.36\,{\rm Mpc}^{-1}\).  The old
\textsc{CAMB} v1.6.6 (the last v1.x release) default was faster only because it used the historical low
lensing cutoff \(\max(k\eta_0)=10375\).  When the old code is run with the same
high-\(k\) support as the new default, it is somewhat slower than the current
code and still substantially less accurate at high $\ell$
(\cref{fig:lensed-cl-validation,sec:validation}).

\begin{table}[t]
\centering
\resizebox{\columnwidth}{!}{%
\begin{tabular}{lrr}
\hline
Case & CPU [s] & wall [s] \\
\hline
\multicolumn{3}{c}{Flat \(\ell_{\max}=4000\), Planck-2018 model} \\
\textsc{CAMB} current default & 4.2 & 1.1 \\
\textsc{CAMB} v1.6.6, same high-\(k\) support & 5.0 & 1.3 \\
\textsc{CAMB} v1.6.6, historical low-\(k\) default & 2.8 & 0.70 \\
\textsc{CLASS} default, \(\ell=4000\) & 13 & 4.3 \\
\textsc{CLASS} \code{cl_permille} default, \(\ell=4000\) & 14 & 4.6 \\
\textsc{CLASS} default, high-\(k\), \(\ell=6000\to4000\) & 17 & 5.3 \\
\textsc{CLASS} \code{cl_permille}, high-\(k\), \(\ell=6000\to4000\) & 20 & 6.1 \\
\hline
\multicolumn{3}{c}{Non-flat 10-model geometric mean} \\
\textsc{CAMB} current default & 5.8 & 1.5 \\
\textsc{CAMB} v1.6.6, same high-\(k\) support & 98 & 25 \\
\textsc{CLASS} default & 18 & 4.4 \\
\textsc{CLASS} \code{cl_permille} defaults & 32 & 7.6 \\
\hline
\end{tabular}
}
\caption{Representative four-thread timings for lensed CMB calculations on a
typical laptop.  The default \textsc{CLASS} rows use \textsc{CLASS}'s own default
CMB source cutoff and, for the flat case, no lensed-spectrum margin.  The
high-\(k\) flat \textsc{CLASS} rows raise the \textsc{CLASS} CMB source cutoff to match
the current \textsc{CAMB} default, and compute lensed spectra to \(\ell=6000\)
before truncating to the requested \(\ell_{\max}=4000\).  This changes the \textsc{CLASS}
CMB source calculation, not the output matter-power grid.  The non-flat rows
are geometric means over the curvature sweep described in
\cref{sec:bessel-curvature}. For non-flat models the \textsc{CLASS} timing results do not
include hyperspherical-Bessel table precomputation.
}
\label{tab:performance-timings}
\end{table}

The \textsc{CLASS} timing depends sensitively on whether the calculation is set
up as a converged high-\(\ell\) comparison.  If \textsc{CLASS} is run only to the requested
\(\ell_{\max}=4000\), the lensed-spectrum convolution edge dominates the residual:
for \code{cl_permille} at the same high source cutoff, stopping at \(4000\)
rather than computing to \(6000\) and truncating changes the plotted spectra by
\(1.8\times10^{-2}\) at \(\ell=4000\).  Once this edge is removed, changing from
\textsc{CLASS}'s default source cutoff to the high-\(k\) cutoff used in the table changes
the \code{cl_permille} spectra by \(2.2\times10^{-3}\) in the full \(TT,EE,TE\)
envelope at \(\ell=4000\).  These lensing-related settings mainly affect the last
part of the damping tail, \(\ell\gtrsim3000\), but they are needed for a clean
high-\(\ell\) accuracy comparison.

The main conclusion is that the new \textsc{CAMB} defaults give substantially
more accurate lensed spectra than the previous version with little timing cost
at matched lensing support.

For the non-flat models, the revised
hyperspherical-Bessel calculation scheme also removes the large cost of running the
previous version:
across the ten-model curvature sweep of \cref{sec:bessel-curvature} the
geometric-mean CPU time falls from $98.4\,$s for \textsc{CAMB} v1.6.6 to
$5.79\,$s at $\ell_{\max}=4000$, a factor of $17$ at higher target accuracy, and
the near-flat $|\Omega_K|\le10^{-3}$ models ($5.06\,$s geometric mean) cost only
modestly more than the exactly flat case ($4.19\,$s).

\section{Conclusions}
\label{sec:summary}

We have described an update to \textsc{CAMB} aimed at the accuracy and speed
required by the next generation of CMB and large-scale-structure surveys, focused
on the lensed CMB and matter power spectra. The central technical development is a
hyperspherical Bessel scheme based on a flat-Bessel comparison: it is exact in the
flat limit, smooth through the turning point, reduces to a constant rescaling of
the flat argument in near-flat models, and supplies robust starting values for a
cheap Numerov propagation of neighbouring source points. We have also added
faster and more accurate recombination and neutrino-background evaluation, a stabilized PPF
dark-energy evolution, and recalibrated lensing and curvature accuracy, all
validated by a deterministic harness across a large grid of models. Although we
have concentrated on the lensed CMB and matter power, the same line-of-sight and
accuracy machinery carries over to the cross-correlations, number-count and
lensing observables that \textsc{CAMB} also computes.
The unboosted defaults are numerically converged to the $10^{-3}$ level relative to the boosted reference over
interior multipoles $600 \le \ell \le 3500$ and the
quasi-linear matter power.

\Cref{app:accuracy} documents the accuracy parameters and default tolerances. The
defaults target next-generation CMB experiments up to Advanced SO precision, and
where higher accuracy is needed we provide tools to tune them easily.

We have also used this work as a worked example of custom-precision numerical
methods developed with AI agents under human supervision. We expect this mode of
development -- bespoke approximations, gated by physically motivated tests and
validated deterministically -- to become an increasingly automated way to meet the
accuracy and computational demands of physics codes.

\begin{acknowledgments}
I am supported by the UK STFC grant ST/X001040/1. I thank Jens Chluba and
Yacine Ali-Haimoud for recombination discussions, Simons Observatory collaborators,
and CAMB GitHub issue and pull-request
contributors for reporting errors and contributing code or patches. The
algorithms and code described here were co-developed with LLM chats and AI coding
agents (Claude Code with Opus 4.8/Fable, Codex with GPT 5.5/5.6, and Copilot with GPT 5.4/Claude Sonnet) under my supervision, with some cross-checks using Gemini 3.1 Pro; I am responsible
for the physics, the validation design, and the final review of all changes.
\end{acknowledgments}

\appendix

\section{Hyperspherical Bessel details}
\label{app:bessel}

This appendix collects the formulae behind \cref{sec:bessel}.
\Cref{app:bessel-action} gives the closed-form flat action, the curved action
integral, the numerical
inversion of the map, and the size of the term the leading-order map drops;
\cref{app:bessel-smallchi} the small-$\chi$ expansion behind the near-flat
approximations; \cref{app:bessel-gates} the precise accuracy gates and fallback
order; and \cref{app:bessel-recur} the exact recurrence and the \textsc{CLASS}
comparison.

\subsection{Action integrals, the inverse map, and the leading error}
\label{app:bessel-action}

With $f_K(\chi)=\alpha^2-\SK^{-2}(\chi)$, the curved action measured from the
turning point is
\begin{equation}
  Q_K(\chi)=
  \begin{cases}
  \displaystyle\int_\chi^{\chi_t}\sqrt{-f_K}\,\dd s, & \chi<\chi_t,\\[1.4ex]
  \displaystyle\int_{\chi_t}^{\chi}\sqrt{f_K}\,\dd s, & \chi>\chi_t,
  \end{cases}
  \label{eq:curved-action}
\end{equation}
with $Q_K(\chi_t)=0$. For evaluating the curved action analytically it is useful
to define $x_K=\alpha\SK(\chi)$. The corresponding dimensionless flat variable is
$x_0=\alpha z$, in terms of which the flat action is
\begin{equation}
  I_0(x_0)=
  \begin{cases}
  \log\!\dfrac{1+\sqrt{1-x_0^2}}{x_0}-\sqrt{1-x_0^2}, & x_0<1,\\[1.4ex]
  \sqrt{x_0^2-1}-\arccos(1/x_0), & x_0>1.
  \end{cases}
  \label{eq:flat-action}
\end{equation}
with $I_0(1)=0$; $x_0<1$ is the evanescent side and $x_0>1$ the oscillatory side. The
curved integral \cref{eq:curved-action} has closed logarithm/$\atanTwo$ forms in
$x_K$ on each side of the turning point (the same integrals that appear in the
Langer/WKB
treatments~\cite{Kosowsky:1998nc,tram2013}), evaluated without quadrature; the
explicit branch-stable expressions, implemented in \code{qintegral_exact}, are
given in the implementation note~\cite{cambhypersphericalbessel}.

The map \cref{eq:action-map} is inverted for $x_0=\alpha z$ branch by branch. Below
the turning point write $x_0=\sech t$, so $I_0=t-\tanh t$; above it write
$x_0=1/\cos\theta$, so $I_0=\tan\theta-\theta$. Near the turning point both give
$I_0\simeq t^3/3$, so the code forms $p=(3Q_K)^{1/3}$ and uses a fitted polynomial
in $p^2$ for $x_0$ (avoiding cancellation in the nearly-equal logarithmic or
trigonometric terms); away from the turning point it uses the natural asymptotic
inverses, $t\simeq Q_K+1$ on the evanescent side and $\theta\to\pi/2$ on the
oscillatory side. Finally $z=z_t x_0=x_0/\alpha$.

The residual of the leading-order map \cref{eq:basic-olver} follows from an exact
substitution. Writing $u=A\,V(z)$, with
$A=(z')^{-1/2}$ the amplitude of \cref{eq:amplitude-general} and $V$ a
function of $z$ to be determined, \cref{eq:reduced} becomes
\begin{equation}
  \frac{\dd^2V}{\dd z^2}
  +\big[\hat\ell^2 f_0(z)+\Psi(z)\big]V=0,\qquad
  \Psi=-\frac{1}{2(z')^2}\{z,\chi\},
  \label{eq:olver-residual}
\end{equation}
where
$\{z,\chi\}=z'''/z'-\tfrac32(z''/z')^2$ is the Schwarzian derivative of the
map $z(\chi)$, with primes denoting $\chi$-derivatives. The residual $\Psi$
vanishes in the flat case $z=\chi$ and measures the nonlinear,
curvature-dependent part of the coordinate map.

The leading-order approximation drops $\Psi$, the only term not multiplied by
the large parameter $\hat\ell^2$. The transformed equation then reduces to the
flat comparison equation \cref{eq:flat-comparison}, whose regular solution is
$V=v_0=z\jl(\nu z)$. Substituting this solution into $u=AV$ reproduces
\cref{eq:basic-olver}. In the near-flat,
large-$\alpha$ regime, the small-$\chi$ expansion gives, for either
$K=\pm1$,
\begin{equation}
  \Psi(\chi)=\frac{1}{30\alpha^2}
  +\order(\alpha^{-4})+\order(\chi^2).
\end{equation}
In the overlapping oscillatory region $1/\alpha\ll\chi\ll1$, where
$f_0\simeq\alpha^2$, the residual is therefore smaller than the leading
coefficient by
\begin{equation}
  \frac{\Psi}{\hat\ell^2 f_0}
  \simeq\frac{1}{30\hat\ell^2\alpha^4}.
\end{equation}
The additional $1/\hat\ell^2$ suppression explains why the accepted minimum
$\alpha$ can decrease with increasing $l$. This asymptotic scaling, together
with the numerical calibration in \cref{app:bessel-gates}, motivates the broad
$\alpha\simeq{\rm few}$ raw-Olver boundary at moderate $l$.

\subsection{Small-\texorpdfstring{$\chi$}{chi} expansion}
\label{app:bessel-smallchi}

Writing $z=\chi F$, $a=K\chi^2$ and $h=K/\alpha^2$, the differentiated map
$(\dd z/\dd\chi)^2(\alpha^2-z^{-2})=\alpha^2-\SK^{-2}$ has the cubic-order solution
\begin{align}
  F&=1-h\left[\frac16+\frac{4a+13h}{360}+\frac{48a^2+148ha+737h^2}{45360}\right],
  \label{eq:F-smallchi}\\
  D&\equiv\frac{\dd z}{\dd\chi}=1 \nonumber\\
  &\quad -h\left[\frac16+\frac{12a+13h}{360}+\frac{240a^2+444ha+737h^2}{45360}\right],
  \label{eq:D-smallchi}
\end{align}
giving the local approximation
\begin{equation}
\phi_l^\nu\simeq (\chi F/\SK)\,D^{-1/2}\jl(\nu\chi F).
\end{equation}
Setting $a=0$ recovers
$F_0=D_0$ of \cref{eq:F0}, the constant-map (shifted-$\nu$) limit.

The Liouville equation determines the amplitude only up to a constant,
$A=C(z')^{-1/2}$. The transformed equation \cref{eq:olver-residual} is
independent of $C$, which only determines how the overall normalization of
$u=AV$ is divided between the amplitude and the comparison solution. Once the
comparison solution is fixed to be the conventionally normalized
$v_0(z)=z\jl(\nu z)$, however, $C$ fixes the normalization of the resulting
approximation.

To compare the canonical choice $C=1$ with the exact normalization of the
regular solution at the origin, define
\begin{equation}
  \lambda\equiv\lim_{\chi\to0}\frac{z(\chi)}{\chi}
  =\lim_{\chi\to0}z'(\chi).
\end{equation}
The expansion above gives
$\lambda=F_0+\order(h^4)$. Since $\SK(\chi)\sim\chi$ and
$\jl(\nu z)\sim(\nu z)^l/(2l+1)!!$, the Olver approximation gives
\begin{equation}
  \phi_l^\nu(\chi)\sim
  C\,\frac{\nu^l\lambda^{l+1/2}}{(2l+1)!!}\,\chi^l.
  \label{eq:olver-origin-normalization}
\end{equation}
The exact regular normalization instead has coefficient
$\prod_{j=1}^l b_j/(2l+1)!!$. Enforcing this normalization exactly within the
leading comparison solution would require
\begin{equation}
  C_{\rm origin}
  =\frac{\prod_{j=1}^l b_j}{\nu^l\lambda^{l+1/2}}.
  \label{eq:olver-origin-C}
\end{equation}
We do not apply this finite-$l$ factor. It would remove the normalization error
at the origin, but would not include the accompanying correction to the
solution shape generated by the neglected Schwarzian term $\Psi$. It is
therefore not a systematic next-order Olver correction, and need not reduce the
error near the turning point and in the oscillatory region where the radial
function contributes most strongly. We instead retain $C=1$ as part of the
leading comparison construction. This choice preserves the Wronskian, is exact
in flat space, and gives errors bounded empirically by the calibrated gates and
recurrence fallback.

In the near-flat limit $h\to0$ at fixed $l$, the canonical and exact
normalizations agree asymptotically. With
$h=Kl(l+1)/\nu^2$ and $\hat\ell^2=l(l+1)$, the exact product has the expansion
\begin{align}
  \log\left(\frac{\prod_{j=1}^l b_j}{\nu^l}\right)
  &=
  \frac12\sum_{j=1}^l
  \log\left(1-\frac{h j^2}{\hat\ell^2}\right)
  \nonumber\\
  &=
  -\left(l+\frac12\right)\frac{h}{6}
  -\left(l+\frac12\right)\frac{h^2}{60}
  \left(3-\frac{1}{\hat\ell^2}\right) \nonumber\\
  &+\order\left(lh^3\right),
  \label{eq:exact-origin-expansion}
\end{align}
where we used
$\sum_{j=1}^l j^2=\hat\ell^2(2l+1)/6$ and
$\sum_{j=1}^l j^4=\hat\ell^2(2l+1)(3\hat\ell^2-1)/30$.
On the Olver side, \cref{eq:F0} gives
\begin{equation}
  \log\lambda
  =-\frac{h}{6}-\frac{h^2}{20}+\order(h^3),
\end{equation}
and hence
\begin{equation}
  \log\lambda^{l+1/2}
  =-\left(l+\frac12\right)
  \left(\frac{h}{6}+\frac{h^2}{20}\right)
  +\order\left(lh^3\right).
\end{equation}
The two expressions agree exactly at $\order(h)$ and at leading order in $l$ at
$\order(h^2)$. Their finite-$l$ logarithmic difference is
\begin{equation}
  \log C_{\rm origin}
  =\frac{(l+\tfrac12)h^2}{60\hat\ell^2}
  +\order\left(lh^3\right)
  \sim\frac{1}{60l\alpha^4},
  \label{eq:origin-normalization-mismatch}
\end{equation}
where $h^2=\alpha^{-4}$ for $K=\pm1$. The agreement provides a non-trivial
consistency check of the $13/360$ coefficient in $F_0$ against the exact
$\sum_j j^4$ contribution. The remaining normalization mismatch has the same
parametric suppression as the accumulated effect expected from the neglected
Schwarzian residual and forms part of the leading-order error controlled by the
dispatcher gates.

\subsection{Accuracy gates}
\label{app:bessel-gates}

For a pointwise call, the dispatching routine \code{phi_olver} decides as
follows. For $l\le2$ it uses
\code{phi_recurs}. For $l>2$ it first tries the small-$\chi$ map if
\begin{equation}
  \begin{aligned}
  &\left[\frac{\nu}{l}>2.5\quad{\rm or}\quad
  \left(l\ge50\ {\rm and}\ \frac{\nu}{l}>1\right)\right],\\
  &\frac{l^2\chi^7}{\nu}<5\times10^{-2},
  \end{aligned}
\end{equation}
using $\nu/l$ and $l^2$ only as cheap high-$l$ gate proxies; the map itself uses
the more accurate $\hat\ell=\sqrt{l(l+1)}$ curvature scale. If that branch is not
used, open models accept the raw-Olver value (the full action map of
\cref{eq:action-map}) whenever, with
\(\alpha_g=\nu/l\),
\begin{equation}
  \begin{aligned}
  &\alpha_g\ge\alpha_{\rm open}(l),\\
  &\alpha_{\rm open}(l)=\max\!\left[
  0.095,\,
  0.12\left(\frac{500}{l}\right)^{p(l)}
  \right],
  \end{aligned}
  \label{eq:open-alpha-gate}
\end{equation}
with $p(l)=0.70$ for $l<500$ and $p(l)=0.14$ for $l\ge500$. Below that cutoff,
open models accept the raw-Olver value only in the endpoint corner
\begin{equation}
  \frac{\chi}{2\nu}\le3.0\times10^{-3}.
\label{eq:open-endpoint-gate}
\end{equation}
Closed models accept the raw-Olver value only when
\begin{equation}
  \frac{\chi}{2(\nu-l)}\le7.0\times10^{-3}.
\label{eq:closed-endpoint-gate}
\end{equation}
Whenever these raw-Olver gates fail, the fallback sequence first
tries the second-order Airy one-point approximation for validated high-$l$ points
($l\ge100$ open, $l\ge150$ closed) and, in open models, a narrow small-$\nu$
approximation before calling \code{phi_recurs}. These thresholds are calibrated
on peak-normalized error, keeping the raw Olver envelope near $10^{-4}$ on the
validation grid.

\begin{figure*}[t]
\centering
  \includegraphics[width=0.96\textwidth]{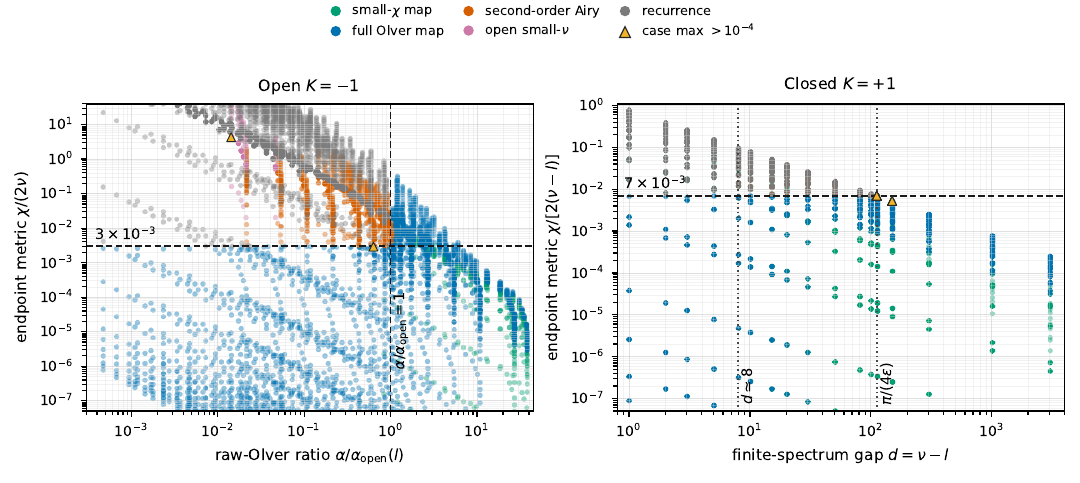}
\caption{Hyperspherical-Bessel pointwise-evaluation domains in the current dispatcher.
Each dot is one sampled call \((K,l,\nu,\chi)\), coloured by the branch selected
by \code{phi_olver}; errors are measured against \code{phi_recurs} and
normalized by the first peak amplitude. The axes project the full call onto the
variables controlling the leading gates. In the open panel the horizontal
coordinate is \(\alpha_g/\alpha_{\rm open}(l)\), where
\(\alpha_g=\nu/l\) is the code's high-\(l\) proxy for
\(\alpha=\nu/\hat\ell\) in \cref{eq:turning}, and
\(\alpha_{\rm open}\) is defined in \cref{eq:open-alpha-gate}; the vertical
coordinate is the endpoint metric \(\chi/(2\nu)\), with the dashed lines showing
the raw-Olver gates of \cref{eq:open-alpha-gate,eq:open-endpoint-gate}. In the
closed panel the horizontal coordinate is the finite-spectrum gap
\(d\equiv\nu-l\) and the vertical coordinate is \(\chi/[2(\nu-l)]\); the dashed
line is the raw-Olver metric gate \cref{eq:closed-endpoint-gate}. The dotted
closed-panel guides mark the large-\(l\) Airy onset \(d\simeq8\) and the
\(d\simeq112\) value above which \cref{eq:closed-endpoint-gate} covers the full
folded interval \(0\le\chi\le\pi/2\). Because \(l\), \(\nu\), and \(\chi\) are
not uniquely encoded by these two projected coordinates, different calls can
land on the same plotted point; overlapping Airy and recurrence colours therefore
mean different hidden cases selected different fallback branches, not that one
call used multiple branches. Yellow triangles mark cases whose maximum sampled
error lies between \(10^{-4}\) and \(2\times10^{-4}\); no sampled case exceeds
\(2\times10^{-4}\). The open small-\(\nu\) branch applies only in the open
panel.}
\label{fig:bessel-accuracy-domains}
\end{figure*}

The Airy and small-$\nu$ branches in \cref{fig:bessel-accuracy-domains} are
therefore fallback labels, not separate user-facing algorithms. The Airy branch
uses the same reduced radial equation but maps it to an Airy comparison equation,
including the second-order local correction to the leading Langer form. It is
mainly a high-$l$ rescue path in the open low-\(\nu/l\) tail and in closed models
away from the origin of the folded interval $0\le\chi\le\pi/2$ but not too
close to the finite spectral endpoint.
The open small-$\nu$ branch is rarer: after the Airy gate fails, it uses an
action-matched Macdonald-function approximation for the remaining very low-\(\nu\)
open tail. The full derivations and calibration scans are given in the
hyperspherical-Bessel implementation note~\cite{cambhypersphericalbessel}; here
we only record the dispatcher gates needed to interpret the figure.

It is useful to translate the closed gates into the distance from the finite
closed-spectrum endpoint. Writing \(d=\nu-l\), so that the Gegenbauer degree is
\(n=d-1\), and \(\lambda=l+1/2\), the closed Airy fallback's base validity gate is
\begin{equation}
  \lambda\left[\left(\frac{\nu}{\lambda}\right)^2-1\right]
  =\frac{(l+d)^2-\lambda^2}{\lambda}\ge15 .
\end{equation}
For large \(l\) this is approximately \(2d-1\ge15\), i.e.\ \(d\gtrsim8\). By
contrast, the raw-Olver endpoint gate \cref{eq:closed-endpoint-gate} is
\(\chi/(2d)\le7.0\times10^{-3}\).
Since closed calls are folded into \(0\le\chi\le\pi/2\), this covers the whole
folded interval only once
\(d\gtrsim\pi/[4(7.0\times10^{-3})]\simeq112\). Thus high-\(l\) closed modes
with \(8\lesssim d\lesssim112\) use the raw-Olver value near the folded origin
and the Airy fallback over the rest of the interval; modes closer to the finite
endpoint, or with \(l<150\), fall through to recurrence when the raw-Olver gate
does not apply.

The entire source-integration range can be filled from shifted flat Bessels when the small-$\chi$ map is accurate over a whole range. With $\alpha_g=\nu/l$ and
$m_\chi=l^2\chi_{\max}^7/\nu$ this is when
\begin{equation}
  \bigl[\alpha_g>2.5\ {\rm or}\ (l\ge50\ {\rm and}\ \alpha_g>1)\bigr],
  \quad
  m_\chi<0.1/B_{\rm acc},
\end{equation}
with $B_{\rm acc}\ge1$ the non-flat source accuracy boost. The pointwise
\code{phi_olver} small-$\chi$ gate is separate and uses the tighter current-point
metric $l^2\chi^7/\nu<5\times10^{-2}$. The shifted-$\nu$ filler additionally
requires the constant-map error estimate
$\epsilon_{\rm shift}<10^{-3}/B_{\rm acc}$, where, with $t=\alpha\chi$,
\begin{equation}
  \epsilon_{\rm shift}=\frac12\frac{\hat\ell\,t_{\max}^3}{90\alpha^4}+\frac{t_{\max}^2}{180\alpha^4}
  \label{eq:shift-error}
\end{equation}
combines the leading omitted argument shift and amplitude correction at the range
endpoint. A one-sided bookkeeping sampling-ratio test
(\(\peakscale\ge1-0.03\), with \(\peakscale\) defined in
\cref{sec:bessel-lsampling}) gates whether the flat-table machinery is reused at
all.

\Cref{fig:bessel-accuracy-domains} shows the pointwise-evaluation domains and flags the worst-case validation errors over a
wide region of hyperspherical-Bessel parameter/argument space. Small $\nu/l$ and other domains included here are rarely encountered in the CMB calculation, but the \code{phi_olver} function is validated across the full domain shown for completeness.

\subsection{Recurrence and the \textsc{CLASS} comparison}
\label{app:bessel-recur}

For closed models ($K=+1$) any $\chi$ is first folded into $0\le\chi\le\pi/2$
using the parity relations
\begin{align}
  \phi_l^\nu(2\pi-\chi)
  &=(-1)^l\,\phi_l^\nu(\chi),\\
  \phi_l^\nu(\pi-\chi)
  &=(-1)^{\nu-l-1}\,\phi_l^\nu(\chi).
  \label{eq:closed-parity}
\end{align}
which follow immediately from the Gegenbauer form below. The resulting parity
factor is the closed-space sign applied to all $K=+1$ evaluations, including
the Olver map of \cref{eq:basic-olver}.

The exact seeds are
\begin{eqnarray}
\phi_0^\nu&=&\sin(\nu\chi)/(\nu\SK) \\
\phi_1^\nu&=&[\cotK\chi\,\sin(\nu\chi)/\nu-\cos(\nu\chi)]/(\SK\,b_1)
\end{eqnarray}
 (with Taylor
safeguards for small arguments), reducing to $j_{0,1}(\nu\chi)$ for $K=0$. The
three-term recurrence is
\begin{equation}
  b_j\phi_j^\nu=(2j-1)\cotK\chi\,\phi_{j-1}^\nu-b_{j-1}\phi_{j-2}^\nu,
  \label{eq:up-recurrence}
\end{equation}
used upward only in the safely oscillatory region
$|\cotK\chi|<b_l/\max(1,l)$, with an open-space refinement using
$y_{\rm open}=\nu\sinh\chi/l$.  For $K=-1$ upward recurrence is also used in the
near-turning layer $y_{\rm open}\ge1-5\times10^{-3}$, and for the very low
$\nu/l$ family when $\nu/l\le2.0\times10^{-3}$ and $y_{\rm open}\ge0.8$.
These extra open gates were calibrated with peak-normalized forward/Miller
comparisons, keeping tiny-$\chi$ low-$\nu/l$ cases on Miller.  Elsewhere the
recurrence is run downward (Miller),
\begin{equation}
  \phi_{j-1}^\nu=\frac{(2j+1)\cotK\chi\,\phi_j^\nu-b_{j+1}\phi_{j+1}^\nu}{b_j},
  \label{eq:down-recurrence}
\end{equation}
from a top boundary condition and normalized to the exact seeds. For flat and open
models the start uses the continued-fraction logarithmic derivative
\begin{equation}
  D_l=l\cotK\chi-\cfrac{b_{l+1}^2}{(2l+3)\cotK\chi-\cfrac{b_{l+2}^2}{(2l+5)\cotK\chi-\cdots}} .
\end{equation}
For closed models ($K=+1$, integer $\nu$) the regular solution is the finite
Gegenbauer form $\phi_l^\nu\propto\sin^l\chi\,C_{\nu-l-1}^{l+1}(\cos\chi)$; the
code starts Miller recurrence either from the finite endpoint $b_\nu\phi_\nu=0$
(when $\nu-l>64$) or directly from $G=C_n^{l+1}(\cos\chi)$ and its derivative near
the endpoint, computed together by the Gegenbauer recurrence and rescaled only
where $G$ is safely non-zero. A direct comparison with \textsc{CLASS} v3.3.4 shows
the same seeds, the same upward/Miller recurrences, the same continued-fraction
derivative and the same Gegenbauer start; the differences are organizational
(\textsc{CLASS} tabulates over $l$ and $\chi$ while \code{phi_recurs} is a
pointwise reference; the normalization seed and the precise closed-start switch
differ).

\section{Reionization heating and the low-redshift linear matter power}
\label{app:reion-heating}

The ionization history enters the late-time matter calculation through the
baryon temperature and sound speed. This is a less clean problem than the CMB
optical depth. Reionization is intrinsically an inhomogeneous, non-linear
radiative-transfer process, but a Boltzmann code evolves a single background
ionization fraction and linear perturbations about it. Treating $x_e(z)$ as a
homogeneous background is already an effective description of the volume-averaged
free-electron density. The gas temperature is even less directly described by a
single number: ionized regions are photo-heated to roughly $10^4\,{\rm K}$, but
the heating is patchy and coupled to sources, sinks, shocks, and the evolving
thermal state of the intergalactic medium.

This matters because the high-$k$ ``linear'' matter power spectrum at low
redshift is not a uniquely physical observable once strongly non-linear
astrophysical processes have acted on the baryons. The ambiguity shows up in
practical calculations at very high $k$, where apparently small choices about
which late-time quantities are evolved can change the nominally linear matter
power at $z=0$. The underlying problem is not that one choice should change the
physical linear solution, but that the baryon sound speed after reionization is
itself an effective convention in a calculation that has deliberately omitted the
non-linear physics that set the ionization and temperature fields.

This ambiguity means that no homogeneous prescription provides a unique physical prediction 
for the post-reionization linear baryon transfer function on sufficiently small scales. 
\textsc{CAMB} therefore adopts a convention that avoids inserting a sharp, model-dependent 
photoheating feature into the default linear transfer functions. The photon opacity and CMB 
anisotropies use the full reionization history, but the baryon sound-speed terms retain the 
residual pre-reionization ionization fraction rather than interpreting the homogeneous reionization step as a homogeneous thermal history.
An optional, disabled-by-default diagnostic
alternative can instead include a crude homogeneous heating prescription. When
\code{include_heating} is enabled, the code defines a clipped heating weight
\begin{equation}
  y_{\rm heat} =
  \operatorname{clip}\!\left[\frac{x_e-x_e^{\rm rec}}{f-x_e^{\rm rec}},\,0,\,1\right],
\end{equation}
where $x_e^{\rm rec}$ is the recombination residual at the same time and $f=1+f_{\rm He}$
is the main-step plateau of \cref{eq:reion-tanh}, corresponding to ionized
hydrogen plus singly ionized helium. Normalizing to $f$ rather than to
unity makes the weight reach one only when the hydrogen and first-helium reionization step
is complete, so the heating smoothly tracks the same $\tanh$ history rather than saturating
partway through it. It then
interpolates the gas temperature from the recombination value to an input
$T_{\rm reion}$, defaulting to $10^4\,{\rm K}$, and interpolates the baryon
sound speed from the usual \textsc{CAMB} expression to the ideal-gas value
\begin{equation}
  c_s^2 = \frac{5}{3}\,\frac{k_{\rm B}T_{\rm gas}}{\mu m_p c^2},
\end{equation}
so $c_s^2$ is dimensionless (the squared sound speed in units of $c^2$, matching the
SI constants used elsewhere). Here $\mu$ is the dimensionless mean molecular weight in
units of $m_p$, set by the ionization state as
$\mu^{-1}=(1-\tfrac34 Y_{\rm He})+(1-Y_{\rm He})\,x_e$. This construction follows the
shape of the homogeneous reionization history, not the physics of patchy heating.
It is intentionally crude.

\begin{figure}[t]
  \centering
  \includegraphics[width=\columnwidth]{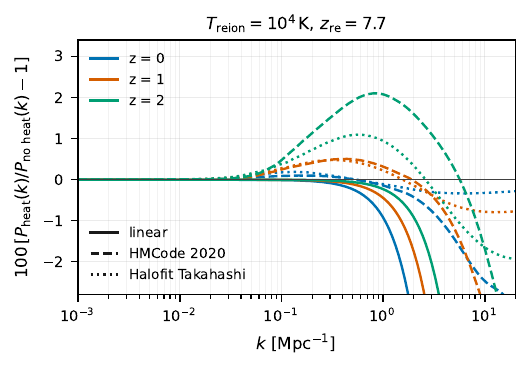}
  \caption{Percentage change in the $z=0,1,2$ total matter power spectrum from enabling
  the crude homogeneous reionization-heating option with
  $T_{\rm reion}=10^4\,{\rm K}$ and $z_{\rm re}=7.7$, for linear theory,
  HMCode 2020~\cite{Mead:2020vgs} and Takahashi Halofit~\cite{Takahashi:2012em}.
  The calculation is a diagnostic of the size of
  the convention-dependent high-$k$ effect, not a physical model of the
  reionized intergalactic medium. In particular, halo-model curves should not be
  interpreted as physical predictions for reionization feedback. The vertical range
  is restricted to show the quasi-linear behaviour, so the linear curves leave the
  bottom of the panel at $k\gtrsim$ a few $\,{\rm Mpc}^{-1}$.}
  \label{fig:reion-heating}
\end{figure}

\Cref{fig:reion-heating} illustrates the scale of the effect in a simple
$\Lambda$CDM example. The linear calculation shows the expected suppression of
small-scale baryon clustering from the larger post-reionization sound speed, which
persists to low redshift. This suppression is small on large scales, around a
percent at $k\sim1\,{\rm Mpc}^{-1}$ at $z=0$, but grows rapidly towards smaller
scales and is notionally very large, reaching several tens of percent by
$k\sim10$--$20\,{\rm Mpc}^{-1}$, where the linear curves run off the bottom of the
panel. These are scales on which the linear spectrum is not itself a physical
observable, so the precise size is only indicative.

The halo-model curves are shown as a warning about propagation through common
non-linear post-processing prescriptions. Fitting functions such as these are
calibrated on $N$-body simulations with standard $\Lambda$CDM linear input, and
are not calibrated for a linear spectrum carrying a localized high-$k$ feature of
this kind; their response is therefore not a physical prediction of reionization
feedback, and in detail depends on which linear model the non-linear prescription
was tuned against. The most striking artefact is not the mild small-scale
suppression but a percent-level \emph{increase} of power on quasi-linear scales
($k\sim0.3$--$1\,{\rm Mpc}^{-1}$, most pronounced at the higher redshifts shown),
where the heating should physically remove rather than add power. Since these are
among the scales most often used in practice, this spurious quasi-linear gain is
arguably a more relevant caution than the small-scale suppression. More
generally, a homogeneous heating prescription makes an additional low-redshift
contribution to the scale dependence of growth, so it is not consistent with the
scale-independent growth approximation often used to rescale linear spectra or
combine transfer functions outside the code.

The purpose of the option is therefore modest. It is not a model of reionization
heating, and it should not be used to infer the thermal history of the
intergalactic medium. It is a way to estimate the order of magnitude of one
definition-dependent effect in what users call the low-redshift linear matter
power spectrum. It is also a reminder that when a calculation asks for linear
results after non-linear astrophysics has changed the baryon state, the exact
question being answered needs to be specified carefully.

\section{Accuracy parameters}
\label{app:accuracy}

The user-facing accuracy is set by a small number of boost parameters, all
defaulting to one, with larger values tightening the result towards the converged
limit. They form a hierarchy rather than a set of independent line-by-line
tolerances, and three high-level controls do most of the work. \code{AccuracyBoost}
is the broad umbrella factor: it multiplies most of the lower-level controls at
once -- time and background sampling, integration tolerances, and the various
source, transfer and lensing $k$-sampling densities -- so raising it tightens the
whole calculation together. \code{lSampleBoost} and \code{lAccuracyBoost} are
largely orthogonal to it and to each other: the first sets the density of
explicitly computed multipoles before the output $C_\ell$ are filled in by
interpolation, the second the depth of the Boltzmann hierarchies (the number of
multipoles evolved).

Beneath \code{AccuracyBoost} sit a number of narrower sub-controls -- among them
\code{IntTolBoost} (integration tolerances), \code{BackgroundTimeStepBoost} (the
background and recombination grid), and the source, transfer and non-flat
$k$-sampling boosts -- each of which is \emph{also} multiplied by \code{AccuracyBoost}
and can be raised on its own for targeted tuning without inflating the global
factor. Standing outside this boost hierarchy, \code{lens_potential_accuracy} sets
the maximum wavenumber $k_{\max}$ for the lensing and transfer integration
(\cref{sec:lensing}); left unset it is chosen automatically from the requested
$\ell_{\max}$ as $\max(4,(\ell_{\max}-1500)/500)$ rather than defaulting to one.

A useful workflow is therefore to use the high-level boosts in \code{check_accuracy}
to expose instabilities and coupled failures, then drill down to the smallest
sub-control or code change that removes the problem. Production analyses should
rarely need $\code{AccuracyBoost} >1$; where the defaults are insufficient, targeted
sub-boosts usually give much better performance than raising the global factor.

The validation reference used by \code{check_accuracy} (\cref{sec:validation})
by default doubles \code{AccuracyBoost}, \code{lSampleBoost}, \code{lAccuracyBoost} and
\code{IntTolBoost} together. A single doubling does not by itself prove absolute
convergence -- it measures the distance to a more converged run, not to the exact
result -- but the default-to-reference differences are stable when the reference is
tightened further, for example to the stricter \code{--strict-reference} preset
(\code{AccuracyBoost}=\code{lSampleBoost}=\code{lAccuracyBoost}=3 with denser
high-\(\ell\) sampling, used for \cref{fig:lensed-cl-validation}). That stability
indicates the doubled reference is already close to converged, so the
default-to-reference difference is a good estimate of the actual error in the
default calculation.
\Cref{tab:tol} lists
the validated acceptance tolerances: these are maximum absolute fractional errors
within each range, not RMS errors over the range. The actual errors are therefore
usually much smaller than the quoted numbers, except for isolated points that
happen to set the maximum.

\begin{table}[t]
\centering
\small
\begin{tabular}{lll}
\toprule
Output & Range & Tolerance \\
\midrule
Lensed $TT,EE$ & $2\le\ell<600$ & $3\times10^{-3}$ \\
(and normalized $TE$) & $600\le\ell<3500$ & $1\times10^{-3}$ \\
 & $3500\le\ell<6000$ & $3\times10^{-3}$ \\
 & $\ell\ge6000$ & $2\times10^{-2}$ \\
\midrule
Lensing potential & $2\le L<5$ & $9\times10^{-3}$ \\
$C_L^{\phi\phi}$ & $5\le L<2500$ & $5\times10^{-3}$ \\
 & $2500\le L<4000$ & $1\times10^{-2}$ \\
 & $L\ge4000$ & $2\times10^{-2}$ \\
\midrule
$BB$ & $2\le\ell<1000$ & $5\times10^{-3}$ \\
 & $1000\le\ell<6000$ & $1\times10^{-2}$ \\
 & $6000\le\ell<8000$ & $2\times10^{-2}$ \\
 & $\ell\ge8000$ & $1\times10^{-1}$ \\
\midrule
Matter power & $k_h<5\times10^{-3}$ & $3\times10^{-3}$ \\
(linear) & $5\times10^{-3}<k_h<2$ & $1\times10^{-3}$ \\
 & $k_h>2$ & $3\times10^{-3}$ \\
\midrule
Matter power & $k_h<5\times10^{-3}$ & $3\times10^{-3}$ \\
(nonlinear) & $5\times10^{-3}<k_h<2$ & $1\times10^{-3}$ \\
 & $2<k_h<10$ & $3\times10^{-3}$ \\
 & $k_h>10$ & $1\times10^{-2}$ \\
\bottomrule
\end{tabular}
\caption{Validated acceptance tolerances for the default settings, applied by
\code{check_accuracy} as maximum absolute fractional differences against the
boosted reference, not RMS differences over the range; here
$k_h\equiv k/(h\,\mathrm{Mpc}^{-1})$. The $10^{-3}$ entries -- the interior lensed CMB and the
quasi-linear matter power -- are the design target for Advanced SO and the
matter-power surveys; the looser entries are at the spectrum ends, in the
high-$\ell$ lensed tail, and in the deeply nonlinear regime where the nonlinear
model sets the floor. The matter-power tolerances are those applied with
\code{Transfer.high_precision} set, which should always be enabled when
accurate matter power spectra are needed.}
\label{tab:tol}
\end{table}

To map spectrum-level differences to a likelihood scale, the accuracy-checking
code uses a simple SO-like Gaussian mean likelihood: at each multipole the
covariance is that expected from cosmic variance on sky fraction
$f_{\rm sky}=0.6$ with a 1.4 arcmin beam and white noise levels of 2.6 and
3.67 $\mu{\rm K}$-arcmin in temperature and polarization, and the quadratic
spectral differences are summed over $2\le \ell\le4000$. This is a deliberately 
compact and stringent numerical-accuracy proxy for the more complete Simons Observatory
forecasting likelihoods~\cite{SO:2018sim,SOEnhancedLAT:2025}; the same
diagnostic quantifies the recombination-model comparison in
\cref{sec:recomb-model}. To put the maximum-error tolerances of \cref{tab:tol}
on this scale, we evaluated this likelihood with the Planck-2018 best-fit
spectra as the fiducial model. This is a pessimistic envelope calculation, not
a model for the usual error distribution. If the errors in $TT$, $EE$, normalized $TE$ and
$BB$ at every multipole were independent random numbers uniformly distributed up
to the maximum values in \cref{tab:tol}, the expectation would be
$\Delta\chi^2\simeq9.6$ (with approximately 2.4, 5.0 and 2.2 from the three
multipole ranges $2\le\ell<600$, $600\le\ell<3500$ and
$3500\le\ell\le4000$). If instead every one of these spectra sat coherently at
the maximum-error boundary, the same calculation would give
$\Delta\chi^2\simeq34$; choosing the global sign combination that maximizes the
distance changes this only slightly, to $\Delta\chi^2\simeq35$. Real numerical
differences are much smaller than this envelope: for the same Planck-2018
fiducial model, the actual default calculation compared to the boosted
\code{check_accuracy} reference gives $\Delta\chi^2\simeq0.46$.
The lensed $BB$ spectrum contributes approximately  $\Delta\chi^2\simeq0.01$,
compared to $\Delta\chi^2\simeq1.6$ for the independent random envelope, or
$\Delta\chi^2\simeq4.9$ for a coherent residual at the $BB$ tolerance boundary.

In real parameter analyses the required numerical accuracy is much less severe:
foreground residuals, calibration, beams and other instrumental nuisance
parameters, and nonlinear lensing modelling uncertainties generally set the
practical floor before numerical errors at this conservative envelope would move
parameters appreciably.

For the lensing-potential we use the public SO MV $N_L^{\kappa\kappa}$ curve,
$f_{\rm sky}=0.4$, and the scalar Gaussian likelihood for $C_L^{\kappa\kappa}$
over $2\le L\le3000$. The same maximum-error envelope gives
$\Delta\chi^2\simeq0.14$ for independent random residuals and
$\Delta\chi^2\simeq0.43$ for a coherent residual at the tolerance boundary. The
result is dominated by $5\le L<2500$; extending the same proxy to $L=4000$
changes the numbers by less than one per cent. The matter-power tolerances still
need a separate large-scale-structure likelihood to map them to a likelihood
distance.

\bibliographystyle{apsrev4-2}

\begin{thebibliography}{75}%
\makeatletter
\providecommand \@ifxundefined [1]{%
 \@ifx{#1\undefined}
}%
\providecommand \@ifnum [1]{%
 \ifnum #1\expandafter \@firstoftwo
 \else \expandafter \@secondoftwo
 \fi
}%
\providecommand \@ifx [1]{%
 \ifx #1\expandafter \@firstoftwo
 \else \expandafter \@secondoftwo
 \fi
}%
\providecommand \natexlab [1]{#1}%
\providecommand \enquote  [1]{``#1''}%
\providecommand \bibnamefont  [1]{#1}%
\providecommand \bibfnamefont [1]{#1}%
\providecommand \citenamefont [1]{#1}%
\providecommand \href@noop [0]{\@secondoftwo}%
\providecommand \href [0]{\begingroup \@sanitize@url \@href}%
\providecommand \@href[1]{\@@startlink{#1}\@@href}%
\providecommand \@@href[1]{\endgroup#1\@@endlink}%
\providecommand \@sanitize@url [0]{\catcode `\\12\catcode `\$12\catcode
  `\&12\catcode `\#12\catcode `\^12\catcode `\_12\catcode `\%12\relax}%
\providecommand \@@startlink[1]{}%
\providecommand \@@endlink[0]{}%
\providecommand \url  [0]{\begingroup\@sanitize@url \@url }%
\providecommand \@url [1]{\endgroup\@href {#1}{\urlprefix }}%
\providecommand \urlprefix  [0]{URL }%
\providecommand \Eprint [0]{\href }%
\providecommand \doibase [0]{https://doi.org/}%
\providecommand \selectlanguage [0]{\@gobble}%
\providecommand \bibinfo  [0]{\@secondoftwo}%
\providecommand \bibfield  [0]{\@secondoftwo}%
\providecommand \translation [1]{[#1]}%
\providecommand \BibitemOpen [0]{}%
\providecommand \bibitemStop [0]{}%
\providecommand \bibitemNoStop [0]{.\EOS\space}%
\providecommand \EOS [0]{\spacefactor3000\relax}%
\providecommand \BibitemShut  [1]{\csname bibitem#1\endcsname}%
\let\auto@bib@innerbib\@empty
%</preamble>
\bibitem [{\citenamefont {Lewis}\ \emph {et~al.}(2000)\citenamefont {Lewis},
  \citenamefont {Challinor},\ and\ \citenamefont {Lasenby}}]{Lewis:1999bs}%
  \BibitemOpen
  \bibfield  {author} {\bibinfo {author} {\bibfnamefont {A.}~\bibnamefont
  {Lewis}}, \bibinfo {author} {\bibfnamefont {A.}~\bibnamefont {Challinor}},\
  and\ \bibinfo {author} {\bibfnamefont {A.}~\bibnamefont {Lasenby}},\ }\href
  {https://doi.org/10.1086/309179} {\bibfield  {journal} {\bibinfo  {journal}
  {\apj}\ }\textbf {\bibinfo {volume} {538}},\ \bibinfo {pages} {473} (\bibinfo
  {year} {2000})},\ \Eprint {https://arxiv.org/abs/astro-ph/9911177}
  {arXiv:astro-ph/9911177 [astro-ph]} \BibitemShut {NoStop}%
%%CITATION = ASTRO-PH/9911177;%%
\bibitem [{\citenamefont {Blas}\ \emph {et~al.}(2011)\citenamefont {Blas},
  \citenamefont {Lesgourgues},\ and\ \citenamefont {Tram}}]{Blas:2011rf}%
  \BibitemOpen
  \bibfield  {author} {\bibinfo {author} {\bibfnamefont {D.}~\bibnamefont
  {Blas}}, \bibinfo {author} {\bibfnamefont {J.}~\bibnamefont {Lesgourgues}},\
  and\ \bibinfo {author} {\bibfnamefont {T.}~\bibnamefont {Tram}},\ }\href
  {https://doi.org/10.1088/1475-7516/2011/07/034} {\bibfield  {journal}
  {\bibinfo  {journal} {\jcap}\ }\textbf {\bibinfo {volume} {1107}},\ \bibinfo
  {pages} {034} (\bibinfo {year} {2011})},\ \Eprint
  {https://arxiv.org/abs/1104.2933} {arXiv:1104.2933 [astro-ph.CO]}
  \BibitemShut {NoStop}%
%%CITATION = 1104.2933;%%
\bibitem [{\citenamefont {Seljak}\ and\ \citenamefont
  {Zaldarriaga}(1996)}]{Seljak:1996is}%
  \BibitemOpen
  \bibfield  {author} {\bibinfo {author} {\bibfnamefont {U.}~\bibnamefont
  {Seljak}}\ and\ \bibinfo {author} {\bibfnamefont {M.}~\bibnamefont
  {Zaldarriaga}},\ }\href {https://doi.org/10.1086/177793} {\bibfield
  {journal} {\bibinfo  {journal} {\apj}\ }\textbf {\bibinfo {volume} {469}},\
  \bibinfo {pages} {437} (\bibinfo {year} {1996})},\ \Eprint
  {https://arxiv.org/abs/astro-ph/9603033} {arXiv:astro-ph/9603033}
  \BibitemShut {NoStop}%
\bibitem [{\citenamefont {Ma}\ and\ \citenamefont
  {Bertschinger}(1995)}]{Ma:1995ey}%
  \BibitemOpen
  \bibfield  {author} {\bibinfo {author} {\bibfnamefont {C.-P.}\ \bibnamefont
  {Ma}}\ and\ \bibinfo {author} {\bibfnamefont {E.}~\bibnamefont
  {Bertschinger}},\ }\href@noop {} {\bibfield  {journal} {\bibinfo  {journal}
  {\apj}\ }\textbf {\bibinfo {volume} {455}},\ \bibinfo {pages} {7} (\bibinfo
  {year} {1995})},\ \Eprint {https://arxiv.org/abs/astro-ph/9506072}
  {astro-ph/9506072} \BibitemShut {NoStop}%
%%CITATION = ASTRO-PH 9506072;%%
\bibitem [{\citenamefont {Bertschinger}(1995)}]{Bertschinger:1995er}%
  \BibitemOpen
  \bibfield  {author} {\bibinfo {author} {\bibfnamefont {E.}~\bibnamefont
  {Bertschinger}},\ }\href@noop {} {\bibinfo {title} {{COSMICS: cosmological
  initial conditions and microwave anisotropy codes}}} (\bibinfo {year}
  {1995}),\ \Eprint {https://arxiv.org/abs/astro-ph/9506070}
  {arXiv:astro-ph/9506070} \BibitemShut {NoStop}%
\bibitem [{\citenamefont {Hu}\ \emph {et~al.}(1998)\citenamefont {Hu},
  \citenamefont {Seljak}, \citenamefont {White},\ and\ \citenamefont
  {Zaldarriaga}}]{Hu:1998mn}%
  \BibitemOpen
  \bibfield  {author} {\bibinfo {author} {\bibfnamefont {W.}~\bibnamefont
  {Hu}}, \bibinfo {author} {\bibfnamefont {U.}~\bibnamefont {Seljak}}, \bibinfo
  {author} {\bibfnamefont {M.~J.}\ \bibnamefont {White}},\ and\ \bibinfo
  {author} {\bibfnamefont {M.}~\bibnamefont {Zaldarriaga}},\ }\href@noop {}
  {\bibfield  {journal} {\bibinfo  {journal} {\prd}\ }\textbf {\bibinfo
  {volume} {57}},\ \bibinfo {pages} {3290} (\bibinfo {year} {1998})},\ \Eprint
  {https://arxiv.org/abs/astro-ph/9709066} {astro-ph/9709066} \BibitemShut
  {NoStop}%
%%CITATION = ASTRO-PH 9709066;%%
\bibitem [{\citenamefont {Zaldarriaga}\ \emph {et~al.}(1998)\citenamefont
  {Zaldarriaga}, \citenamefont {Seljak},\ and\ \citenamefont
  {Bertschinger}}]{Zaldarriaga:1997va}%
  \BibitemOpen
  \bibfield  {author} {\bibinfo {author} {\bibfnamefont {M.}~\bibnamefont
  {Zaldarriaga}}, \bibinfo {author} {\bibfnamefont {U.}~\bibnamefont
  {Seljak}},\ and\ \bibinfo {author} {\bibfnamefont {E.}~\bibnamefont
  {Bertschinger}},\ }\href {https://doi.org/10.1086/305223} {\bibfield
  {journal} {\bibinfo  {journal} {\apj}\ }\textbf {\bibinfo {volume} {494}},\
  \bibinfo {pages} {491} (\bibinfo {year} {1998})},\ \Eprint
  {https://arxiv.org/abs/astro-ph/9704265} {arXiv:astro-ph/9704265 [astro-ph]}
  \BibitemShut {NoStop}%
%%CITATION = ASTRO-PH/9704265;%%
\bibitem [{\citenamefont {Challinor}(2000)}]{Challinor:2000as}%
  \BibitemOpen
  \bibfield  {author} {\bibinfo {author} {\bibfnamefont {A.}~\bibnamefont
  {Challinor}},\ }\href {https://doi.org/10.1103/PhysRevD.62.043004} {\bibfield
   {journal} {\bibinfo  {journal} {\prd}\ }\textbf {\bibinfo {volume} {62}},\
  \bibinfo {pages} {043004} (\bibinfo {year} {2000})},\ \Eprint
  {https://arxiv.org/abs/astro-ph/9911481} {arXiv:astro-ph/9911481 [astro-ph]}
  \BibitemShut {NoStop}%
%%CITATION = ASTRO-PH/9911481;%%
\bibitem [{\citenamefont {Howlett}\ \emph {et~al.}(2012)\citenamefont
  {Howlett}, \citenamefont {Lewis}, \citenamefont {Hall},\ and\ \citenamefont
  {Challinor}}]{Howlett:2012mh}%
  \BibitemOpen
  \bibfield  {author} {\bibinfo {author} {\bibfnamefont {C.}~\bibnamefont
  {Howlett}}, \bibinfo {author} {\bibfnamefont {A.}~\bibnamefont {Lewis}},
  \bibinfo {author} {\bibfnamefont {A.}~\bibnamefont {Hall}},\ and\ \bibinfo
  {author} {\bibfnamefont {A.}~\bibnamefont {Challinor}},\ }\href
  {https://doi.org/10.1088/1475-7516/2012/04/027} {\bibfield  {journal}
  {\bibinfo  {journal} {\jcap}\ }\textbf {\bibinfo {volume} {1204}},\ \bibinfo
  {pages} {027} (\bibinfo {year} {2012})},\ \Eprint
  {https://arxiv.org/abs/1201.3654} {arXiv:1201.3654 [astro-ph.CO]}
  \BibitemShut {NoStop}%
%%CITATION = ARXIV:1201.3654;%%
\bibitem [{\citenamefont {Ade}\ \emph {et~al.}(2019)\citenamefont {Ade} \emph
  {et~al.}}]{SO:2018sim}%
  \BibitemOpen
  \bibfield  {author} {\bibinfo {author} {\bibfnamefont {P.}~\bibnamefont
  {Ade}} \emph {et~al.} (\bibinfo {collaboration} {Simons Observatory}),\
  }\href {https://doi.org/10.1088/1475-7516/2019/02/056} {\bibfield  {journal}
  {\bibinfo  {journal} {JCAP}\ }\textbf {\bibinfo {volume} {02}},\ \bibinfo
  {pages} {056}},\ \Eprint {https://arxiv.org/abs/1808.07445} {arXiv:1808.07445
  [astro-ph.CO]} \BibitemShut {NoStop}%
\bibitem [{\citenamefont {Lewis}\ and\ \citenamefont
  {Challinor}(2006)}]{Lewis:2006fu}%
  \BibitemOpen
  \bibfield  {author} {\bibinfo {author} {\bibfnamefont {A.}~\bibnamefont
  {Lewis}}\ and\ \bibinfo {author} {\bibfnamefont {A.}~\bibnamefont
  {Challinor}},\ }\href {https://doi.org/10.1016/j.physrep.2006.03.002}
  {\bibfield  {journal} {\bibinfo  {journal} {Phys. Rept.}\ }\textbf {\bibinfo
  {volume} {429}},\ \bibinfo {pages} {1} (\bibinfo {year} {2006})},\ \Eprint
  {https://arxiv.org/abs/astro-ph/0601594} {arXiv:astro-ph/0601594 [astro-ph]}
  \BibitemShut {NoStop}%
%%CITATION = ASTRO-PH/0601594;%%
\bibitem [{\citenamefont {Kosowsky}(1998)}]{Kosowsky:1998nc}%
  \BibitemOpen
  \bibfield  {author} {\bibinfo {author} {\bibfnamefont {A.}~\bibnamefont
  {Kosowsky}},\ }\href@noop {} {\bibinfo {title} {Efficient computation of
  hyperspherical {Bessel} functions}} (\bibinfo {year} {1998}),\ \Eprint
  {https://arxiv.org/abs/astro-ph/9805173} {astro-ph/9805173} \BibitemShut
  {NoStop}%
%%CITATION = ASTRO-PH 9805173;%%
\bibitem [{\citenamefont {Tram}(2017)}]{tram2013}%
  \BibitemOpen
  \bibfield  {author} {\bibinfo {author} {\bibfnamefont {T.}~\bibnamefont
  {Tram}},\ }\href {https://doi.org/10.4208/cicp.OA-2016-0071} {\bibfield
  {journal} {\bibinfo  {journal} {Commun. Comput. Phys.}\ }\textbf {\bibinfo
  {volume} {22}},\ \bibinfo {pages} {852} (\bibinfo {year} {2017})},\ \Eprint
  {https://arxiv.org/abs/1311.0839} {arXiv:1311.0839 [astro-ph.CO]}
  \BibitemShut {NoStop}%
\bibitem [{\citenamefont {Lesgourgues}\ and\ \citenamefont
  {Tram}(2014)}]{lesgourgues-tram2013}%
  \BibitemOpen
  \bibfield  {author} {\bibinfo {author} {\bibfnamefont {J.}~\bibnamefont
  {Lesgourgues}}\ and\ \bibinfo {author} {\bibfnamefont {T.}~\bibnamefont
  {Tram}},\ }\href {https://doi.org/10.1088/1475-7516/2014/09/032} {\bibfield
  {journal} {\bibinfo  {journal} {JCAP}\ }\textbf {\bibinfo {volume} {09}},\
  \bibinfo {pages} {032}},\ \Eprint {https://arxiv.org/abs/1312.2697}
  {arXiv:1312.2697 [astro-ph.CO]} \BibitemShut {NoStop}%
\bibitem [{\citenamefont {Lewis}(2004)}]{Lewis:2004ef}%
  \BibitemOpen
  \bibfield  {author} {\bibinfo {author} {\bibfnamefont {A.}~\bibnamefont
  {Lewis}},\ }\href@noop {} {\bibfield  {journal} {\bibinfo  {journal} {\prd}\
  }\textbf {\bibinfo {volume} {70}},\ \bibinfo {pages} {043011} (\bibinfo
  {year} {2004})},\ \Eprint {https://arxiv.org/abs/astro-ph/0406096}
  {astro-ph/0406096} \BibitemShut {NoStop}%
%%CITATION = ASTRO-PH 0406096;%%
\bibitem [{\citenamefont {Lewis}()}]{camb_notes}%
  \BibitemOpen
  \bibfield  {author} {\bibinfo {author} {\bibfnamefont {A.}~\bibnamefont
  {Lewis}},\ }\bibinfo {note}
  {\url{https://cosmologist.info/notes/CAMB.pdf}}\BibitemShut {NoStop}%
\bibitem [{\citenamefont {Bucher}\ \emph {et~al.}(2000)\citenamefont {Bucher},
  \citenamefont {Moodley},\ and\ \citenamefont {Turok}}]{Bucher:1999re}%
  \BibitemOpen
  \bibfield  {author} {\bibinfo {author} {\bibfnamefont {M.}~\bibnamefont
  {Bucher}}, \bibinfo {author} {\bibfnamefont {K.}~\bibnamefont {Moodley}},\
  and\ \bibinfo {author} {\bibfnamefont {N.}~\bibnamefont {Turok}},\
  }\href@noop {} {\bibfield  {journal} {\bibinfo  {journal} {\prd}\ }\textbf
  {\bibinfo {volume} {62}},\ \bibinfo {pages} {083508} (\bibinfo {year}
  {2000})},\ \Eprint {https://arxiv.org/abs/astro-ph/9904231}
  {astro-ph/9904231} \BibitemShut {NoStop}%
%%CITATION = ASTRO-PH 9904231;%%
\bibitem [{\citenamefont {Cyr-Racine}\ and\ \citenamefont
  {Sigurdson}(2011)}]{CyrRacine:2010bk}%
  \BibitemOpen
  \bibfield  {author} {\bibinfo {author} {\bibfnamefont {F.-Y.}\ \bibnamefont
  {Cyr-Racine}}\ and\ \bibinfo {author} {\bibfnamefont {K.}~\bibnamefont
  {Sigurdson}},\ }\href {https://doi.org/10.1103/PhysRevD.83.103521} {\bibfield
   {journal} {\bibinfo  {journal} {\prd}\ }\textbf {\bibinfo {volume} {83}},\
  \bibinfo {pages} {103521} (\bibinfo {year} {2011})},\ \Eprint
  {https://arxiv.org/abs/1012.0569} {arXiv:1012.0569 [astro-ph.CO]}
  \BibitemShut {NoStop}%
%%CITATION = 1012.0569;%%
\bibitem [{\citenamefont {Dormand}\ and\ \citenamefont
  {Prince}(1980)}]{DormandPrince:1980}%
  \BibitemOpen
  \bibfield  {author} {\bibinfo {author} {\bibfnamefont {J.~R.}\ \bibnamefont
  {Dormand}}\ and\ \bibinfo {author} {\bibfnamefont {P.~J.}\ \bibnamefont
  {Prince}},\ }\href {https://doi.org/10.1016/0771-050X(80)90013-3} {\bibfield
  {journal} {\bibinfo  {journal} {J. Comput. Appl. Math.}\ }\textbf {\bibinfo
  {volume} {6}},\ \bibinfo {pages} {19} (\bibinfo {year} {1980})}\BibitemShut
  {NoStop}%
\bibitem [{\citenamefont {Abbott}\ and\ \citenamefont
  {Schaefer}(1986)}]{abbott-schaefer}%
  \BibitemOpen
  \bibfield  {author} {\bibinfo {author} {\bibfnamefont {L.~F.}\ \bibnamefont
  {Abbott}}\ and\ \bibinfo {author} {\bibfnamefont {R.~K.}\ \bibnamefont
  {Schaefer}},\ }\href {https://doi.org/10.1086/164525} {\bibfield  {journal}
  {\bibinfo  {journal} {Astrophys. J.}\ }\textbf {\bibinfo {volume} {308}},\
  \bibinfo {pages} {546} (\bibinfo {year} {1986})}\BibitemShut {NoStop}%
\bibitem [{\citenamefont {Olver}(1958)}]{olver1958uniform}%
  \BibitemOpen
  \bibfield  {author} {\bibinfo {author} {\bibfnamefont {F.~W.~J.}\
  \bibnamefont {Olver}},\ }\href {https://doi.org/10.1098/rsta.1958.0021}
  {\bibfield  {journal} {\bibinfo  {journal} {Philos. Trans. Roy. Soc. Lond.
  A}\ }\textbf {\bibinfo {volume} {250}},\ \bibinfo {pages} {479} (\bibinfo
  {year} {1958})}\BibitemShut {NoStop}%
\bibitem [{\citenamefont {Olver}(1974)}]{olver}%
  \BibitemOpen
  \bibfield  {author} {\bibinfo {author} {\bibfnamefont {F.~W.~J.}\
  \bibnamefont {Olver}},\ }\href@noop {} {\emph {\bibinfo {title} {{Asymptotics
  and Special Functions}}}}\ (\bibinfo  {publisher} {Academic Press},\ \bibinfo
  {address} {New York},\ \bibinfo {year} {1974})\BibitemShut {NoStop}%
\bibitem [{\citenamefont {Lewis}(2026{\natexlab{a}})}]{cambflatbessel}%
  \BibitemOpen
  \bibfield  {author} {\bibinfo {author} {\bibfnamefont {A.}~\bibnamefont
  {Lewis}}} (\bibinfo {year} {2026}{\natexlab{a}}),\ \bibinfo {note}
  {\url{https://github.com/cmbant/CAMB/blob/master/docs/changelog/flat_bessel_approximations.pdf}}\BibitemShut
  {NoStop}%
\bibitem [{\citenamefont {Numerov}(1924)}]{Numerov:1924}%
  \BibitemOpen
  \bibfield  {author} {\bibinfo {author} {\bibfnamefont {B.~V.}\ \bibnamefont
  {Numerov}},\ }\href {https://doi.org/10.1093/mnras/84.8.592} {\bibfield
  {journal} {\bibinfo  {journal} {Mon. Not. Roy. Astron. Soc.}\ }\textbf
  {\bibinfo {volume} {84}},\ \bibinfo {pages} {592} (\bibinfo {year}
  {1924})}\BibitemShut {NoStop}%
\bibitem [{\citenamefont {Numerov}(1927)}]{Numerov:1927}%
  \BibitemOpen
  \bibfield  {author} {\bibinfo {author} {\bibfnamefont {B.~V.}\ \bibnamefont
  {Numerov}},\ }\href {https://doi.org/10.1002/asna.19272301903} {\bibfield
  {journal} {\bibinfo  {journal} {Astron. Nachr.}\ }\textbf {\bibinfo {volume}
  {230}},\ \bibinfo {pages} {359} (\bibinfo {year} {1927})}\BibitemShut
  {NoStop}%
\bibitem [{\citenamefont {Hamann}\ \emph {et~al.}(2008)\citenamefont {Hamann},
  \citenamefont {Lesgourgues},\ and\ \citenamefont {Mangano}}]{Hamann:2007sb}%
  \BibitemOpen
  \bibfield  {author} {\bibinfo {author} {\bibfnamefont {J.}~\bibnamefont
  {Hamann}}, \bibinfo {author} {\bibfnamefont {J.}~\bibnamefont
  {Lesgourgues}},\ and\ \bibinfo {author} {\bibfnamefont {G.}~\bibnamefont
  {Mangano}},\ }\href {https://doi.org/10.1088/1475-7516/2008/03/004}
  {\bibfield  {journal} {\bibinfo  {journal} {\jcap}\ }\textbf {\bibinfo
  {volume} {0803}},\ \bibinfo {pages} {004} (\bibinfo {year} {2008})},\ \Eprint
  {https://arxiv.org/abs/0712.2826} {arXiv:0712.2826 [astro-ph]} \BibitemShut
  {NoStop}%
%%CITATION = ARXIV:0712.2826;%%
\bibitem [{\citenamefont {Pisanti}\ \emph {et~al.}(2008)\citenamefont
  {Pisanti}, \citenamefont {Cirillo}, \citenamefont {Esposito}, \citenamefont
  {Iocco}, \citenamefont {Mangano}, \citenamefont {Miele},\ and\ \citenamefont
  {Serpico}}]{Pisanti:2007hk}%
  \BibitemOpen
  \bibfield  {author} {\bibinfo {author} {\bibfnamefont {O.}~\bibnamefont
  {Pisanti}}, \bibinfo {author} {\bibfnamefont {A.}~\bibnamefont {Cirillo}},
  \bibinfo {author} {\bibfnamefont {S.}~\bibnamefont {Esposito}}, \bibinfo
  {author} {\bibfnamefont {F.}~\bibnamefont {Iocco}}, \bibinfo {author}
  {\bibfnamefont {G.}~\bibnamefont {Mangano}}, \bibinfo {author} {\bibfnamefont
  {G.}~\bibnamefont {Miele}},\ and\ \bibinfo {author} {\bibfnamefont {P.~D.}\
  \bibnamefont {Serpico}},\ }\href {https://doi.org/10.1016/j.cpc.2008.02.015}
  {\bibfield  {journal} {\bibinfo  {journal} {Comput. Phys. Commun.}\ }\textbf
  {\bibinfo {volume} {178}},\ \bibinfo {pages} {956} (\bibinfo {year}
  {2008})},\ \Eprint {https://arxiv.org/abs/0705.0290} {arXiv:0705.0290
  [astro-ph]} \BibitemShut {NoStop}%
%%CITATION = ARXIV:0705.0290;%%
\bibitem [{\citenamefont {Consiglio}\ \emph {et~al.}(2018)\citenamefont
  {Consiglio}, \citenamefont {de~Salas}, \citenamefont {Mangano}, \citenamefont
  {Miele}, \citenamefont {Pastor},\ and\ \citenamefont
  {Pisanti}}]{Consiglio:2017pot}%
  \BibitemOpen
  \bibfield  {author} {\bibinfo {author} {\bibfnamefont {R.}~\bibnamefont
  {Consiglio}}, \bibinfo {author} {\bibfnamefont {P.~F.}\ \bibnamefont
  {de~Salas}}, \bibinfo {author} {\bibfnamefont {G.}~\bibnamefont {Mangano}},
  \bibinfo {author} {\bibfnamefont {G.}~\bibnamefont {Miele}}, \bibinfo
  {author} {\bibfnamefont {S.}~\bibnamefont {Pastor}},\ and\ \bibinfo {author}
  {\bibfnamefont {O.}~\bibnamefont {Pisanti}},\ }\href
  {https://doi.org/10.1016/j.cpc.2018.06.022} {\bibfield  {journal} {\bibinfo
  {journal} {Comput. Phys. Commun.}\ }\textbf {\bibinfo {volume} {233}},\
  \bibinfo {pages} {237} (\bibinfo {year} {2018})},\ \Eprint
  {https://arxiv.org/abs/1712.04378} {arXiv:1712.04378 [astro-ph.CO]}
  \BibitemShut {NoStop}%
\bibitem [{\citenamefont {Pitrou}\ \emph {et~al.}(2018)\citenamefont {Pitrou},
  \citenamefont {Coc}, \citenamefont {Uzan},\ and\ \citenamefont
  {Vangioni}}]{Pitrou:2018cgg}%
  \BibitemOpen
  \bibfield  {author} {\bibinfo {author} {\bibfnamefont {C.}~\bibnamefont
  {Pitrou}}, \bibinfo {author} {\bibfnamefont {A.}~\bibnamefont {Coc}},
  \bibinfo {author} {\bibfnamefont {J.-P.}\ \bibnamefont {Uzan}},\ and\
  \bibinfo {author} {\bibfnamefont {E.}~\bibnamefont {Vangioni}},\ }\href
  {https://doi.org/10.1016/j.physrep.2018.04.005} {\bibfield  {journal}
  {\bibinfo  {journal} {Phys. Rept.}\ }\textbf {\bibinfo {volume} {754}},\
  \bibinfo {pages} {1} (\bibinfo {year} {2018})},\ \Eprint
  {https://arxiv.org/abs/1801.08023} {arXiv:1801.08023 [astro-ph.CO]}
  \BibitemShut {NoStop}%
\bibitem [{\citenamefont {Trotta}\ and\ \citenamefont
  {Hansen}(2004)}]{Trotta:2003xg}%
  \BibitemOpen
  \bibfield  {author} {\bibinfo {author} {\bibfnamefont {R.}~\bibnamefont
  {Trotta}}\ and\ \bibinfo {author} {\bibfnamefont {S.~H.}\ \bibnamefont
  {Hansen}},\ }\href@noop {} {\bibfield  {journal} {\bibinfo  {journal} {\prd}\
  }\textbf {\bibinfo {volume} {69}},\ \bibinfo {pages} {023509} (\bibinfo
  {year} {2004})},\ \Eprint {https://arxiv.org/abs/astro-ph/0306588}
  {astro-ph/0306588} \BibitemShut {NoStop}%
%%CITATION = ASTRO-PH 0306588;%%
\bibitem [{\citenamefont {Aghanim}\ \emph {et~al.}(2020)\citenamefont {Aghanim}
  \emph {et~al.}}]{PCP2018}%
  \BibitemOpen
  \bibfield  {author} {\bibinfo {author} {\bibfnamefont {N.}~\bibnamefont
  {Aghanim}} \emph {et~al.} (\bibinfo {collaboration} {Planck}),\ }\href
  {https://doi.org/10.1051/0004-6361/201833910} {\bibfield  {journal} {\bibinfo
   {journal} {\aap}\ }\textbf {\bibinfo {volume} {641}},\ \bibinfo {pages} {A6}
  (\bibinfo {year} {2020})},\ \Eprint {https://arxiv.org/abs/1807.06209}
  {arXiv:1807.06209 [astro-ph.CO]} \BibitemShut {NoStop}%
\bibitem [{\citenamefont {Pitrou}\ \emph {et~al.}(2021)\citenamefont {Pitrou},
  \citenamefont {Coc}, \citenamefont {Uzan},\ and\ \citenamefont
  {Vangioni}}]{Pitrou:2020etk}%
  \BibitemOpen
  \bibfield  {author} {\bibinfo {author} {\bibfnamefont {C.}~\bibnamefont
  {Pitrou}}, \bibinfo {author} {\bibfnamefont {A.}~\bibnamefont {Coc}},
  \bibinfo {author} {\bibfnamefont {J.-P.}\ \bibnamefont {Uzan}},\ and\
  \bibinfo {author} {\bibfnamefont {E.}~\bibnamefont {Vangioni}},\ }\href
  {https://doi.org/10.1093/mnras/stab135} {\bibfield  {journal} {\bibinfo
  {journal} {Mon. Not. Roy. Astron. Soc.}\ }\textbf {\bibinfo {volume} {502}},\
  \bibinfo {pages} {2474} (\bibinfo {year} {2021})},\ \Eprint
  {https://arxiv.org/abs/2011.11320} {arXiv:2011.11320 [astro-ph.CO]}
  \BibitemShut {NoStop}%
\bibitem [{\citenamefont {Peebles}(1968)}]{Peebles:1968ja}%
  \BibitemOpen
  \bibfield  {author} {\bibinfo {author} {\bibfnamefont {P.~J.~E.}\
  \bibnamefont {Peebles}},\ }\href {https://doi.org/10.1086/149628} {\bibfield
  {journal} {\bibinfo  {journal} {Astrophys. J.}\ }\textbf {\bibinfo {volume}
  {153}},\ \bibinfo {pages} {1} (\bibinfo {year} {1968})}\BibitemShut {NoStop}%
\bibitem [{\citenamefont {Zeldovich}\ \emph {et~al.}(1968)\citenamefont
  {Zeldovich}, \citenamefont {Kurt},\ and\ \citenamefont
  {Sunyaev}}]{Zeldovich:1968}%
  \BibitemOpen
  \bibfield  {author} {\bibinfo {author} {\bibfnamefont {Y.~B.}\ \bibnamefont
  {Zeldovich}}, \bibinfo {author} {\bibfnamefont {V.~G.}\ \bibnamefont
  {Kurt}},\ and\ \bibinfo {author} {\bibfnamefont {R.~A.}\ \bibnamefont
  {Sunyaev}},\ }\href@noop {} {\bibfield  {journal} {\bibinfo  {journal} {Zh.
  Eksp. Teor. Fiz.}\ }\textbf {\bibinfo {volume} {55}},\ \bibinfo {pages} {278}
  (\bibinfo {year} {1968})},\ \bibinfo {note} {[Sov. Phys. JETP \textbf{28},
  146 (1969)]}\BibitemShut {NoStop}%
\bibitem [{\citenamefont {Chluba}\ and\ \citenamefont
  {Thomas}(2011)}]{Chluba:2010ca}%
  \BibitemOpen
  \bibfield  {author} {\bibinfo {author} {\bibfnamefont {J.}~\bibnamefont
  {Chluba}}\ and\ \bibinfo {author} {\bibfnamefont {R.~M.}\ \bibnamefont
  {Thomas}},\ }\href {https://doi.org/10.1111/j.1365-2966.2010.17940.x}
  {\bibfield  {journal} {\bibinfo  {journal} {\mnras}\ }\textbf {\bibinfo
  {volume} {412}},\ \bibinfo {pages} {748} (\bibinfo {year} {2011})},\ \Eprint
  {https://arxiv.org/abs/1010.3631} {arXiv:1010.3631 [astro-ph.CO]}
  \BibitemShut {NoStop}%
%%CITATION = 1010.3631;%%
\bibitem [{\citenamefont {Ali-Haimoud}\ and\ \citenamefont
  {Hirata}(2010)}]{Ali-Haimoud:2010tlj}%
  \BibitemOpen
  \bibfield  {author} {\bibinfo {author} {\bibfnamefont {Y.}~\bibnamefont
  {Ali-Haimoud}}\ and\ \bibinfo {author} {\bibfnamefont {C.~M.}\ \bibnamefont
  {Hirata}},\ }\href {https://doi.org/10.1103/PhysRevD.82.063521} {\bibfield
  {journal} {\bibinfo  {journal} {Phys. Rev. D}\ }\textbf {\bibinfo {volume}
  {82}},\ \bibinfo {pages} {063521} (\bibinfo {year} {2010})},\ \Eprint
  {https://arxiv.org/abs/1006.1355} {arXiv:1006.1355 [astro-ph.CO]}
  \BibitemShut {NoStop}%
\bibitem [{\citenamefont {Ali-Haimoud}\ and\ \citenamefont
  {Hirata}(2011)}]{AliHaimoud:2010dx}%
  \BibitemOpen
  \bibfield  {author} {\bibinfo {author} {\bibfnamefont {Y.}~\bibnamefont
  {Ali-Haimoud}}\ and\ \bibinfo {author} {\bibfnamefont {C.~M.}\ \bibnamefont
  {Hirata}},\ }\href {https://doi.org/10.1103/PhysRevD.83.043513} {\bibfield
  {journal} {\bibinfo  {journal} {\prd}\ }\textbf {\bibinfo {volume} {83}},\
  \bibinfo {pages} {043513} (\bibinfo {year} {2011})},\ \Eprint
  {https://arxiv.org/abs/1011.3758} {arXiv:1011.3758 [astro-ph.CO]}
  \BibitemShut {NoStop}%
\bibitem [{\citenamefont {Lee}\ and\ \citenamefont
  {Ali-Ha\"\i{}moud}(2020)}]{Lee:2020obi}%
  \BibitemOpen
  \bibfield  {author} {\bibinfo {author} {\bibfnamefont {N.}~\bibnamefont
  {Lee}}\ and\ \bibinfo {author} {\bibfnamefont {Y.}~\bibnamefont
  {Ali-Ha\"\i{}moud}},\ }\href {https://doi.org/10.1103/PhysRevD.102.083517}
  {\bibfield  {journal} {\bibinfo  {journal} {Phys. Rev. D}\ }\textbf {\bibinfo
  {volume} {102}},\ \bibinfo {pages} {083517} (\bibinfo {year} {2020})},\
  \Eprint {https://arxiv.org/abs/2007.14114} {arXiv:2007.14114 [astro-ph.CO]}
  \BibitemShut {NoStop}%
\bibitem [{\citenamefont {Chluba}\ \emph {et~al.}(2012)\citenamefont {Chluba},
  \citenamefont {Fung},\ and\ \citenamefont {Switzer}}]{Chluba:2011hw}%
  \BibitemOpen
  \bibfield  {author} {\bibinfo {author} {\bibfnamefont {J.}~\bibnamefont
  {Chluba}}, \bibinfo {author} {\bibfnamefont {J.}~\bibnamefont {Fung}},\ and\
  \bibinfo {author} {\bibfnamefont {E.~R.}\ \bibnamefont {Switzer}},\ }\href
  {https://doi.org/10.1111/j.1365-2966.2012.21110.x} {\bibfield  {journal}
  {\bibinfo  {journal} {Mon. Not. Roy. Astron. Soc.}\ }\textbf {\bibinfo
  {volume} {423}},\ \bibinfo {pages} {3227} (\bibinfo {year} {2012})},\ \Eprint
  {https://arxiv.org/abs/1110.0247} {arXiv:1110.0247 [astro-ph.CO]}
  \BibitemShut {NoStop}%
\bibitem [{\citenamefont {Seager}\ \emph {et~al.}(1999)\citenamefont {Seager},
  \citenamefont {Sasselov},\ and\ \citenamefont {Scott}}]{Seager:1999bc}%
  \BibitemOpen
  \bibfield  {author} {\bibinfo {author} {\bibfnamefont {S.}~\bibnamefont
  {Seager}}, \bibinfo {author} {\bibfnamefont {D.~D.}\ \bibnamefont
  {Sasselov}},\ and\ \bibinfo {author} {\bibfnamefont {D.}~\bibnamefont
  {Scott}},\ }\href {https://doi.org/10.1086/312250} {\bibfield  {journal}
  {\bibinfo  {journal} {Astrophys. J. Lett.}\ }\textbf {\bibinfo {volume}
  {523}},\ \bibinfo {pages} {L1} (\bibinfo {year} {1999})},\ \Eprint
  {https://arxiv.org/abs/astro-ph/9909275} {arXiv:astro-ph/9909275 [astro-ph]}
  \BibitemShut {NoStop}%
\bibitem [{\citenamefont {Seager}\ \emph {et~al.}(2000)\citenamefont {Seager},
  \citenamefont {Sasselov},\ and\ \citenamefont {Scott}}]{Seager:1999km}%
  \BibitemOpen
  \bibfield  {author} {\bibinfo {author} {\bibfnamefont {S.}~\bibnamefont
  {Seager}}, \bibinfo {author} {\bibfnamefont {D.~D.}\ \bibnamefont
  {Sasselov}},\ and\ \bibinfo {author} {\bibfnamefont {D.}~\bibnamefont
  {Scott}},\ }\href@noop {} {\bibfield  {journal} {\bibinfo  {journal} {\apjs}\
  }\textbf {\bibinfo {volume} {128}},\ \bibinfo {pages} {407} (\bibinfo {year}
  {2000})},\ \Eprint {https://arxiv.org/abs/astro-ph/9912182}
  {astro-ph/9912182} \BibitemShut {NoStop}%
%%CITATION = ASTRO-PH 9912182;%%
\bibitem [{\citenamefont {Wong}\ \emph {et~al.}(2008)\citenamefont {Wong},
  \citenamefont {Moss},\ and\ \citenamefont {Scott}}]{Wong:2007ym}%
  \BibitemOpen
  \bibfield  {author} {\bibinfo {author} {\bibfnamefont {W.~Y.}\ \bibnamefont
  {Wong}}, \bibinfo {author} {\bibfnamefont {A.}~\bibnamefont {Moss}},\ and\
  \bibinfo {author} {\bibfnamefont {D.}~\bibnamefont {Scott}},\ }\href
  {https://doi.org/10.1111/j.1365-2966.2008.13092.x} {\bibfield  {journal}
  {\bibinfo  {journal} {\mnras}\ }\textbf {\bibinfo {volume} {386}},\ \bibinfo
  {pages} {1023} (\bibinfo {year} {2008})},\ \Eprint
  {https://arxiv.org/abs/0711.1357} {arXiv:0711.1357 [astro-ph]} \BibitemShut
  {NoStop}%
%%CITATION = ARXIV:0711.1357;%%
\bibitem [{\citenamefont {Shaw}\ and\ \citenamefont
  {Chluba}(2011)}]{Shaw:2011wq}%
  \BibitemOpen
  \bibfield  {author} {\bibinfo {author} {\bibfnamefont {J.~R.}\ \bibnamefont
  {Shaw}}\ and\ \bibinfo {author} {\bibfnamefont {J.}~\bibnamefont {Chluba}},\
  }\href {https://doi.org/10.1111/j.1365-2966.2011.18782.x} {\bibfield
  {journal} {\bibinfo  {journal} {Mon. Not. Roy. Astron. Soc.}\ }\textbf
  {\bibinfo {volume} {415}},\ \bibinfo {pages} {1343} (\bibinfo {year}
  {2011})},\ \Eprint {https://arxiv.org/abs/1102.3683} {arXiv:1102.3683
  [astro-ph.CO]} \BibitemShut {NoStop}%
\bibitem [{\citenamefont {Silk}(1968)}]{Silk:1968}%
  \BibitemOpen
  \bibfield  {author} {\bibinfo {author} {\bibfnamefont {J.}~\bibnamefont
  {Silk}},\ }\href {https://doi.org/10.1086/149449} {\bibfield  {journal}
  {\bibinfo  {journal} {Astrophys. J.}\ }\textbf {\bibinfo {volume} {151}},\
  \bibinfo {pages} {459} (\bibinfo {year} {1968})}\BibitemShut {NoStop}%
\bibitem [{\citenamefont {Rosenbrock}(1963)}]{Rosenbrock:1963}%
  \BibitemOpen
  \bibfield  {author} {\bibinfo {author} {\bibfnamefont {H.~H.}\ \bibnamefont
  {Rosenbrock}},\ }\href {https://doi.org/10.1093/comjnl/5.4.329} {\bibfield
  {journal} {\bibinfo  {journal} {The Computer Journal}\ }\textbf {\bibinfo
  {volume} {5}},\ \bibinfo {pages} {329} (\bibinfo {year} {1963})}\BibitemShut
  {NoStop}%
\bibitem [{\citenamefont {Verwer}\ \emph {et~al.}(1999)\citenamefont {Verwer},
  \citenamefont {Spee}, \citenamefont {Blom},\ and\ \citenamefont
  {Hundsdorfer}}]{Verwer:1999}%
  \BibitemOpen
  \bibfield  {author} {\bibinfo {author} {\bibfnamefont {J.~G.}\ \bibnamefont
  {Verwer}}, \bibinfo {author} {\bibfnamefont {E.~J.}\ \bibnamefont {Spee}},
  \bibinfo {author} {\bibfnamefont {J.~G.}\ \bibnamefont {Blom}},\ and\
  \bibinfo {author} {\bibfnamefont {W.}~\bibnamefont {Hundsdorfer}},\ }\href
  {https://doi.org/10.1137/S1064827597326651} {\bibfield  {journal} {\bibinfo
  {journal} {SIAM J. Sci. Comput.}\ }\textbf {\bibinfo {volume} {20}},\
  \bibinfo {pages} {1456} (\bibinfo {year} {1999})}\BibitemShut {NoStop}%
\bibitem [{\citenamefont {Mortonson}\ and\ \citenamefont
  {Hu}(2008)}]{Mortonson:2007hq}%
  \BibitemOpen
  \bibfield  {author} {\bibinfo {author} {\bibfnamefont {M.~J.}\ \bibnamefont
  {Mortonson}}\ and\ \bibinfo {author} {\bibfnamefont {W.}~\bibnamefont {Hu}},\
  }\href {https://doi.org/10.1086/523958} {\bibfield  {journal} {\bibinfo
  {journal} {Astrophys. J.}\ }\textbf {\bibinfo {volume} {672}},\ \bibinfo
  {pages} {737} (\bibinfo {year} {2008})},\ \Eprint
  {https://arxiv.org/abs/0705.1132} {arXiv:0705.1132 [astro-ph]} \BibitemShut
  {NoStop}%
\bibitem [{\citenamefont {Lewis}(2008)}]{Lewis:2008wr}%
  \BibitemOpen
  \bibfield  {author} {\bibinfo {author} {\bibfnamefont {A.}~\bibnamefont
  {Lewis}},\ }\href {https://doi.org/10.1103/PhysRevD.78.023002} {\bibfield
  {journal} {\bibinfo  {journal} {\prd}\ }\textbf {\bibinfo {volume} {78}},\
  \bibinfo {pages} {023002} (\bibinfo {year} {2008})},\ \Eprint
  {https://arxiv.org/abs/0804.3865} {arXiv:0804.3865 [astro-ph]} \BibitemShut
  {NoStop}%
%%CITATION = 0804.3865;%%
\bibitem [{\citenamefont {Bosman}\ \emph {et~al.}(2022)\citenamefont {Bosman}
  \emph {et~al.}}]{Bosman:2021obf}%
  \BibitemOpen
  \bibfield  {author} {\bibinfo {author} {\bibfnamefont {S.~E.~I.}\
  \bibnamefont {Bosman}} \emph {et~al.} (\bibinfo {collaboration} {XQR-30}),\
  }\href {https://doi.org/10.1093/mnras/stac563} {\bibfield  {journal}
  {\bibinfo  {journal} {Mon. Not. Roy. Astron. Soc.}\ }\textbf {\bibinfo
  {volume} {512}},\ \bibinfo {pages} {1046} (\bibinfo {year} {2022})},\ \Eprint
  {https://arxiv.org/abs/2108.03699} {arXiv:2108.03699 [astro-ph.CO]}
  \BibitemShut {NoStop}%
\bibitem [{\citenamefont {Fan}\ \emph {et~al.}(2006)\citenamefont {Fan},
  \citenamefont {Carilli},\ and\ \citenamefont {Keating}}]{Fan:2006dp}%
  \BibitemOpen
  \bibfield  {author} {\bibinfo {author} {\bibfnamefont {X.}~\bibnamefont
  {Fan}}, \bibinfo {author} {\bibfnamefont {C.~L.}\ \bibnamefont {Carilli}},\
  and\ \bibinfo {author} {\bibfnamefont {B.}~\bibnamefont {Keating}},\ }\href
  {https://doi.org/10.1146/annurev.astro.44.051905.092514} {\bibfield
  {journal} {\bibinfo  {journal} {Ann. Rev. Astron. Astrophys.}\ }\textbf
  {\bibinfo {volume} {44}},\ \bibinfo {pages} {415} (\bibinfo {year} {2006})},\
  \Eprint {https://arxiv.org/abs/astro-ph/0602375} {arXiv:astro-ph/0602375
  [astro-ph]} \BibitemShut {NoStop}%
\bibitem [{\citenamefont {Davies}\ \emph {et~al.}(2018)\citenamefont {Davies}
  \emph {et~al.}}]{Davies:2018yfp}%
  \BibitemOpen
  \bibfield  {author} {\bibinfo {author} {\bibfnamefont {F.~B.}\ \bibnamefont
  {Davies}} \emph {et~al.},\ }\href {https://doi.org/10.3847/1538-4357/aad6dc}
  {\bibfield  {journal} {\bibinfo  {journal} {Astrophys. J.}\ }\textbf
  {\bibinfo {volume} {864}},\ \bibinfo {pages} {142} (\bibinfo {year}
  {2018})},\ \Eprint {https://arxiv.org/abs/1802.06066} {arXiv:1802.06066
  [astro-ph.CO]} \BibitemShut {NoStop}%
\bibitem [{\citenamefont {Mason}\ \emph {et~al.}(2018)\citenamefont {Mason}
  \emph {et~al.}}]{Mason:2017eqr}%
  \BibitemOpen
  \bibfield  {author} {\bibinfo {author} {\bibfnamefont {C.~A.}\ \bibnamefont
  {Mason}} \emph {et~al.},\ }\href {https://doi.org/10.3847/1538-4357/aab0a7}
  {\bibfield  {journal} {\bibinfo  {journal} {Astrophys. J.}\ }\textbf
  {\bibinfo {volume} {856}},\ \bibinfo {pages} {2} (\bibinfo {year} {2018})},\
  \Eprint {https://arxiv.org/abs/1709.05356} {arXiv:1709.05356 [astro-ph.CO]}
  \BibitemShut {NoStop}%
\bibitem [{\citenamefont {Qin}\ \emph {et~al.}(2020)\citenamefont {Qin},
  \citenamefont {Poulin}, \citenamefont {Mesinger}, \citenamefont {Greig},
  \citenamefont {Murray},\ and\ \citenamefont {Park}}]{Qin:2020xrg}%
  \BibitemOpen
  \bibfield  {author} {\bibinfo {author} {\bibfnamefont {Y.}~\bibnamefont
  {Qin}}, \bibinfo {author} {\bibfnamefont {V.}~\bibnamefont {Poulin}},
  \bibinfo {author} {\bibfnamefont {A.}~\bibnamefont {Mesinger}}, \bibinfo
  {author} {\bibfnamefont {B.}~\bibnamefont {Greig}}, \bibinfo {author}
  {\bibfnamefont {S.}~\bibnamefont {Murray}},\ and\ \bibinfo {author}
  {\bibfnamefont {J.}~\bibnamefont {Park}},\ }\href
  {https://doi.org/10.1093/mnras/staa2797} {\bibfield  {journal} {\bibinfo
  {journal} {Mon. Not. Roy. Astron. Soc.}\ }\textbf {\bibinfo {volume} {499}},\
  \bibinfo {pages} {550} (\bibinfo {year} {2020})},\ \Eprint
  {https://arxiv.org/abs/2006.16828} {arXiv:2006.16828 [astro-ph.CO]}
  \BibitemShut {NoStop}%
\bibitem [{\citenamefont {Adam}\ \emph {et~al.}(2016)\citenamefont {Adam} \emph
  {et~al.}}]{Planck:2016mks}%
  \BibitemOpen
  \bibfield  {author} {\bibinfo {author} {\bibfnamefont {R.}~\bibnamefont
  {Adam}} \emph {et~al.} (\bibinfo {collaboration} {Planck}),\ }\href
  {https://doi.org/10.1051/0004-6361/201628897} {\bibfield  {journal} {\bibinfo
   {journal} {Astron. Astrophys.}\ }\textbf {\bibinfo {volume} {596}},\
  \bibinfo {pages} {A108} (\bibinfo {year} {2016})},\ \Eprint
  {https://arxiv.org/abs/1605.03507} {arXiv:1605.03507 [astro-ph.CO]}
  \BibitemShut {NoStop}%
\bibitem [{\citenamefont {Douspis}\ \emph {et~al.}(2015)\citenamefont
  {Douspis}, \citenamefont {Aghanim}, \citenamefont {Ili\'c},\ and\
  \citenamefont {Langer}}]{Douspis:2015nca}%
  \BibitemOpen
  \bibfield  {author} {\bibinfo {author} {\bibfnamefont {M.}~\bibnamefont
  {Douspis}}, \bibinfo {author} {\bibfnamefont {N.}~\bibnamefont {Aghanim}},
  \bibinfo {author} {\bibfnamefont {S.}~\bibnamefont {Ili\'c}},\ and\ \bibinfo
  {author} {\bibfnamefont {M.}~\bibnamefont {Langer}},\ }\href
  {https://doi.org/10.1051/0004-6361/201526543} {\bibfield  {journal} {\bibinfo
   {journal} {Astron. Astrophys.}\ }\textbf {\bibinfo {volume} {580}},\
  \bibinfo {pages} {L4} (\bibinfo {year} {2015})},\ \Eprint
  {https://arxiv.org/abs/1509.02785} {arXiv:1509.02785 [astro-ph.CO]}
  \BibitemShut {NoStop}%
\bibitem [{\citenamefont {Lewis}\ and\ \citenamefont
  {Challinor}(2002)}]{Lewis:2002nc}%
  \BibitemOpen
  \bibfield  {author} {\bibinfo {author} {\bibfnamefont {A.}~\bibnamefont
  {Lewis}}\ and\ \bibinfo {author} {\bibfnamefont {A.}~\bibnamefont
  {Challinor}},\ }\href@noop {} {\bibfield  {journal} {\bibinfo  {journal}
  {\prd}\ }\textbf {\bibinfo {volume} {66}},\ \bibinfo {pages} {023531}
  (\bibinfo {year} {2002})},\ \Eprint {https://arxiv.org/abs/astro-ph/0203507}
  {astro-ph/0203507} \BibitemShut {NoStop}%
%%CITATION = ASTRO-PH 0203507;%%
\bibitem [{\citenamefont {Lesgourgues}\ and\ \citenamefont
  {Tram}(2011)}]{Lesgourgues:2011rh}%
  \BibitemOpen
  \bibfield  {author} {\bibinfo {author} {\bibfnamefont {J.}~\bibnamefont
  {Lesgourgues}}\ and\ \bibinfo {author} {\bibfnamefont {T.}~\bibnamefont
  {Tram}},\ }\href {https://doi.org/10.1088/1475-7516/2011/09/032} {\bibfield
  {journal} {\bibinfo  {journal} {JCAP}\ }\textbf {\bibinfo {volume} {09}},\
  \bibinfo {pages} {032}},\ \Eprint {https://arxiv.org/abs/1104.2935}
  {arXiv:1104.2935 [astro-ph.CO]} \BibitemShut {NoStop}%
\bibitem [{\citenamefont {Hu}(2008)}]{Hu:2008zd}%
  \BibitemOpen
  \bibfield  {author} {\bibinfo {author} {\bibfnamefont {W.}~\bibnamefont
  {Hu}},\ }\href {https://doi.org/10.1103/PhysRevD.77.103524} {\bibfield
  {journal} {\bibinfo  {journal} {Phys. Rev. D}\ }\textbf {\bibinfo {volume}
  {77}},\ \bibinfo {pages} {103524} (\bibinfo {year} {2008})},\ \Eprint
  {https://arxiv.org/abs/0801.2433} {arXiv:0801.2433 [astro-ph]} \BibitemShut
  {NoStop}%
\bibitem [{\citenamefont {Fang}\ \emph {et~al.}(2008)\citenamefont {Fang},
  \citenamefont {Hu},\ and\ \citenamefont {Lewis}}]{Fang:2008sn}%
  \BibitemOpen
  \bibfield  {author} {\bibinfo {author} {\bibfnamefont {W.}~\bibnamefont
  {Fang}}, \bibinfo {author} {\bibfnamefont {W.}~\bibnamefont {Hu}},\ and\
  \bibinfo {author} {\bibfnamefont {A.}~\bibnamefont {Lewis}},\ }\href
  {https://doi.org/10.1103/PhysRevD.78.087303} {\bibfield  {journal} {\bibinfo
  {journal} {Phys. Rev. D}\ }\textbf {\bibinfo {volume} {78}},\ \bibinfo
  {pages} {087303} (\bibinfo {year} {2008})},\ \Eprint
  {https://arxiv.org/abs/0808.3125} {arXiv:0808.3125 [astro-ph]} \BibitemShut
  {NoStop}%
\bibitem [{\citenamefont {Poulin}\ \emph {et~al.}(2018)\citenamefont {Poulin},
  \citenamefont {Smith}, \citenamefont {Grin}, \citenamefont {Karwal},\ and\
  \citenamefont {Kamionkowski}}]{Poulin:2018dzj}%
  \BibitemOpen
  \bibfield  {author} {\bibinfo {author} {\bibfnamefont {V.}~\bibnamefont
  {Poulin}}, \bibinfo {author} {\bibfnamefont {T.~L.}\ \bibnamefont {Smith}},
  \bibinfo {author} {\bibfnamefont {D.}~\bibnamefont {Grin}}, \bibinfo {author}
  {\bibfnamefont {T.}~\bibnamefont {Karwal}},\ and\ \bibinfo {author}
  {\bibfnamefont {M.}~\bibnamefont {Kamionkowski}},\ }\href
  {https://doi.org/10.1103/PhysRevD.98.083525} {\bibfield  {journal} {\bibinfo
  {journal} {Phys. Rev. D}\ }\textbf {\bibinfo {volume} {98}},\ \bibinfo
  {pages} {083525} (\bibinfo {year} {2018})},\ \Eprint
  {https://arxiv.org/abs/1806.10608} {arXiv:1806.10608 [astro-ph.CO]}
  \BibitemShut {NoStop}%
\bibitem [{\citenamefont {Smith}\ \emph {et~al.}(2020)\citenamefont {Smith},
  \citenamefont {Poulin},\ and\ \citenamefont {Amin}}]{Smith:2019ihp}%
  \BibitemOpen
  \bibfield  {author} {\bibinfo {author} {\bibfnamefont {T.~L.}\ \bibnamefont
  {Smith}}, \bibinfo {author} {\bibfnamefont {V.}~\bibnamefont {Poulin}},\ and\
  \bibinfo {author} {\bibfnamefont {M.~A.}\ \bibnamefont {Amin}},\ }\href
  {https://doi.org/10.1103/PhysRevD.101.063523} {\bibfield  {journal} {\bibinfo
   {journal} {Phys. Rev. D}\ }\textbf {\bibinfo {volume} {101}},\ \bibinfo
  {pages} {063523} (\bibinfo {year} {2020})},\ \Eprint
  {https://arxiv.org/abs/1908.06995} {arXiv:1908.06995 [astro-ph.CO]}
  \BibitemShut {NoStop}%
\bibitem [{\citenamefont {Challinor}\ and\ \citenamefont
  {Lewis}(2005)}]{Challinor:2005jy}%
  \BibitemOpen
  \bibfield  {author} {\bibinfo {author} {\bibfnamefont {A.}~\bibnamefont
  {Challinor}}\ and\ \bibinfo {author} {\bibfnamefont {A.}~\bibnamefont
  {Lewis}},\ }\href {https://doi.org/10.1103/PhysRevD.71.103010} {\bibfield
  {journal} {\bibinfo  {journal} {\prd}\ }\textbf {\bibinfo {volume} {71}},\
  \bibinfo {pages} {103010} (\bibinfo {year} {2005})},\ \Eprint
  {https://arxiv.org/abs/astro-ph/0502425} {arXiv:astro-ph/0502425 [astro-ph]}
  \BibitemShut {NoStop}%
%%CITATION = ASTRO-PH/0502425;%%
\bibitem [{\citenamefont {Hadzhiyska}\ \emph {et~al.}(2018)\citenamefont
  {Hadzhiyska}, \citenamefont {Spergel},\ and\ \citenamefont
  {Dunkley}}]{Hadzhiyska:2017nqe}%
  \BibitemOpen
  \bibfield  {author} {\bibinfo {author} {\bibfnamefont {B.}~\bibnamefont
  {Hadzhiyska}}, \bibinfo {author} {\bibfnamefont {D.}~\bibnamefont
  {Spergel}},\ and\ \bibinfo {author} {\bibfnamefont {J.}~\bibnamefont
  {Dunkley}},\ }\href {https://doi.org/10.1103/PhysRevD.97.043521} {\bibfield
  {journal} {\bibinfo  {journal} {Phys. Rev. D}\ }\textbf {\bibinfo {volume}
  {97}},\ \bibinfo {pages} {043521} (\bibinfo {year} {2018})},\ \Eprint
  {https://arxiv.org/abs/1711.03168} {arXiv:1711.03168 [astro-ph.CO]}
  \BibitemShut {NoStop}%
\bibitem [{\citenamefont {Lewis}\ \emph {et~al.}(2017)\citenamefont {Lewis},
  \citenamefont {Hall},\ and\ \citenamefont {Challinor}}]{Lewis:2017ans}%
  \BibitemOpen
  \bibfield  {author} {\bibinfo {author} {\bibfnamefont {A.}~\bibnamefont
  {Lewis}}, \bibinfo {author} {\bibfnamefont {A.}~\bibnamefont {Hall}},\ and\
  \bibinfo {author} {\bibfnamefont {A.}~\bibnamefont {Challinor}},\ }\href
  {https://doi.org/10.1088/1475-7516/2017/08/023} {\bibfield  {journal}
  {\bibinfo  {journal} {JCAP}\ }\textbf {\bibinfo {volume} {08}},\ \bibinfo
  {pages} {023}},\ \Eprint {https://arxiv.org/abs/1706.02673} {arXiv:1706.02673
  [astro-ph.CO]} \BibitemShut {NoStop}%
\bibitem [{\citenamefont {Pratten}\ and\ \citenamefont
  {Lewis}(2016)}]{Pratten:2016dsm}%
  \BibitemOpen
  \bibfield  {author} {\bibinfo {author} {\bibfnamefont {G.}~\bibnamefont
  {Pratten}}\ and\ \bibinfo {author} {\bibfnamefont {A.}~\bibnamefont
  {Lewis}},\ }\href {https://doi.org/10.1088/1475-7516/2016/08/047} {\bibfield
  {journal} {\bibinfo  {journal} {JCAP}\ }\textbf {\bibinfo {volume} {08}},\
  \bibinfo {pages} {047}},\ \Eprint {https://arxiv.org/abs/1605.05662}
  {arXiv:1605.05662 [astro-ph.CO]} \BibitemShut {NoStop}%
\bibitem [{\citenamefont {Lewis}\ \emph {et~al.}(2011)\citenamefont {Lewis},
  \citenamefont {Challinor},\ and\ \citenamefont {Hanson}}]{Lewis:2011fk}%
  \BibitemOpen
  \bibfield  {author} {\bibinfo {author} {\bibfnamefont {A.}~\bibnamefont
  {Lewis}}, \bibinfo {author} {\bibfnamefont {A.}~\bibnamefont {Challinor}},\
  and\ \bibinfo {author} {\bibfnamefont {D.}~\bibnamefont {Hanson}},\ }\href
  {https://doi.org/10.1088/1475-7516/2011/03/018} {\bibfield  {journal}
  {\bibinfo  {journal} {\jcap}\ }\textbf {\bibinfo {volume} {1103}},\ \bibinfo
  {pages} {018} (\bibinfo {year} {2011})},\ \Eprint
  {https://arxiv.org/abs/1101.2234} {arXiv:1101.2234 [astro-ph.CO]}
  \BibitemShut {NoStop}%
%%CITATION = 1101.2234;%%
\bibitem [{\citenamefont {Takahashi}\ \emph {et~al.}(2012)\citenamefont
  {Takahashi}, \citenamefont {Sato}, \citenamefont {Nishimichi}, \citenamefont
  {Taruya},\ and\ \citenamefont {Oguri}}]{Takahashi:2012em}%
  \BibitemOpen
  \bibfield  {author} {\bibinfo {author} {\bibfnamefont {R.}~\bibnamefont
  {Takahashi}}, \bibinfo {author} {\bibfnamefont {M.}~\bibnamefont {Sato}},
  \bibinfo {author} {\bibfnamefont {T.}~\bibnamefont {Nishimichi}}, \bibinfo
  {author} {\bibfnamefont {A.}~\bibnamefont {Taruya}},\ and\ \bibinfo {author}
  {\bibfnamefont {M.}~\bibnamefont {Oguri}},\ }\href
  {https://doi.org/10.1088/0004-637X/761/2/152} {\bibfield  {journal} {\bibinfo
   {journal} {\apj}\ }\textbf {\bibinfo {volume} {761}},\ \bibinfo {pages}
  {152} (\bibinfo {year} {2012})},\ \Eprint {https://arxiv.org/abs/1208.2701}
  {arXiv:1208.2701 [astro-ph.CO]} \BibitemShut {NoStop}%
%%CITATION = ARXIV:1208.2701;%%
\bibitem [{\citenamefont {Mead}\ \emph {et~al.}(2021)\citenamefont {Mead},
  \citenamefont {Brieden}, \citenamefont {Tr{\"o}ster},\ and\ \citenamefont
  {Heymans}}]{Mead:2020vgs}%
  \BibitemOpen
  \bibfield  {author} {\bibinfo {author} {\bibfnamefont {A.~J.}\ \bibnamefont
  {Mead}}, \bibinfo {author} {\bibfnamefont {S.}~\bibnamefont {Brieden}},
  \bibinfo {author} {\bibfnamefont {T.}~\bibnamefont {Tr{\"o}ster}},\ and\
  \bibinfo {author} {\bibfnamefont {C.}~\bibnamefont {Heymans}},\ }\href
  {https://doi.org/10.1093/mnras/stab082} {\bibfield  {journal} {\bibinfo
  {journal} {Mon. Not. Roy. Astron. Soc.}\ }\textbf {\bibinfo {volume} {502}},\
  \bibinfo {pages} {1401} (\bibinfo {year} {2021})},\ \Eprint
  {https://arxiv.org/abs/2009.01858} {arXiv:2009.01858 [astro-ph.CO]}
  \BibitemShut {NoStop}%
\bibitem [{\citenamefont {Salcido}\ \emph {et~al.}(2023)\citenamefont
  {Salcido}, \citenamefont {McCarthy}, \citenamefont {Kwan}, \citenamefont
  {Upadhye},\ and\ \citenamefont {Font}}]{Salcido:2023etq}%
  \BibitemOpen
  \bibfield  {author} {\bibinfo {author} {\bibfnamefont {J.}~\bibnamefont
  {Salcido}}, \bibinfo {author} {\bibfnamefont {I.~G.}\ \bibnamefont
  {McCarthy}}, \bibinfo {author} {\bibfnamefont {J.}~\bibnamefont {Kwan}},
  \bibinfo {author} {\bibfnamefont {A.}~\bibnamefont {Upadhye}},\ and\ \bibinfo
  {author} {\bibfnamefont {A.~S.}\ \bibnamefont {Font}},\ }\href
  {https://doi.org/10.1093/mnras/stad1474} {\bibfield  {journal} {\bibinfo
  {journal} {Mon. Not. Roy. Astron. Soc.}\ }\textbf {\bibinfo {volume} {523}},\
  \bibinfo {pages} {2247} (\bibinfo {year} {2023})},\ \Eprint
  {https://arxiv.org/abs/2305.09710} {arXiv:2305.09710 [astro-ph.CO]}
  \BibitemShut {NoStop}%
\bibitem [{\citenamefont {Lesgourgues}(2011)}]{Lesgourgues:2011rg}%
  \BibitemOpen
  \bibfield  {author} {\bibinfo {author} {\bibfnamefont {J.}~\bibnamefont
  {Lesgourgues}},\ }\href@noop {} {\bibinfo {title} {{The Cosmic Linear
  Anisotropy Solving System (CLASS) III: Comparision with CAMB for LambdaCDM}}}
  (\bibinfo {year} {2011}),\ \Eprint {https://arxiv.org/abs/1104.2934}
  {arXiv:1104.2934 [astro-ph.CO]} \BibitemShut {NoStop}%
\bibitem [{\citenamefont {McCarthy}\ \emph {et~al.}(2022)\citenamefont
  {McCarthy}, \citenamefont {Hill},\ and\ \citenamefont
  {Madhavacheril}}]{McCarthy:2021lfp}%
  \BibitemOpen
  \bibfield  {author} {\bibinfo {author} {\bibfnamefont {F.}~\bibnamefont
  {McCarthy}}, \bibinfo {author} {\bibfnamefont {J.~C.}\ \bibnamefont {Hill}},\
  and\ \bibinfo {author} {\bibfnamefont {M.~S.}\ \bibnamefont
  {Madhavacheril}},\ }\href {https://doi.org/10.1103/PhysRevD.105.023517}
  {\bibfield  {journal} {\bibinfo  {journal} {Phys. Rev. D}\ }\textbf {\bibinfo
  {volume} {105}},\ \bibinfo {pages} {023517} (\bibinfo {year} {2022})},\
  \Eprint {https://arxiv.org/abs/2103.05582} {arXiv:2103.05582 [astro-ph.CO]}
  \BibitemShut {NoStop}%
\bibitem [{\citenamefont {Casas}\ \emph {et~al.}(2024)\citenamefont {Casas}
  \emph {et~al.}}]{Euclid:2023pxu}%
  \BibitemOpen
  \bibfield  {author} {\bibinfo {author} {\bibfnamefont {S.}~\bibnamefont
  {Casas}} \emph {et~al.} (\bibinfo {collaboration} {Euclid}),\ }\href
  {https://doi.org/10.1051/0004-6361/202346772} {\bibfield  {journal} {\bibinfo
   {journal} {Astron. Astrophys.}\ }\textbf {\bibinfo {volume} {682}},\
  \bibinfo {pages} {A90} (\bibinfo {year} {2024})},\ \Eprint
  {https://arxiv.org/abs/2303.09451} {arXiv:2303.09451 [astro-ph.CO]}
  \BibitemShut {NoStop}%
\bibitem [{\citenamefont {Bolliet}\ \emph {et~al.}(2024)\citenamefont
  {Bolliet}, \citenamefont {Spurio~Mancini}, \citenamefont {Hill},
  \citenamefont {Madhavacheril}, \citenamefont {Jense}, \citenamefont
  {Calabrese},\ and\ \citenamefont {Dunkley}}]{Bolliet:2023sst}%
  \BibitemOpen
  \bibfield  {author} {\bibinfo {author} {\bibfnamefont {B.}~\bibnamefont
  {Bolliet}}, \bibinfo {author} {\bibfnamefont {A.}~\bibnamefont
  {Spurio~Mancini}}, \bibinfo {author} {\bibfnamefont {J.~C.}\ \bibnamefont
  {Hill}}, \bibinfo {author} {\bibfnamefont {M.}~\bibnamefont {Madhavacheril}},
  \bibinfo {author} {\bibfnamefont {H.~T.}\ \bibnamefont {Jense}}, \bibinfo
  {author} {\bibfnamefont {E.}~\bibnamefont {Calabrese}},\ and\ \bibinfo
  {author} {\bibfnamefont {J.}~\bibnamefont {Dunkley}},\ }\href
  {https://doi.org/10.1093/mnras/stae1201} {\bibfield  {journal} {\bibinfo
  {journal} {Mon. Not. Roy. Astron. Soc.}\ }\textbf {\bibinfo {volume} {531}},\
  \bibinfo {pages} {1351} (\bibinfo {year} {2024})},\ \Eprint
  {https://arxiv.org/abs/2303.01591} {arXiv:2303.01591 [astro-ph.CO]}
  \BibitemShut {NoStop}%
\bibitem [{\citenamefont
  {Lewis}(2026{\natexlab{b}})}]{cambhypersphericalbessel}%
  \BibitemOpen
  \bibfield  {author} {\bibinfo {author} {\bibfnamefont {A.}~\bibnamefont
  {Lewis}}} (\bibinfo {year} {2026}{\natexlab{b}}),\ \bibinfo {note}
  {\url{https://github.com/cmbant/CAMB/blob/master/docs/changelog/hyperspherical_bessels.pdf}}\BibitemShut
  {NoStop}%
\bibitem [{\citenamefont {Abitbol}\ \emph {et~al.}(2025)\citenamefont {Abitbol}
  \emph {et~al.}}]{SOEnhancedLAT:2025}%
  \BibitemOpen
  \bibfield  {author} {\bibinfo {author} {\bibfnamefont {M.}~\bibnamefont
  {Abitbol}} \emph {et~al.} (\bibinfo {collaboration} {Simons Observatory}),\
  }\href@noop {} {\bibfield  {journal} {\bibinfo  {journal} {JCAP}\ }\textbf
  {\bibinfo {volume} {07}},\ \bibinfo {pages} {034}},\ \Eprint
  {https://arxiv.org/abs/2503.00636} {arXiv:2503.00636 [astro-ph.CO]}
  \BibitemShut {NoStop}%
\end{thebibliography}

\end{document}